\documentclass[sigconf]{acmart}

\usepackage[utf8]{inputenc}

\usepackage{float}

\usepackage{listings}
\usepackage{color}
\definecolor{keywordcolor}{rgb}{0.56, 0.13, 0.00}
\definecolor{ndkeywordcolor}{rgb}{0.05, 0.46, 0.17}
\definecolor{commentcolor}{rgb}{0.41, 0.64, 0.70}
\definecolor{stringcolor}{rgb}{0.25, 0.44, 0.63}

\lstdefinelanguage{TypeScript}{
  keywords={typeof, new, true, false, catch, function, return, null, catch, switch, var, if, in, while, do, else, case, break, boolean},
  morekeywords={[2]{class, export, throw, implements, import, this}},
  identifierstyle=\color{black},
  sensitive=false,
  comment=[l]{//},
  morecomment=[s]{/*}{*/},
  commentstyle=\color{commentcolor}\ttfamily,
  stringstyle=\color{stringcolor}\ttfamily,
  morestring=[b]',
  morestring=[b]"
}

\usepackage{multirow}

\usepackage{enumitem}

\newcommand{\mycircle}[1]{\raisebox{.5pt}{\textcircled{\raisebox{-.9pt} {#1}}}}

\copyrightyear{2022} 
\acmYear{2022} 
\setcopyright{acmlicensed}
\acmConference[CHI '22]{CHI Conference on Human Factors in Computing Systems}{April 29-May 5, 2022}{New Orleans, LA, USA}
\acmBooktitle{CHI Conference on Human Factors in Computing Systems (CHI '22), April 29-May 5, 2022, New Orleans, LA, USA}
\acmPrice{15.00}
\acmDOI{10.1145/3491102.3517612}
\acmISBN{978-1-4503-9157-3/22/04}

\acmSubmissionID{9037}

\begin{document}

\title[]{OneLabeler: A Flexible System for Building Data Labeling Tools}

\author{Yu Zhang}
\email{yu.zhang@cs.ox.ac.uk}
\orcid{0000-0002-9035-0463}
\affiliation{
  \institution{University of Oxford}
  \city{Oxford}
  \postcode{OX1 3QD}
  \country{The United Kingdom}
}

\author{Yun Wang}
\email{wangyun@microsoft.com}
\orcid{0000-0003-0468-4043}
\affiliation{
  \institution{Microsoft Research Asia}
  \city{Beijing}
  \postcode{100080}
  \country{China}
}

\author{Haidong Zhang}
\email{haidong.zhang@microsoft}
\affiliation{
  \institution{Microsoft Research Asia}
  \city{Beijing}
  \postcode{100080}
  \country{China}
}

\author{Bin Zhu}
\email{binzhu@microsoft.com}
\orcid{0000-0002-3571-7808}
\affiliation{
  \institution{Microsoft Research Asia}
  \city{Beijing}
  \postcode{100080}
  \country{China}
}

\author{Siming Chen}
\email{simingchen@fudan.edu.cn}
\orcid{0000-0002-2690-3588}
\affiliation{
  \institution{Fudan University}
  \city{Shanghai}
  \postcode{200433}
  \country{China}
}

\author{Dongmei Zhang}
\email{dongmeiz@microsoft.com}
\affiliation{
  \institution{Microsoft Research Asia}
  \city{Beijing}
  \postcode{100080}
  \country{China}
}

\renewcommand{\shortauthors}{Zhang et al.}

\begin{abstract}
  Labeled datasets are essential for supervised machine learning.
  Various data labeling tools have been built to collect labels in different usage scenarios.
  However, developing labeling tools is time-consuming, costly, and expertise-demanding on software development.
  In this paper, we propose a conceptual framework for data labeling and OneLabeler based on the conceptual framework to support easy building of labeling tools for diverse usage scenarios.
  The framework consists of common modules and states in labeling tools summarized through coding of existing tools.
  OneLabeler supports configuration and composition of common software modules through visual programming to build data labeling tools.
  A module can be a human, machine, or mixed computation procedure in data labeling.
  We demonstrate the expressiveness and utility of the system through ten example labeling tools built with OneLabeler.
  A user study with developers provides evidence that OneLabeler supports efficient building of diverse data labeling tools.
\end{abstract}

\ccsdesc[500]{Human-centered computing~Interactive systems and tools}
\ccsdesc[500]{Information systems~Multimedia information systems}

\keywords{data labeling, framework, toolkit, interactive machine learning, visual programming}

\maketitle

\section{Introduction}
\label{sec:introduction}

Labeled datasets play an essential role in supervised machine learning.
Modern neural networks are generally trained with large labeled datasets.
Data labeling typically involves human annotators.
Many data labeling tools have been built to enable annotators to label data objects of various types for various learning tasks (e.g., classification~\cite{Paiva2015Approach,Kucher2017Active}, segmentation~\cite{Andriluka2018Fluid}) in different application domains (e.g., image~\cite{Cui2007EasyAlbum,Suh2007Semi}, text~\cite{Settles2011Closing,Choi2019AILA}, video~\cite{Rooij2010MediaTable,Liao2016Visualization}).
These tools are generally built for a specific labeling task.
Different types of labeling tasks require individually built labeling tools.

Building customized labeling tools for various applications is challenging.
Programming a labeling tool generally requires cross-disciplinary knowledge on interaction techniques and visual design, algorithmic techniques such as active learning and semi-automatic labeling, and software development skills to implement and compose all these modules into a labeling tool.
It is time-consuming and costly.
Although many labeling tools have been built, most of them are monolithic applications with limited usage scenarios and are not designed to support further editing and customization.
Adapting an existing tool to fulfill a new labeling task with new requirements is generally difficult.

Meanwhile, labeling tools for different labeling tasks share commonalities.
To alleviate the burden of building labeling tools, we advocate a modular composable design that extracts the commonalities among different labeling tools to enable easy customization and extension.
In this paper, we present a conceptual data labeling framework with a modular composable design.
We have identified eight types of common modules through inductive coding of modules in existing data labeling tools.
The framework consists of common conceptual modules and constraints in composing these modules.
In the framework, a data labeling tool is modeled as a graph denoting its workflow.
In the graph, modules are nodes, and the execution order of the modules is encoded with directed edges.

Based on the framework, we develop OneLabeler, a system for building data labeling tools.
OneLabeler is designed with reusability and flexibility in mind.
It enables visual programming to compose and configure software modules.
Developers can build a data labeling tool by creating a workflow graph using built-in implementations of the conceptual modules.
In a created labeling tool, interface modules and algorithm modules can be instantiated as human, machine, or mixed computation procedures.
The constraints on composing modules are embedded in a static program checker to check the workflow graph and verify the feasibility of the created labeling tool.
A labeling tool built with OneLabeler can be exported as an installer of the tool for sharing.

To demonstrate the expressiveness of the proposed framework and system, we present a case study of ten tools built with OneLabeler.
These tools cover various usage scenarios, including a set of typical labeling tools for different data types and labeling tasks, a classification tool for a customized webpage data type, a machine-aided multi-task text labeling tool, a mixed-initiative image classification tool, and prototyping an interactive machine learning system for chart image reverse engineering.
To examine OneLabeler's usability, we conduct a user study in which developers are asked to accomplish four tasks on building data labeling tools with OneLabeler.
The study results indicate that OneLabeler is easy to learn and use, enabling developers to efficiently build different data labeling tools.

This paper has the following major contributions:
\begin{itemize}[leftmargin=*]
    \item We propose a conceptual framework that illuminates common conceptual modules and composition constraints in data labeling.
    \item We develop the OneLabeler system based on the framework to support easy building of diverse data labeling tools.
    \item We conduct an extensive case study to demonstrate the capability of OneLabeler in creating diverse data labeling tools, as well as a user study to validate the usability of OneLabeler.
\end{itemize}

OneLabeler's source code can be found at \url{https://github.com/microsoft/OneLabeler}.

\section{Related Work}
\label{sec:related-work}

\subsection{Data Labeling Tools}

Due to the importance of labeled data for supervised machine learning, various labeling tools have been proposed for different applications.
For example, LabelMe~\cite{Russell2008LabelMe} is a labeling tool for image object detection, wherein an annotator can label objects through bounding boxes and polygons.
VoTT~\cite{Microsoft2017VoTT} supports bounding box and polygon annotation in images and video frames.
Labelbox~\cite{L2018Labelbox} enables annotation for classification, segmentation, and object detection in images and videos.
VIA~\cite{Dutta2019VIA} allows annotators to label spatial regions and temporal segments in images, audios, and videos.
These tools require intensive human labeling efforts.

Various techniques focus on reducing human efforts, typically through integration of machine assistance for semi-automatic labeling.
ISSE~\cite{Bryan2014ISSE} algorithmically suggests refined segmentation to help annotators separate sound into its respective sources by painting on time-frequency visualizations.
Fluid Annotation~\cite{Andriluka2018Fluid} uses a pre-trained model to propose a set of possibly-overlapping segments to help annotate image segmentation.
V-Awake~\cite{GarciaCaballero2019V} focuses on time series segmentation.
It uses LSTM to assign tentative labels and visualizes the model information to help annotators diagnose and correct model predictions.

Another family of techniques commonly used in labeling tools is active learning that focuses on prioritizing user efforts.
It has been used for labeling documents~\cite{Settles2011Closing,Kucher2017Active} and geometric objects~\cite{Zhang2021MI3}.
In a similar vein, Deng et al. propose a strategy to select data objects to label that maximize annotation's utility-to-cost ratio in multi-label classification~\cite{Deng2014Scalable}.
While active learning focuses on algorithmic selection strategies, the selection may also involve user feedback~\cite{Paiva2015Approach}.
For example, projection scatterplots with iso-contours visualizing label uncertainty have been used to assist users in data selection~\cite{Liao2016Visualization}.

Aside from semi-automatic labeling and active learning addressing algorithmic aspects, novel interaction and visual design are proposed to support efficient labeling.
Clustering is frequently used in data labeling tools~\cite{Tang2013Towards,Cui2007EasyAlbum}.
Similar data objects that likely share the same label can be grouped to batch their labeling~\cite{Suh2007Semi}.
MediaTable~\cite{Rooij2010MediaTable} combines automatic content analysis and a tabular interface providing \textit{focus+context} in image and video labeling.
Choi et al.~\cite{Choi2019AILA} highlight keywords that imply document sentiment, identified by the attention mechanism of LSTM, to aid annotators.

Another branch of study focuses on designing tasks assigned to annotators.
A prominent example is gaming with a purpose (GWAP)~\cite{Ahn2008Designing} in which annotators are instructed to play games to implicitly contribute data labels.
Techniques in this category, such as ESP game~\cite{Ahn2004Labeling} and Peekaboom~\cite{Ahn2006Peekaboom}, focus on the gameplay design that strives to motivate annotators.
Another example is embedding labeling tasks in human/bot verification~\cite{Ahn2008reCAPTCHA,Yang2008Comprehensive}.
A large body of work on crowdsourcing annotations relates to task design, which typically focuses on task assignment schemes to collect high-quality labels with low cost and latency~\cite{Chai2020Crowdsourcing}.

Despite the diversity in labeling tools, common themes exist, such as using semi-automatic labeling and active learning to save annotation efforts.
These commonalities lay the foundation for our efforts described in Section~\ref{sec:framework} to summarize shared conceptual modules across data labeling tools.

\subsection{Workflows in Data Labeling}
\label{sec:related-workflows}

Research efforts have also been directed to design patterns for labeling tools, typically in the form of generic workflows composed of conceptual modules.
Settles' active learning survey describes a typical pool-based active learning cycle with algorithmic query selection, human annotation, and model training~\cite{Settles2009Active}.
Wang and Hua survey the use of active learning in multimedia data annotation and retrieval, and summarize three schemes: conventional active learning, multiple-instance active learning, and large-scale interactive annotation~\cite{Wang2011Active}.
H{\"{o}}ferlin et al. introduce a workflow that extends the conventional active learning workflow with user selection and model manipulation~\cite{Hoeferlin2012Inter}.
Bernard et al.'s workflow integrates interactive visualization components~\cite{Bernard2018VIAL}.
Zhang et al. propose a workflow that features algorithmic sampling and default labeling~\cite{Zhang2021MI3}.

Similar to the literature on generic workflows, we aim to summarize design patterns in labeling tools.
Instead of a single workflow, our solution is a generative framework for building workflows.
We believe that a single workflow cannot be optimal for all usage scenarios.
Meanwhile, our framework builds on existing workflows, as they are included in the corpus for summarizing common modules through coding (in Section~\ref{sec:framework}).

Our work falls into the family of toolkit research in HCI~\cite{Ledo2018Evaluation} that contributes techniques for building new tools.
Examples in this body of research related to our work are the ones that assist the development of visual interfaces~\cite{Bostock2011D3,Ren2014iVisDesigner,Mendez2016iVoLVER}.
Meanwhile, to the best of our knowledge, there exists no literature on extensible systems for building data labeling tools.
A relevant thread of work that also aims to save the cost of developing labeling tools is data labeling platforms and systems that support multiple labeling applications.
Examples include the project templates of Amazon Mechanical Turk~\cite{Amazon2005Amazon} and Prodigy~\cite{Explosion2017Prodigy}.
While the template-based approach can address a number of common use cases, extending to new usage scenarios requires implementing new templates from scratch, and reuse of software modules is not supported.
LabelStudio~\cite{Tkachenko2020Label} also provides a collection of templates but with more flexibility.
While it allows the user to use active learning and pre-labeling in the labeling interface, customizing interface modules in LabelStudio is difficult.
OneLabeler also aims to promote reuse as these systems but focuses on reuse at the module level instead of template level.
Additionally, building on the conceptual framework, OneLabeler supports a more extensive coverage of data labeling techniques.

\section{Design Requirements}
\label{sec:requirement}

Various scenarios require deploying different labeling tools to collect labels, but developing an effective labeling tool is generally challenging and costly.
We see an opportunity to alleviate developers' efforts through modular composable design for reuse, as there are commonalities in labeling tools.
We aim to develop a flexible system to enable easy building of various data labeling tools.
To better understand pain points in developing labeling tools, we interviewed three experts with expertise in developing labeling tools as an informal pilot study.

\subsection{Pilot Study}

The three experts we interviewed have developed labeling tools for a company-owned research institute, two with the machine learning (ML) background (E1, E2) and one with the HCI background (E3).
They have 8 to 12 years of programming experience.
E1 has developed labeling tools for image object detection and binary classification.
E2 has developed text labeling tools for relation detection between tables and documents.
E3 has developed an image labeling tool for multi-label classification.
The study started with a structured interview with questions concerning participants' experience of developing labeling tools.
Example questions we asked include ``Why did you decide to implement a data labeling software instead of using existing ones?'' and ``What functionalities do you think are critical for data labeling software?''
Then, we conducted a walkthrough demonstration of an early prototype of OneLabeler with rudimentary functionalities.

Through the interview, we observed that no participant ever used public labeling software in their projects due to their application-specific requirements.
E1 mentioned that he needed to integrate a customized rule to forbid labeling the same data object twice in an image, which could not be achieved in existing tools.
E2 mentioned that the text datasets for the labeling tools he developed were stored in different types of structured data (e.g., document, table, webpage) and required different types of structured labeling.
We conclude that customized labeling tools are needed to address application-specific requirements.

Developing a data labeling tool is time-consuming.
It takes 7 to 30 man-days for an experienced engineer to develop one according to the participants' experience.
Regarding the three labeling tools he had developed, E2 commented:
\textit{``... the interface modules for displaying data points are similar.
They all display the table to be annotated, and the supported interactions for annotating the table are similar.''}
This comment implies that labeling tools share commonalities, but facilitating reuse requires a careful software design.
Like other software development, iteration is needed for labeling tools.
E1 commented: \textit{``The customized rules typically need to go through many iterations as the ML project progresses.''}
Similarly, E3 said, \textit{``I didn't know the label categories at the beginning and was not sure about the labeling task that I needed to carry out. Thus, I need to be able to refine the data labeling interface correspondingly as I gradually get more annotations.''}
Finishing a functional labeling tool is not the end of development, as it frequently needs to undergo further editing.
We conclude that building labeling tools is time-consuming and needs iterations.
Meanwhile, there is an opportunity to save the development cost, as labeling tools share commonalities.

While E1 and E2, both from the ML background, regarded the annotation interface design and implementation of annotation interaction as the hardest part, E3 from the HCI background regarded algorithm modules as the hardest to implement.
We conclude that grasping all required techniques for implementing a data labeling tool can be difficult for labeling tool developers since it may involve cross-disciplinary knowledge.
Visual programming can help reduce the skill barrier on labeling tool development in this case~\cite{Kelleher2005Lowering}.

All of the three experts said they would be happy to use OneLabeler if their labeling tasks were supported.
They worked on label tasks of image, video, and text.
We conclude that the system should build in various modules to support different label tasks for good coverage of usage scenarios.

\subsection{Design Principles and Requirements}

The interview confirms the need for a system to support easily building and modifying labeling tools for different labeling tasks and application-specific requirements.
To achieve this goal, we settle on the following design principles for the system:

\begin{itemize}[leftmargin=*]
    \item \textbf{DP1: Modular composable design and reuse at multiple levels.}
        To exploit shared commonalities among labeling tasks and to facilitate reuse and modification, the system should be based on a modular composable design to enable building labeling tools with composable primitives.
        Primitives should have low coupling.
        Reuse low-level implementations as composable primitives, while reuse high-level implementations as editable templates.
        These reuses can significantly reduce the development time and efforts.
    \item \textbf{DP2: Ease building and guide the development process.}
        To reduce the skill barrier, the system should facilitate easy building of labeling tools through visual configuration and guiding developers towards feasible solutions.
\end{itemize}

\noindent Based on the design principles, we propose the following specific requirements for our OneLabeler system.
\textbf{R1}, \textbf{R2}, and \textbf{R3} meet \textbf{DP1}, while \textbf{R4} and \textbf{R5} meet \textbf{DP2}.

\begin{itemize}[leftmargin=*]
    \item \textbf{R1: Unified module APIs.}
        Software modules (i.e., interface modules and algorithm modules) should implement unified APIs.
        Each API should represent a family of common modules in labeling tools.
        Unified APIs allow substituting a software module in a labeling tool or extending the system with a new module without changing other modules (i.e., separation of concerns).
        In this way, unified APIs make it easier to extend and customize software modules.
    \item \textbf{R2: Module composition.}
        The system should support composition of modules to allow a rich space of labeling tools to be created (i.e., good coverage of usage scenarios) with a small number of primitives.
        Especially, algorithm and interface modules should be composable, as labeling tools typically feature joint efforts of humans and machines.
    \item \textbf{R3: Reuse modules and tools.}
        The system should build in common software modules to be reused as primitives and example composed tools to be reused as boilerplate to scaffold building of labeling tools, which can significantly ease building commonly used labeling tools and enable extending and customizing built-in label tools to fit application-specific requirements.
    \item \textbf{R4: Visual programming.}
        The system should support visual programming to enable developers to reuse software modules and compose workflows with no or minimal textual coding.
        Labeling tool development requires cross-disciplinary knowledge.
        A developer may not be proficient with every part of the required technology stacks.
        Visual programming can reduce the development barrier in such scenarios.
        It has become a common practice in enterprise machine learning services (e.g., Microsoft Azure\footnote{https://azure.microsoft.com/services/machine-learning} and Amazon Web Services\footnote{https://aws.amazon.com/sagemaker/}).
        Moreover, the diagram in visual programming provides an overview of the developed system that assists the development process.
    \item \textbf{R5: Static checking.}
        The system should support static checking to detect infeasible configurations and recommend their fixes to guide editing actions during workflow creation.
        It enables a developer to spot and localize bugs at an early stage without running the program and guides the developer to reach a feasible solution quickly.
\end{itemize}

\noindent A critical step to fulfill these design requirements is exploitation of commonalities in various labeling tools.
This is discussed in the next section, where we identify common APIs (\textbf{R1}), common software modules for reuse (\textbf{R3}), and common constraints for composing modules (\textbf{R2}, \textbf{R5}).

\section{Decomposing Labeling Tools}
\label{sec:framework}

To facilitate fast and easy development of labeling tools, our system should build on a modular design.
To design APIs with expressiveness and a suitable level of abstraction, we consult decomposition of labeling tool modules in the literature.
In the following, we describe the common conceptual modules and states identified through inductive coding of the literature.
Each module, together with its input and output states, defines a unified API for a family of techniques (\textbf{R1}).
For each module, we give examples of its instances that can serve as software modules in labeling tools (\textbf{R3}).
An instance of the module corresponds to an implementation satisfying the API.
The constraints in composing the modules in building labeling tools are described at the end of the section.
The constraints guide the module composition (\textbf{R2}) and facilitate static checking (\textbf{R5}).

\subsection{Methodology}

To identify common conceptual modules, we use data labeling flowcharts in the literature as the dataset and summarize common themes of modules and states of labeling tools.
The rationale is that flowcharts depict cautious decomposition of software into executable modules by the authors.
Commonalities in the decompositions likely imply modules with high cohesion and low coupling.

We start with 76 papers on labeling tools and methods published between 2003 and 2021, collected from venues in human-computer interaction, visualization, multimedia, and machine learning.
The full list of the papers is provided in the supplementary material.
For each paper, we extract flowchart figures describing data labeling workflows, which can be application-specific or generic.
We discard the papers without flowchart figures, and obtain a dataset of 36 flowchart figures extracted from 33 papers (2 of them contain multiple flowchart figures).

We conduct an inductive coding of the extracted 36 flowcharts with two stages.
In the first stage, we extract the tagged phrases from each flowchart and categorize them into ``module'' (human, machine, and mixed computation process) and ``state'' (input and output).
The phrases form preliminary sets of module codes and state codes for that flowchart.
For example, Crayons~\cite{Fails2003Interactive} supports image pixel-level segmentation with a flowchart containing five phrases.
Four of them (``train'', ``classify'', ``feedback to designer'', and ``manual correction'') are categorized as preliminary codes for modules because they describe actions and are placed inside blocks in the flowchart figure, while one of them (``interactive use'') is excluded as it is a modifier of another phrase.

In the second stage, we group the preliminary codes collected from all the flowcharts into themes.
The grouping is not mutually exclusive.
We further remove the themes outside the scope of data labeling (e.g., ``model understanding'' is removed), and merge themes when necessary to synthesize a final code for each theme.
The final module/state codes are generated from representative preliminary codes and normalized into short noun phrases describing an action/variable.
For example, ``classify'' in Crayons~\cite{Fails2003Interactive} is finally grouped into the theme ``default labeling'', and this theme also serves as its final code.

\subsection{Results}

After the first stage of the coding process, we obtain 188 and 163 preliminary codes for module and state, respectively.
For each coded flowchart figure, the number of preliminary module codes ranges between 1 and 11 ($ mean = 5.22 $, $ SD = 2.25 $), and the number of preliminary state codes ranges between 0 and 10 ($ mean = 4.53 $, $ SD = 2.61 $).
After the second stage, we identify five final codes for states (sorted by frequency as marked):
\begin{itemize}[leftmargin=*]
    \item \textbf{Data Objects}: the list of entities to be labeled (59/163).
    \item \textbf{Labels}: the list of annotations assigned to entities (57/163).
    \item \textbf{Samples}: an entity subset annotators handle at a time (18/163).
    \item \textbf{Model}: one or multiple machine-learned models (18/163).
    \item \textbf{Features}: the list of feature representations of entities (13/163).
\end{itemize}
Similarly, we identify eight final codes for modules: \textbf{interactive labeling}, \textbf{data object selection}, \textbf{model training}, \textbf{feature extraction}, \textbf{default labeling}, \textbf{quality assurance}, \textbf{stoppage analysis}, and \textbf{label ideation}, which are to be introduced in details in the next subsection.
The occurrences of the final codes in the 36 flowcharts are summarized in Appendix~\ref{sec:supp-coding} (Table~\ref{table:coding-states} and Table~\ref{table:coding-modules}), and the detailed coding results are included in supplementary material.

During the coding process, we have excluded several themes of states and modules regarded as irrelevant to data labeling (see Appendix~\ref{sec:themes}).
For example, understanding the machine-learned model does not directly benefit labeling, and thus we exclude the ``model understanding'' theme, whose occurrences include the phrase ``understand'' in H{\"{o}}ferlin et al.~\cite{Hoeferlin2012Inter}'s flowchart.
Similarly, the ``data collection'' theme refers to the process of collecting or enlarging the dataset to be labeled.
Its occurrences include the phrase ``video crawler'' in Hua and Qi~\cite{Hua2008Online}'s flowchart.
We exclude it because of the low frequency (4 occurrences) and the ambiguity of whether it should be regarded as a part of data labeling or a preparation step before data labeling.

\begin{figure}[htb]
    \centering
    \includegraphics[width=\linewidth]{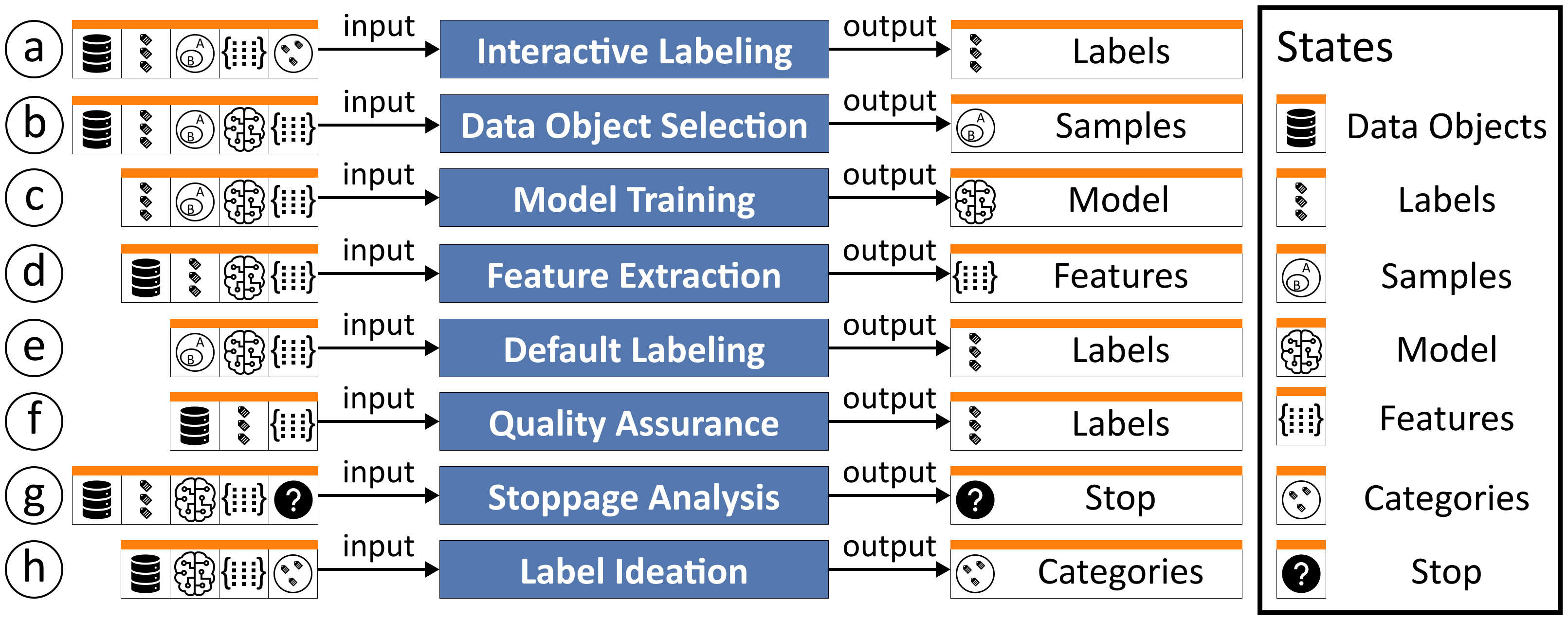}
    \caption{
        Common APIs in data labeling tools.
    }
    \Description[
        This figure shows the inputs and outputs of eight common conceptual modules in data labeling.
        Interactive labeling takes data objects, labels, samples, features, and categories as inputs, and outputs labels.
        Data object selection takes data objects, labels, samples, model, and features as inputs, and outputs samples.
        Model training takes labels, samples, model, and features as inputs, and outputs model.
        Feature extraction takes data objects, labels, model, and features as inputs, and outputs features.
        Default labeling takes samples, model, and features as inputs, and outputs labels.
        Quality assurance takes data objects, labels, and features as inputs, and outputs labels.
        Stoppage analysis takes data objects, labels, model, features, and stop as inputs, and outputs stop.
        Label ideation takes data objects, model, features, and categories as inputs, and outputs categories.
    ]{}
    \label{fig:modules-and-states}
\end{figure}

Combining the modules and states finally devised from the coding process, we further identified the common APIs in data labeling tools, as depicted by Fig.~\ref{fig:modules-and-states}.
To ensure that every API has an output, we add two additional states ``categories'' (output of ``label ideation'') and ``stop'' (output of ``stoppage analysis'').
API inputs are optional, as a software module implementing an API may utilize some or even none of the inputs available.
These API definitions are integrated in OneLabeler (described in Section~\ref{sec:system}).

\subsection{Common Modules in Labeling Tools}
\label{sec:common-modules}

The conceptual modules (sorted by frequency as marked) and their instances as software modules are described below.
An instance can be an algorithm (i.e., a machine computation process), an interface module involving human computation, or a mixed computation process.

\begin{itemize}[leftmargin=*]

\item \textbf{Interactive Labeling} (49/188) is the core module where annotators carry out labeling tasks in an interface (Fig.~\ref{fig:modules-and-states}a).
Different instances of this module typically vary in the design of task, interface, and interaction.
\textit{Task design} concerns the task that annotators are instructed to carry out.
Typical tasks include label assignment and correction tasks for the concerned data type (e.g., image, text) and label task type (e.g., classification, segmentation).
Alternatives include GWAP instructing annotators to play a game~\cite{Ahn2004Labeling,Ahn2006Peekaboom,Ahn2008Designing}.
When strong labels (e.g., segmentation) are needed, using weakly supervised learning may allow annotators to provide weak labels (e.g., classification) that are then algorithmically compiled to strong labels, potentially improving labeling efficiency~\cite{Branson2010Visual}.
Additionally, various research efforts in crowdsourcing concern task scheme design~\cite{Chang2017Revolt}.
\textit{Interface design} typically concerns data objects' layout such as grid matrix and hierarchical layout~\cite{Cui2007EasyAlbum,Zahalka2020II}.
Tasks requiring fine-grained editing may prefer displaying one data object each time~\cite{Russell2008LabelMe,Andriluka2018Fluid}.
\textit{Interaction design} concerns interaction techniques facilitating efficient labeling such as batch edit~\cite{Suh2007Semi} and graph cut~\cite{Boykov2001Fast}.

\item \textbf{Data Object Selection} (34/188) determines the order for data objects to be selected and labeled by annotators (Fig.~\ref{fig:modules-and-states}b).
Instances include active learning strategies that select informative data objects first~\cite{Settles2011Closing}.
The selection may also use clustering algorithms to group similar data objects~\cite{Cui2007EasyAlbum,Liu2009Smart}, which may enable annotators to assign the same label in one go.
Annotators may involve in deciding which data objects to label~\cite{Cakmak2011Mixed,Liao2016Visualization}.

\item \textbf{Model Training} (21/188) trains/updates a learning model (that may serve as input to other modules) with newly gathered labels (Fig.~\ref{fig:modules-and-states}c).
Variations of model training may concern what model is trained and how the training is conducted.
Although any predictive model may serve the purpose, semi-supervised learning and transfer learning techniques better match data labeling scenarios, where unlabeled data objects are abundant (useful for semi-supervised learning), and the cold start issue is prominent (alleviated by transfer learning).
Model training may be conducted by training from scratch or by incremental update methods in online learning.

\item \textbf{Feature Extraction} (20/188) turns data objects into feature representations, typically vectors (Fig.~\ref{fig:modules-and-states}d), facilitating other modules that cannot work with raw data objects (e.g., model training).
Algorithmic instances of feature extraction can be handcrafted (e.g., HoG), unsupervised (e.g., PCA), or supervised (e.g., LDA).
It is also possible to involve annotators in feature extraction~\cite{Cheng2015Flock}.

\item \textbf{Default Labeling} (17/188) assigns tentative labels to data objects, simplifying annotators' work from creating labels to verification and correction (Fig.~\ref{fig:modules-and-states}e).
It may be facilitated by models trained with the model training module, pre-trained models, or rules.

\item \textbf{Quality Assurance} (6/188) reviews label quality and corrects erroneous labels (Fig.~\ref{fig:modules-and-states}f).
Algorithmic relabeling may be used to suggest suspicious labels for annotators to verify~\cite{Brodley1999Identifying,Zhu2003Eliminating,Northcutt2021Confident}.
Annotators may also exploratorily review labels and search for potentially mislabeled data objects to correct~\cite{GarciaCaballero2019V,Xiang2019Interactive,Baeuerle2020Classifier}.
In mission-critical applications, quality assurance may require going through all the labels one by one~\cite{Zhang2021Using}.

\item \textbf{Stoppage Analysis} (4/188) decides whether to keep assigning tasks to annotators or stop (Fig.~\ref{fig:modules-and-states}g).
A common criterion is to check if all the data objects have been labeled once.
Alternative criteria may decide the stoppage time by empirical measures of the label quality~\cite{Deng2014Scalable} or confidence~\cite{Zhu2010Confidence} and stability of models~\cite{Bloodgood2009Method,Zhang2017Stopping} trained with the partially labeled dataset.

\item \textbf{Label Ideation} (3/188) develops the label categories used for labeling (Fig.~\ref{fig:modules-and-states}h).
It may appear as an interface widget allowing an annotator to create new categories ad hoc.
More structured ways to generate categories may involve algorithmic assistance (e.g., topic modeling) in an interface that supports users iteratively propose, verify, and refine categories~\cite{Felix2018Exploratory, Kulesza2014Structured}.

\end{itemize}

Through the coding process, we observe that interactive labeling is by far the most common module in labeling tools.
It appears 49 times among the 188 codes extracted from the 36 flowcharts, meaning that on average more than one module in each flowchart is related to interactive labeling.
Any flowchart that strictly depicts a data labeling workflow with human annotators would need to mention it at least once.

\subsection{Module Composition Constraints}
\label{sec:rules}

Labeling tools can be built with instances of the conceptual modules as building blocks.
However, not all compositions of software modules produce a valid labeling tool.
A labeling tool is represented by a flowchart that contains module nodes, i.e., nodes denoting implementation of common modules, as well as initialization, decision, and exit nodes.
Assuming that states of a labeling tool are stored globally, modules fetch inputs from and register outputs to corresponding global states.

The constraints for composing software modules into a valid labeling tool are listed as follows:
\begin{itemize}[leftmargin=*]
    \item \textbf{Valid Flowchart}:
        The graph should satisfy graph-theoretic constraints that generally hold for a flowchart~\cite{Karp1960Note}.
        For example, all the nodes should be reachable from the initialization node.
    \item \textbf{Input Initialized}:
        For each possible walk on the graph, a node that represents a software module should not be visited until its input parameters are all initialized.
    \item \textbf{No Redundancy}:
        After a module is visited, it should not be revisited until at least one of its inputs has changed its value.
        After a module is visited, its output(s) should be used by a module.
    \item \textbf{Involve Labeling}:
        Interactive labeling should exist in all walks of the graph to ensure it depicts a workflow of a labeling tool.
\end{itemize}

The first two constraints (``valid flowchart'' and ``input initialized'') are derived from the rationale that the flowchart should represent a correct program.
The third constraint concerns efficiency of the software.
Labeling tools often require heavy human computation (e.g., manual annotation) and heavy machine computation (e.g., model training), making performance optimization a critical issue.
The fourth constraint ensures that the represented software is a labeling tool.
These constraints translate to rigid propositions (in Appendix~\ref{sec:supp-workflow-constraints}) and are integrated in OneLabeler's static checking function to validate user-created labeling workflows.

\begin{figure*}[htb]
    \centering
    \includegraphics[width=\linewidth]{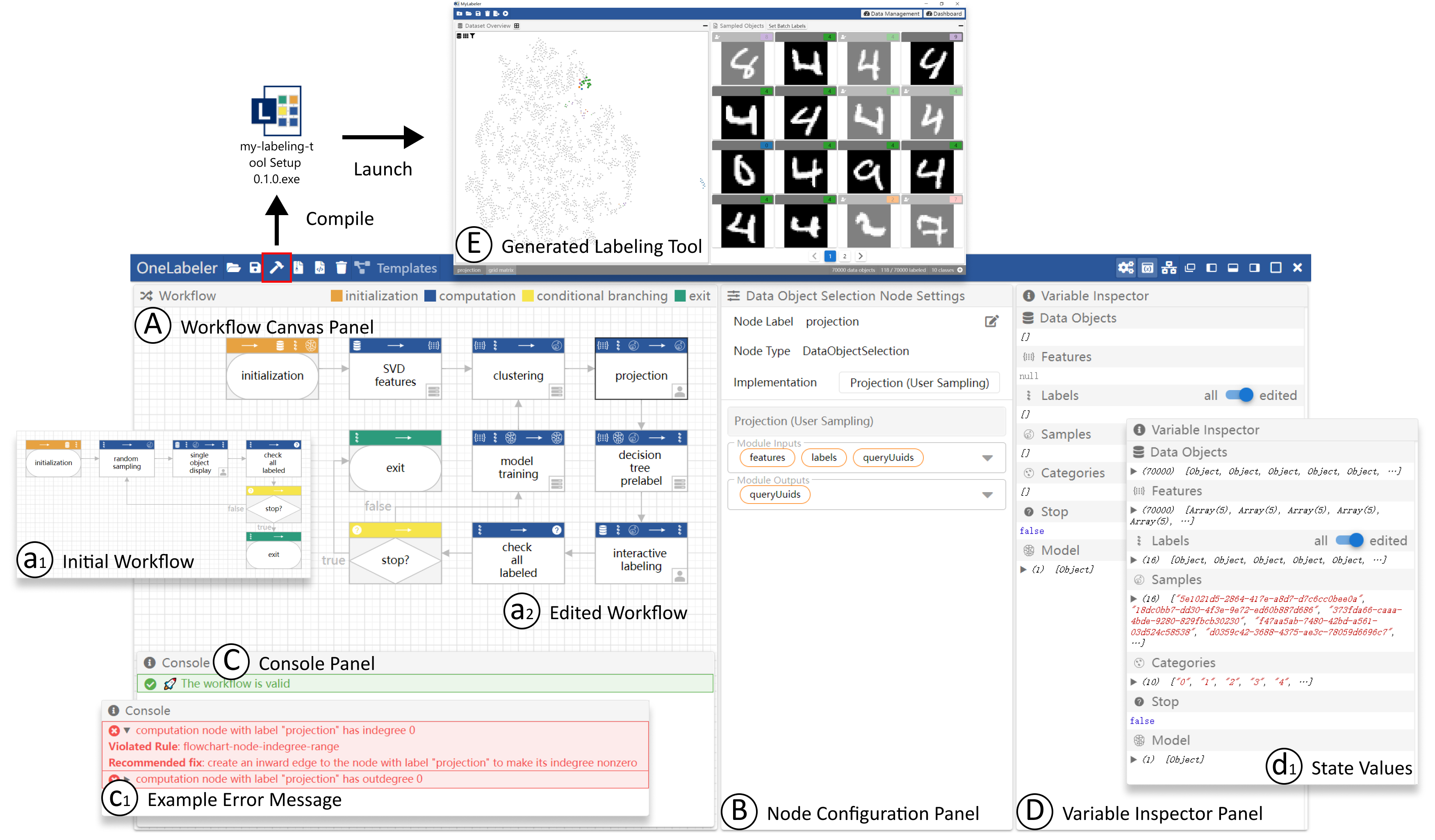}
    \caption{
        OneLabeler's visual programming interface:
        (A) Workflow canvas panel for a developer to create and compose nodes, as well as editing the visual layout of the workflow;
        (B) Node configuration panel for configuring parameters of a selected node;
        (C) Console panel for showing static checking result of whether the workflow denotes a valid labeling tool;
        (D) The variable inspector panel showing state values;
        (E) The generated data labeling tool by OneLabeler according to the workflow.
    }
    \Description[
        This figure shows the usage of OneLabeler's visual programming interface.
        Within the interface, a data labeling workflow is configured.
        The workflow is compiled into an installer.
        By installing the installer, the annotator can use the generated labeling tool.
    ]{}
    \label{fig:visual-programming}
\end{figure*}

\section{{OneLabeler}}
\label{sec:system}

Based on the requirements described in Section~\ref{sec:requirement}, we propose the OneLabeler system for building labeling tools.
OneLabeler enables developers to visually program (\textbf{R4}) data labeling tools by composing software modules into a workflow (\textbf{R2}).
A created labeling tool can be exported as an installer.
The conceptual modules identified in Section~\ref{sec:framework} inform the API design for data labeling modules in OneLabeler (\textbf{R1}).
A developer can reuse a collection of built-in modules and templates (\textbf{R3}) or customize on demand.
OneLabeler's visual programming environment supports static checking (\textbf{R5}) of user-created workflows to assist debugging and guide towards feasible solutions.

\subsection{System Architecture}
\label{sec:system-architecture}

\subsubsection{Module}
\label{sec:module}

The eight types of conceptual modules (interactive labeling, data object selection, model training, feature extraction, default labeling, quality assurance, stoppage analysis, and label ideation) are integrated into OneLabeler as eight API definitions (\textbf{R1}).
An API can be implemented with either an algorithm or interface module.
While an algorithm module can be automatically executed, the execution of an interface module requires human intervention.
In the following, we use the data object selection module as an example for illustration.

\noindent\textbf{Defining a conceptual module as API}:
As shown in Fig.~\ref{fig:modules-and-states}b, the data object selection module conceptually depicts a function that returns a dataset subset, given data objects, labels, features, model, and samples.
The inputs are optional.
Schematically, it translates to an API definition in OneLabeler as:

\begin{lstlisting}[language=TypeScript]
class DataObjectSelection {
  /** Input states. */
  props: ['dataObjects', 'labels', 'samples', 'model', 'features'];
  /** Output state's setter, evoked in .render or .run. */
  methods: ['setSamples'];
  /** Whether the execution of the module is blocking or not. */
  blocking: boolean;
  /** (Optional) Only needed for an interface module. */
  /** Whether to render the module when not executed. */
  persistent?: boolean;
  /** (Optional) Interface module implementation. */
  render?: Function;
  /** (Optional) Algorithm module implementation. */
  run?: Function;
}
\end{lstlisting}

\noindent\textbf{Executing a module implementation}:
Conceptually, executing a module implementation is to send it the input states and wait for the output states to be returned.
For example, the data object selection API defined above can be implemented with an algorithm (e.g., an active learning strategy) or a user interface module (e.g., a projection scatterplot as Fig.~\ref{fig:image-classification}$\mathrm{B_1}$ where an annotator can select manually with lasso).
When the API is implemented as an algorithm, the execution is a straightforward function call.
When the API is implemented as an interface, the execution is regarded as a function call that has a side effect of altering the user interface.
The function returns when a callback function setting the returned states is triggered by user interaction in the interface.
For example, a projection scatterplot (as Fig.~\ref{fig:image-classification}$\mathrm{B_1}$) is regarded as a function that returns the selection after the user finishes a lasso selection in the scatterplot.
Data object selection takes samples as an optional input because when implementing it with a projection scatterplot, the scatterplot may need to highlight the previous samples.

\noindent\textbf{Input and output access}:
Each module fetches inputs and returns outputs through getters and setters exposed by a global data model.
We refer to this storage model as a \emph{blackboard model} as all the modules read from the blackboard and write on the blackboard.
OneLabeler and labeling tools generated with OneLabeler follow the model-view-controller (MVC) design pattern\footnote{To disambiguate, the (data) ``model'' here refers to the part of software responsible for managing states (e.g., states in Fig.~\ref{fig:modules-and-states}), instead of a machine learning ``model''.}.
The seven states of labeling tools in Fig.~\ref{fig:modules-and-states} are stored globally and managed by the data model.
The data fetching and storage are configurable in labeling tools built with OneLabeler.
It allows multiple annotators to use installations of a labeling tool that have data access from the same database, enabling simultaneous annotation by multiple annotators.

\subsubsection{Workflow}
\label{sec:workflow}

To enable developers to create complex tools with a small number of primitives, OneLabeler supports composition of software modules into a workflow declaring a labeling tool (\textbf{R2}).

\noindent\textbf{Defining a workflow}:
Within OneLabeler, a labeling tool is declared by its workflow graph with modules being nodes and execution order specified by directed edges.
The graph can be either created in OneLabeler through visual programming (see Section~\ref{sec:visual-programming}), or uploaded to OneLabeler through a JSON file storing a list of nodes and edges.
A valid workflow graph has to satisfy constraints described in Section~\ref{sec:rules}.
Unlike data flow systems~\cite{Mendez2016iVoLVER,Yu2017VisFlow}, directed edges in OneLabeler only define execution order without implying data transmission.
Each module fetches inputs and returns outputs through the global data model.
Thus, for each directed edge, the source node's output is not necessarily the target node's input.

\noindent\textbf{Executing a workflow}:
To execute a workflow, OneLabeler traverses the graph representing the workflow.
The traversal starts from the initialization node in the graph.
OneLabeler recursively visits subsequent nodes following directed edges.
When a node corresponding to a module is visited, OneLabeler supplies the module's required inputs by reading the blackboard model, and executes the module as a function.
When the module is an interface module, the interface is rendered in a window (e.g., Fig.~\ref{fig:image-classification}$\mathrm{B_1}$ is a rendered window for data object selection using interactive projection).
By default, OneLabeler waits for the function to return.
After the function returns, OneLabeler registers the return value to the global data model and then visits the next node.
Alternatively, the module can be optionally configured to be non-blocking, in which case OneLabeler visits the next node before the function returns.
When the workflow is faithfully executed, the labeling tool's interface may constantly be changing because, after execution, an interface module will no longer show up until the next visit.
The ``persistent'' option addresses this issue of the rapid context switch.
For a ``persistent'' interface module, the interface is persistently rendered in the window even when the node is not executed.
When a decision node is visited, the decision criterion is checked, and the edge to follow is chosen accordingly.
The execution of the labeling workflow terminates when the exit node is visited.

\subsection{Visual Programming of Workflows}
\label{sec:visual-programming}

OneLabeler enables developers to build labeling tools through visual programming of the labeling tools' workflows (\textbf{R4}).
The navigation toolbar provides import/export and compilation functions through button clicking (highlighted in Fig.~\ref{fig:visual-programming} with red).
The labeling tool's workflow can be imported/exported in the JSON format to facilitate reuse and sharing.
A developer can use the built-in templates provided in the template menu as a boilerplate to scaffold development.
When the workflow is finalized, the developer can click the compile button on the navigation bar to compile the workflow into an installer of the corresponding labeling tool.

\subsubsection{Visual Programming and Configuration}
\label{sec:workflow-editing}

In OneLabeler, a data labeling tool is declared by its workflow.
To enable developers to visually configure the workflow without textual programming (\textbf{R4}), OneLabeler provides a visual programming interface (Fig.~\ref{fig:visual-programming}).
Within the interface, the developer can build a labeling tool by interactively configuring its workflow.

\noindent\textbf{Adding a node}:
Within the workflow canvas panel (Fig.~\ref{fig:visual-programming}A), the developer can add nodes through the right-click menu on the canvas.
The menu provides module nodes and control nodes.
The module nodes belong to one of the eight types of conceptual modules (in Section~\ref{sec:framework}).
The control nodes can be initialization, decision, and exit nodes.
After creating a module node, to make it executable, the developer needs to configure the implementation used for the node, as described below.

\noindent\textbf{Composing nodes}:
Two nodes can be composed by creating a directed edge to specify the execution order.
To create an edge between two nodes, the developer can hover on a node serving as the source, which shows the link ports of the node, drag the link port of the node, and release on the link port of another node serving as target.

\noindent\textbf{Configuring a node}:
To configure implementation details of a node, the developer first needs to select a node in the workflow canvas panel through clicking.
The node configuration panel (Fig.~\ref{fig:visual-programming}B) shows the details for the selected node.
The user can select the built-in implementation of the node in the ``implementation'' menu in the configuration panel.
The parameters of the selected implementation can be configured on demand.
For example, for the data object selection module (as shown in Fig.~\ref{fig:visual-programming}B),  the user can configure which selection method to use (e.g., active learning or interactive projection) and the parameters of the selected implementation (e.g., how many data objects to sample each time by the active learning strategy).
Additionally, outputs of the initialization node are configurable.
If a state is configured as initialization's output, it will be initialized to a non-empty value when initialization is executed.

\subsubsection{Static Checking}
\label{sec:static-checking}

To guide a developer smoothly towards a feasible solution (i.e., a workflow that depicts a valid labeling tool), we use the classic idea of static program analysis~\cite{Ayewah2008Using} in software debugging (\textbf{R5}).
The constraints (in Section~\ref{sec:rules}) allow OneLabeler to check the graph-theoretic properties of the workflow created by the developer as a proxy of the feasibility of the created labeling tool.
Besides, we add practical constraints on the graph data structure (e.g., node id should be unique) and on module configuration (i.e., an implementation should be chosen for the module).
OneLabeler integrates these constraints into a checker to statically validate the created labeling tool.
OneLabeler notifies the developer of identified violations of the constraints in the console panel (Fig.~\ref{fig:visual-programming}C), allowing the developer to identify potential mistakes before using the created labeling tool.
Hovering/Clicking an error message in the console panel highlights/selects the node(s) and edge(s) involved in the error in the workflow canvas panel.
Clicking the triangular expand button on the error message shows the error code and the recommended way(s) of fixing the error (Fig.~\ref{fig:visual-programming}$\mathrm{c_1}$).
For example, when a new module node is just created by the developer, there is no edge connecting it to other nodes.
In this case, the ``valid flowchart'' rule (in Section~\ref{sec:rules}) is violated, since this node has indegree and outdegree zero, meaning it is not used.
The error messages \textit{computation node with label ``projection'' has indegree 0} and \textit{computation node with label ``projection'' has outdegree 0} displayed in the console panel remind the developer to resolve this issue.
OneLabeler continuously validates the workflow as the developer edits it.
The messages are ranked by severity.
The low severity errors are hidden from the console panel until high severity ones are resolved to avoid overwhelming the developer with many error messages.
The developer can arrive at a feasible solution, i.e., a labeling tool that works, by iteratively resolving the error messages.

\subsubsection{Labeling Tool Preview}

To help the developer debug the built labeling tool, OneLabeler provides a real-time preview of the built tool.
As the developer finishes the workflow, the developer can try out the preview to examine if it meets the requirement.
During the process, the developer can inspect the state values (e.g., Fig.~\ref{fig:visual-programming}$\mathrm{d_1}$) of the labeling tool through the variable inspector (Fig.~\ref{fig:visual-programming}D).
OneLabeler also enables the developer to manipulate the control flow of the preview for debugging.
Specifically, the developer can conduct single-step debugging for a node or force the control flow to start from a node.
The control flow manipulation functions can be selected in the node's right-click menu.

\subsection{Built-in Modules}
\label{sec:built-in-modules}

OneLabeler provides a collection of built-in implementations for the conceptual modules.
The user can configure the implementations on demand.
Moreover, for interactive labeling modules, the user can configure three dimensions (i.e., data type, label task type, and interface design) to derive various combinations.

\begin{itemize}[leftmargin=*]
    \item For \textbf{interactive labeling}, as it is the most important module in data labeling, OneLabeler splits it into three design dimensions: data type, label task type, and interface design.
    OneLabeler builds in the following implementations for the three dimensions:
    \begin{itemize}
        \item \textbf{Data type}: image, text, video, audio, point cloud.
        \item \textbf{Label task type}: single-/multi-label classification, freeform text annotation, object detection (for image), segmentation (for image and point cloud), text/temporal span tagging (for text, audio and video), span relation (for text, audio and video).
        \item \textbf{Interface design}: single object display (e.g., Fig.~\ref{fig:text-multi-task}$\mathrm{B_1}$) and grid matrix with editable layout (e.g., Fig.~\ref{fig:image-classification}$\mathrm{B_2}$).
    \end{itemize}
    The three dimensions are identified following the rationale that the interface for interactive labeling needs to show data object details (which depends on the \textbf{data type}), provide the interaction for annotating labels (which depends on the \textbf{label task type}), and layout the data objects following an \textbf{interface design}.
    The data types are chosen to cover the data types in the 33 coded papers as well as common data types in machine learning benchmark datasets as indexed by \emph{Papers With Code}\footnote{https://github.com/paperswithcode}, including image, text, video, audio, point cloud, and sequential data.
    We have implemented all these data types except sequential data as it is concerned in only one paper~\cite{Lekschas2020Peax}.
    Similarly, the built-in label task types cover all the tasks in the 33 papers except for video object tracking with one occurrence~\cite{Hoeferlin2012Inter}.
    The two most common interface designs (i.e., grid matrix and single object display) in the 33 papers are already built-in, while others (i.e., thumbnail projection, table with metadata, and thread layout) are excluded for now.
    Those excluded from the current built-in can be implemented in the future.

    \item For \textbf{data object selection}, OneLabeler builds in algorithmic selection and interactive selection implementations.
    For the algorithmic selection, OneLabeler provides active learning techniques (including three entropy-based methods~\cite{Lewis1994Heterogeneous,Brinker2003Incorporating,Xu2007Incorporating}, least confidence, and smallest margin), clustering-based techniques (selection by ranking cluster labels, distance to cluster centroids, and density estimation).
    For interactive selection, OneLabeler provides interactive configurable data projection methods (e.g., Fig.~\ref{fig:image-classification}$\mathrm{B_1}$).
    The projection can be configured to be a scatterplot or heatmap of a T-SNE/PCA/LDA projection of the feature values, where the user can select data objects to label.

    \item For \textbf{model training}, OneLabeler currently builds in several classic supervised/semi-supervised learning algorithms, including decision tree, support vector machine, logistic regression, restricted Boltzmann machine, and graph-based label propagation.
    OneLabeler can be easily extended to support models provided by libraries that follow the scikit-learn library's API design.
  
    \item For \textbf{feature extraction}, OneLabeler builds in three techniques for image feature extraction, including a bag of features technique (handcrafted features describing color, edge, and texture), truncated SVD of raw images (unsupervised), and LDA projection of raw images (supervised).
    For text feature extraction, non-negative matrix factorization of tf-idf is provided.
    
    \item For \textbf{default labeling}, OneLabeler builds in model prediction, where the model can be the ones trained by the model training module(s) or provided by the developer through a prediction API.
    OneLabeler also provides a rule-based default labeling method for text span labeling based on part-of-speech (POS) tagging.
  
    \item For the other three conceptual modules with relatively low frequencies (in Section~\ref{sec:framework}), OneLabeler builds in basic implementations for the time being.
    For \textbf{quality assurance}, modules for interactive labeling can be reused for the annotator to go through data objects one by one.
    For \textbf{stoppage analysis}, OneLabeler builds in a criterion based on the sample rate specifying the rate of data objects to be labeled by annotators before stopping.
    For \textbf{label ideation}, OneLabeler provides an interface widget where label categories can be dynamically added, and for each label, the applicable label task type can be specified.
\end{itemize}

A developer can craft a labeling tool's workflow in the visual programming interface from scratch with the modules described above.
Additionally, OneLabeler builds in a collection of predefined template workflows, covering combinations of built-in data types and label task types.
Specifically, the built-in templates include
image classification,
image segmentation,
text classification,
text span tagging,
video classification,
video temporal segmentation,
audio classification,
audio temporal segmentation,
point cloud classification,
and point cloud segmentation.
The built-in templates are chosen to cover the five built-in data types, each data type with one coarse label task (i.e., classification) and one fine-grained label task (e.g., segmentation).
A developer can start with a predefined template and tailor it towards specific needs.
For example, a developer may start from a template workflow shown in Fig.~\ref{fig:visual-programming}$\mathrm{a_1}$ and add nodes to support default labeling and mixed-initiative sampling and corresponding connections, resulting in a revised workflow shown in Fig.~\ref{fig:visual-programming}$\mathrm{a_2}$.
The workflow (Fig.~\ref{fig:visual-programming}$\mathrm{a_2}$) and the resulting data labeling tool (Fig.~\ref{fig:visual-programming}E) is almost the same as the ones detailed in Section~\ref{sec:image-classification} except for minor differences in the configuration for interactive projection and cluster sampling batch size.

\subsection{Customization}
\label{sec:customization}

OneLabeler's modular architecture supports extensions for module implementations and workflow templates (\textbf{R1}).
If the built-in implementations do not meet a developer's needs, the developer can provide customized implementations.
OneLabeler provides a command-line interface (CLI) to scaffold customization.
Using the CLI, the developer can choose a type of customization as introduced below, and OneLabeler will create corresponding template files for the developer to start with (see our code repository at \url{https://github.com/microsoft/OneLabeler} for details on the CLI).

\subsubsection{General Purpose Module Customization}

To add a customized module to OneLabeler, a developer needs to fill the data structure as specified by the API definition in Section~\ref{sec:module}.
For an algorithm module, a developer needs to implement the ``run'' function that includes the code for running the algorithm and registering the result through setters (e.g., ``setSamples'') when the algorithm returns.
For an interface module, a developer needs to implement the ``render'' function that includes the code for rendering the module and registering the result through setters in the callback function of user interaction events.
For a customized algorithm module implemented as an algorithm server, OneLabeler allows the module to be directly registered in the visual programming interface.

\begin{figure*}[htb]
    \centering
    \includegraphics[width=\linewidth]{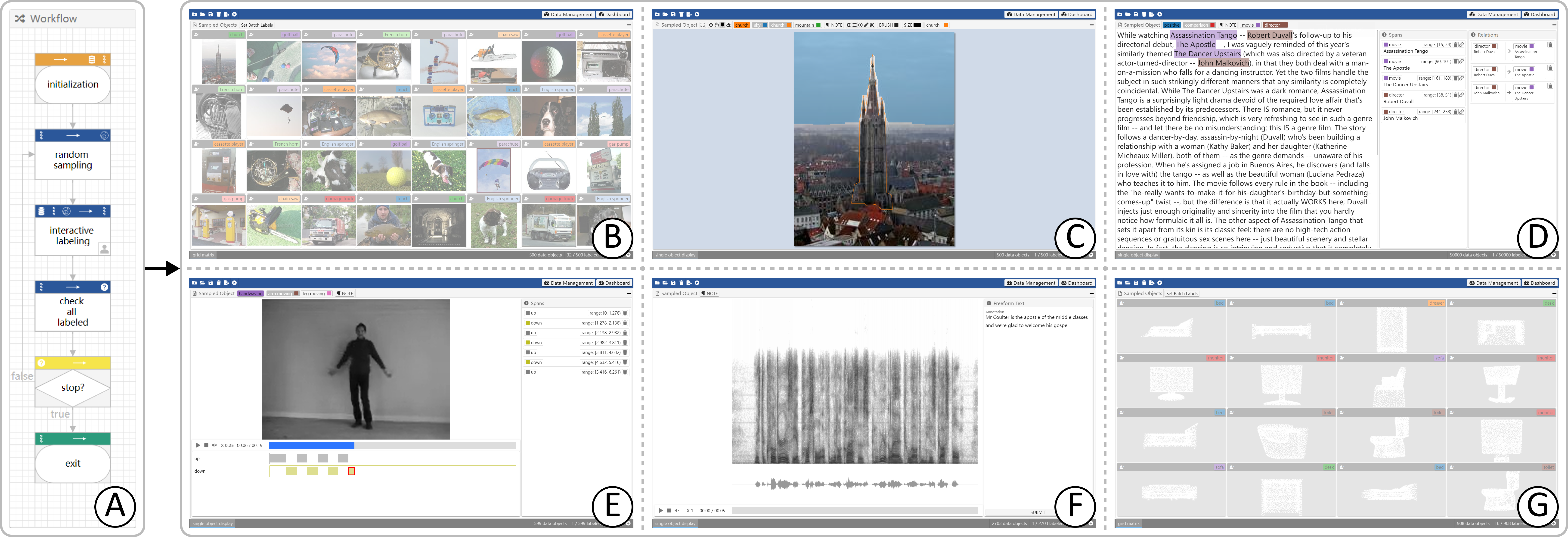}
    \caption{
        A gallery of labeling tools built with OneLabeler using a single workflow (shown in \mycircle{A}) but different configurations for the interactive labeling module.
        The configurations differ in data type, label task type, and interface design.
        The labeling tools \mycircle{C}\mycircle{D}\mycircle{E} all support classification, multi-label classification, and freeform text annotation, while additionally supporting label task types that are specialized to the corresponding data type (e.g., object detection for image).
        (A) A minimal data labeling template that can be used to generate the six labeling tools \mycircle{B}\mycircle{C}\mycircle{D}\mycircle{E}\mycircle{F}\mycircle{G}, with some modifications to node configurations for each labeling tool;
        (B) An image classification tool;
        (C) A multi-task image labeling tool (additionally support object detection, and segmentation);
        (D) A multi-task text labeling tool (additionally support span tagging, and span relation annotation);
        (E) A multi-task video labeling tool (additionally support span tagging);
        (F) An audio freeform text annotation tool;
        (G) A point cloud classification tool.
    }
    \Description[
        This figure shows a workflow created in OneLabeler and the interfaces created by minor tuning of configurations in the workflow.
        Subfigure A shows the workflow with modules named initialization, random sampling, interactive labeling, check all labeled, stop, and exit.
        Subfigure B, C, D, E, F, and G shows created data labeling tools described in the caption.
    ]{}
    \label{fig:gallery}
\end{figure*}

\subsubsection{Customization for Interactive Labeling}

OneLabeler internally represents a built-in interactive labeling module with three submodules: data type, label task type, and interface design.
Customizing interactive labeling modules can be done by following the instructions above for general-purpose customization or by implementing only a customized submodule.
To customize the data type, a developer needs to provide a display function defining how to display a single data object of this data type in the interface design.
This display function takes as input the data object's data structure, together with the width and height of the display area for it.
For example, to extend OneLabeler to support webpage labeling, a developer may specify that a data object of the webpage data type should be displayed with the HTML iframe tag (see details in Section~\ref{sec:webpage-classification}).
To customize the label task type, a developer needs to provide a display function defining the interaction widgets (e.g., buttons and menus).
These widgets are appended to the toolbar in the interface design to support the labeling interactions.
For the label task type that requires altering the visual appearance of the data object (e.g., image segmentation requires displaying a segmentation mask), a developer additionally needs to provide a render function defining how the created annotation of a data object should be displayed.
To customize the interface design, a developer needs to implement a display function defining the annotation interface.
The display function of the annotation interface takes the display function of data type declarations and toolbar and rendering function of label task type declarations as input.

\subsubsection{Template Customization}

To extend OneLabeler with a customized workflow template, a developer can either provide a TypeScript/JavaScript file declaring the workflow as an object or an equivalent JSON file (see Appendix~\ref{sec:customized-workflow-template} for an example).

\subsection{Workflow Compiler}

OneLabeler enables a workflow to be compiled into an installer of the specified labeling tool through button-clicking in the visual programming interface.
The compilation is conducted by the backend of OneLabeler that holds a mirror of OneLabeler's source code.
Upon receiving the compilation request, the backend hard-codes the workflow in the request into the source code mirror.
Then, the backend filters out from the mirror the dead-code irrelevant of the labeling tool declared by the workflow (e.g., the built-in modules of OneLabeler not used by the labeling tool and the code responsible for OneLabeler's workflow editing/compilation).
The modified and filtered version of OneLabeler's source code is then compiled into the declared labeling tool's installer using Electron Packager\footnote{https://github.com/electron/electron-packager}.
The installer installs the labeling tool as desktop software.

\section{Case Study}
\label{sec:case}

OneLabeler builds in various implementations of common modules in data labeling tools as introduced in Section~\ref{sec:system}, which can directly lead to a wide range of data labeling tools with appropriate configurations.
To demonstrate the effectiveness of OneLabeler in supporting the development of data labeling tools, this section presents ten sample data labeling tools built with OneLabeler.

We first introduce a gallery of six basic labeling tools based on built-in data types and label tasks.
Without changing the topology of a simple workflow template in Fig.~\ref{fig:gallery}A, the six example tools can be directly generated and can reproduce major functionalities of popular open-source labeling tools that cover various data types (including image, text, video, audio, and point cloud) and labeling tasks (such as classification, multi-label classification, freeform text annotation, segmentation, span tagging, and span relation).

Moreover, following the API of modules as described in Section~\ref{sec:framework}, developers can also extend the capabilities of OneLabeler by providing new implementations as plugins.
To further demonstrate the expressiveness and extensibility of OneLabeler, we showcase four more examples of advanced data labeling and interactive machine learning applications that integrate additional algorithm and interface modules, requiring more complicated workflows.

\subsection{A Gallery of Basic Labeling Tools}

In Fig.~\ref{fig:gallery}, we present six examples of basic labeling tools without machine assistance, based on the same workflow template in Fig.~\ref{fig:gallery}A.
With corresponding modifications to the configurations of the interactive labeling module, concerning data type, label task type, and interface design, we can achieve specific data labeling tools that target various tasks such as
classification (Fig.~\ref{fig:gallery}B),
segmentation (Fig.~\ref{fig:gallery}C),
span tagging (Fig.~\ref{fig:gallery}D) in different application domains including 
image (Fig.~\ref{fig:gallery}B and~\ref{fig:gallery}C),
text (Fig.~\ref{fig:gallery}D),
video (Fig.~\ref{fig:gallery}E),
audio (Fig.~\ref{fig:gallery}F),
and point cloud (Fig.~\ref{fig:gallery}G).
In Fig.~\ref{fig:gallery}, we load public datasets to the created labeling tools, including Imagenette\footnote{https://www.tensorflow.org/datasets/catalog/imagenette} (Fig.~\ref{fig:gallery}B and Fig.~\ref{fig:gallery}C), IMDb reviews~\cite{Maas2011Learning} (Fig.~\ref{fig:gallery}D), KTH Action~\cite{Schuldt2004Recognizing} (Fig.~\ref{fig:gallery}E), LibriSpeech~\cite{Panayotov2015Librispeech} (Fig.~\ref{fig:gallery}F), and ModelNet~\cite{Wu20153D} (Fig.~\ref{fig:gallery}G).

The example tools can reproduce the capabilities of some existing data-labeling systems in public use.
For example, the labeling tool in Fig.~\ref{fig:gallery}C supports image classification, multi-label classification, freeform text annotation, object detection (with polygon or bounding box), and segmentation.
These functionalities cover the major functionalities of LabelImg~\cite{Lin2015LabelImg} (for object detection with bounding box) and LabelMe~\cite{Russell2008LabelMe} (for object detection with polygon).

Similarly, using the text labeling tool in Fig.~\ref{fig:gallery}D, major features of Doccano~\cite{Nakayama2018doccano} can be reproduced.
Doccano supports labeling of image classification and multiple text-related label tasks, including named entity recognition, sentiment analysis, translation, and text to SQL.
With the text labeling tool in Fig.~\ref{fig:gallery}D, named entity recognition can be done by span tagging, sentiment analysis can be done by classification, while translation and text to SQL can both be done by freeform text labeling.

\subsection{Customizing a Webpage Data Type}
\label{sec:webpage-classification}

OneLabeler can be extended with customized data types.
A customized data type can be implemented as a submodule of the interactive labeling module (Section~\ref{sec:customization}).
Customizing a data type requires specifying how a single data object should be displayed.
For example, to add a new ``webpage'' data type, one can supply the following JavaScript code (by putting the code file in the data type folder of OneLabeler system).
The API design of data type declaration is compatible with Vue.js\footnote{https://vuejs.org/guide/extras/render-function.html}, a popular front-end development framework, enabling developers to utilize their previous programming experience to develop OneLabeler's plugins.

\begin{lstlisting}[language=TypeScript]
/** The declaration of webpage data type. */
export default {
  /** The label of the data type to appear in the menu. */
  label: 'webpage',
  /** Inputs to the display function. */
  props: ['dataObject', 'width', 'height'],
  /** Render a webpage using HTML iframe element. */
  render: (h, { props }) => (
    h('iframe', {
      attrs: {
        width: props.width,
        height: props.height,
        src: props.dataObject.src,
      },
    })
  ),
}
\end{lstlisting}

\begin{figure}[htb]
    \centering
    \includegraphics[width=\linewidth]{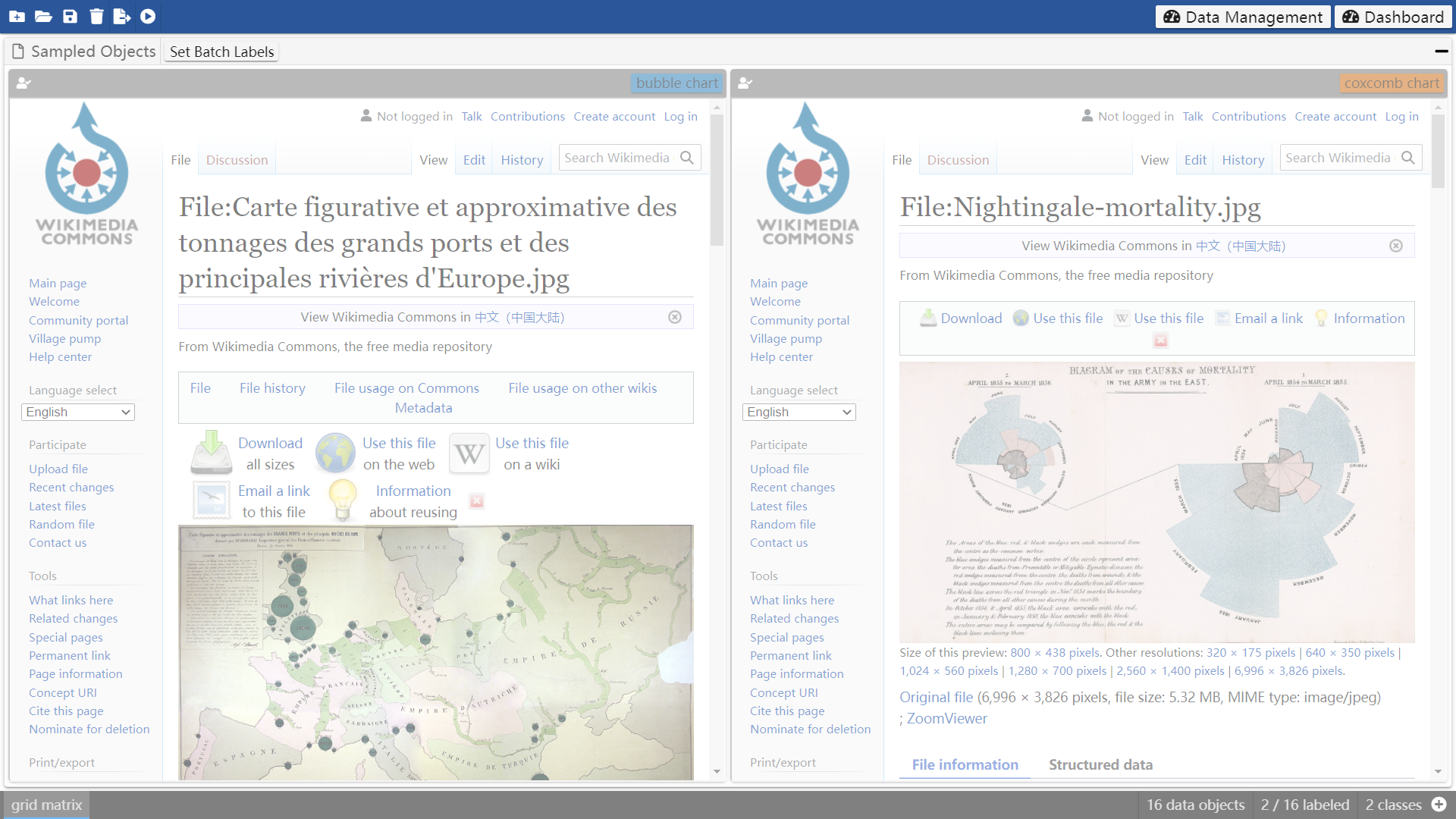}
    \caption{A classification tool for the customized webpage data type.}
    \Description[
        This figure shows a webpage classification tool currently displaying two webpages.
    ]{}
    \label{fig:webpage-classification}
\end{figure}

With the corresponding data type declaration, OneLabeler can support a new data type with tasks independent of data types, such as classification, multi-label classification, and freeform-text annotation.
For example, using the webpage data type, we can build a basic webpage classification tool as shown in Fig~\ref{fig:webpage-classification}.
This tool origins from the same workflow as Fig.~\ref{fig:gallery}A except that the data type is set to the newly customized webpage data type.

Developers can further develop data-type-specific label tasks with additional code.
For example, span tagging is originally targeted at text, video, and audio data types.
By specifying how to select a text ``span'' in HTML text nodes inside a webpage, the developer can make span tagging applicable to webpages.

\subsection{Machine-Aided Multi-Task Text Labeling}
\label{sec:multi-task-text}

\begin{figure}[htb]
    \centering
    \includegraphics[width=\linewidth]{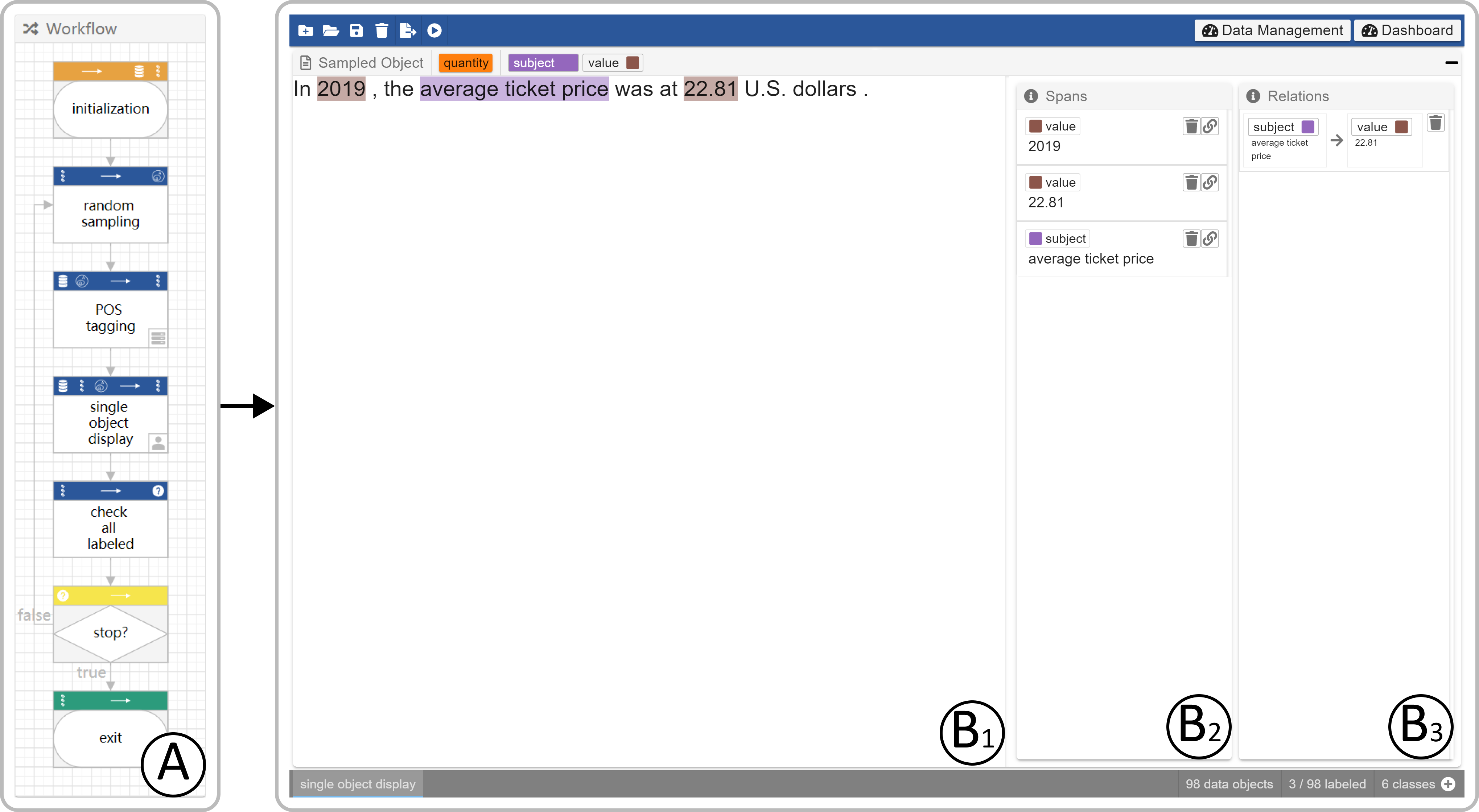}
    \caption{
        Building a machine-aided multi-task text labeling tool:
        (A) User-configured workflow for the tool.
        The interface of generated multi-task text labeling tool:
        ($\mathbf{B_1}$) A labeling panel showing the sentence for the annotator to conduct \textbf{interactive labeling};
        ($\mathbf{B_2}$) A list of created named entities;
        ($\mathbf{B_3}$) A list of created entity relations.
    }
    \Description[
        This figure shows a data labeling workflow and the created multi-task text labeling tool.
        Subfigure A shows the workflow with modules named initialization, random sampling, POS tagging, single object display, check all labeled, stop, and exit.
        Subfigures B1, B2, and B3 show the interface components of the text labeling tool created with the workflow.
    ]{}
    \label{fig:text-multi-task}
\end{figure}

\noindent Here we introduce a configured text labeling tool that incorporates machine assistance.
This tool is motivated by a data labeling scenario concerned in Cui et al.~\cite{Cui2020Text}'s supervised technique that learns from a human-annotated text dataset.
The data objects are quantitative statements.
The label tasks require annotating the statement type (i.e., document classification), the subject and value words in each sentence (i.e., span tagging), and the correspondence between subject and value words (i.e., span relation detection).
In this usage scenario, the three types of text labels described above are required to be assigned to sentences of quantitative statements.
Therefore, we demonstrate the creation of a tool that enables annotators to provide three types of labels: document classification, span tagging, and span relation detection.

In this application, the concerned data objects are text (i.e., strings), and the label tasks are to
(1) assign one of four class categories (``proportion'', ``quantity'', ``change'', and ``rank'') to the text to mark what type of statement it belongs to,
(2) extract named entities from text and associate with one of two class categories (``subject'' and ``value''), and
(3) link pairs of named entities if they are related subject and value.

Semi-automatic labeling~\cite{Andriluka2018Fluid,GarciaCaballero2019V,Zhang2021MI3} is a common technique in the literature for efficient annotation.
To incorporate this idea in the labeling tool, we can start from the simple labeling workflow template as shown in Fig.~\ref{fig:gallery}A, add a default labeling module, and configure it to be implemented with the built-in POS tagger.
Specifically, the developer may decide to use a predictive model to detect tentative text spans that belong to the subject and value words.
In this way, the annotator may save many efforts when the predictive model is accurate, as the annotator only needs to delete false positive detections and create the missing false negative spans.
This functionality can be achieved with a POS tagger built inside OneLabeler that automatically detect numeric values.
Moreover, if the developer is not satisfied with the built-in POS tagger, the developer can also implement a customized default labeling module.
Appendix~\ref{sec:customized-algorithm-module} shows the needed code for the developer to customize a default labeling module for text span detection with a noun detector based on POS tagging.

Fig.~\ref{fig:text-multi-task}A shows the visually programmed workflow for this labeling tool.
The resulting workflow of the generated labeling tool starts from a random sampling of data objects (\textbf{Data Object Selection}).
To save the efforts of annotating text spans, a machine-learned POS tagger is applied to extract phrases denoting values from the sentence (\textbf{Default Labeling}).
Within a labeling panel that displays a single sentence, the user can edit its class category (Fig.~\ref{fig:text-multi-task}$\mathrm{B_1}$), use brush to create named entities (Fig.
\ref{fig:text-multi-task}$\mathrm{B_2}$), and link entities to create entity relation annotation (Fig.~\ref{fig:text-multi-task}$\mathrm{B_3}$) (\textbf{Interactive Labeling}).

\subsection{Mixed-Initiative Image Classification}
\label{sec:image-classification}

\begin{figure}[htb]
    \centering
    \includegraphics[width=\linewidth]{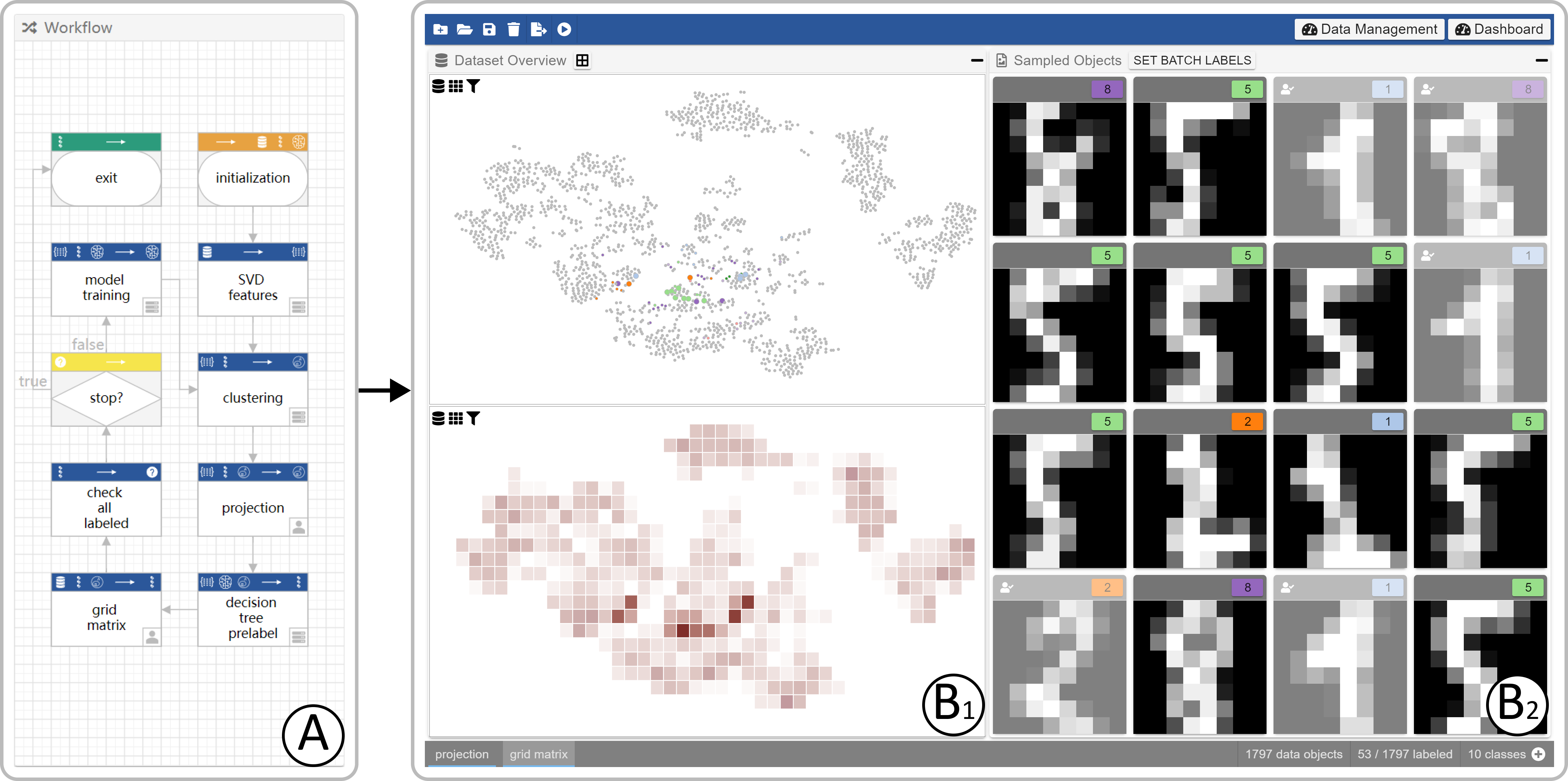}
    \caption{
        Building a mixed-initiative image classification tool:
        (A) User-configured workflow for the tool.
        The interface of generated image classification tool:
        ($\mathbf{B_1}$) A projection of data object features allowing annotators to use lasso selection for \textbf{data object selection};
        ($\mathbf{B_2}$) A labeling panel showing images for an annotator to conduct \textbf{interactive labeling}.
    }
    \Description[
        This figure shows a data labeling workflow and the created mixed-initiative image classification tool.
        Subfigure A shows the workflow with modules named initialization, SVD features, clustering, projection, decision tree prelabel, grid matrix, check all labeled, stop, model training, and exit.
        Subfigures B1 and B2 show the interface components of the image labeling tool created with the workflow.
    ]{}
    \label{fig:image-classification}
\end{figure}

Image classification is a common data labeling application scenario.
We demonstrate that OneLabeler allows fast prototyping of an image classification tool that features mixed-initiative sampling (with clustering and interactive projection) and default labeling (with decision tree).

To support efficient annotation, aside from semi-automatic labeling mentioned above, cluster-based labeling (e.g.,~\cite{Tang2013Towards,Cui2007EasyAlbum}) is also a common technique that involves algorithm and interface design.
In the labeling tool, the developer may decide to incorporate these two ideas by integrating: (1) default labeling so that annotators only need to do correction (i.e., semi-automatic labeling), and (2) mixed-initiative sampling for selecting similar data objects so that multiple data objects sharing the same label may be selected together, allowing one label to be assigned to them simultaneously.
The mixed-initiative sampling may be accomplished by joining a data object selection module implemented with a clustering algorithm and a projection view that enables annotators to initiate cluster selection~\cite{Liao2016Visualization,Bernard2018Comparing}.

To build a labeling tool that incorporates these two ideas, the developer may first start from the simple labeling workflow template shown in Fig.~\ref{fig:gallery}A.
The developer can further edit the workflow and leverage OneLabeler's built-in interface and algorithm modules for data object selection and default labeling.

To support default labeling, the developer adds a default labeling module and uses model prediction as its implementation.
To facilitate model prediction, a model training module is needed where the developer may choose a decision tree as the trained model.
Additionally, feature extraction is required as a decision tree is not an end-to-end model, and the developer may choose SVD as the feature extraction implementation.
To support mixed-initiative sampling, the developer may first add a data object selection module and choose clustering by k-means as an algorithmic implementation for it so that data objects are sampled cluster by cluster, potentially grouping similar data objects.
To allow annotators to involve in the selection process, the developer can add another data object selection module implemented with interactive projection.

Fig.~\ref{fig:image-classification}A shows the finalized visually programmed workflow of this labeling tool and the resulting labeling tool with the UCI handwritten digits dataset~\cite{Dua2017UCI} loaded.
In this application, the concerned data objects are images of handwritten digits, and the task is to assign digit labels (0, 1, ..., 9) to the images.

The resulting workflow of the generated labeling tool starts from singular value decomposition (SVD) as the feature extraction method for processing images (\textbf{Feature Extraction}).
Mixed-initiative sampling is accomplished by sampling data objects by clustering and then allowing the annotator to revise the samples if needed (\textbf{Data Object Selection}).
Specifically, a clustering with k-means is run to group and sort data objects where a batch of 16 data objects is sampled each time.
The interface displays a projection of the data objects, allowing the annotator to use lasso selection to manually sample data objects (Fig.~\ref{fig:image-classification}$\mathrm{B_1}$).
A decision tree is used to assign default labels for the sampled data objects by machine and annotator (\textbf{Default Labeling}).
Within a labeling panel with grid matrix layout, the user can edit the labels of sampled data objects (Fig.~\ref{fig:image-classification}$\mathrm{B_2}$) (\textbf{Interactive Labeling}).
Each time the user confirms the current samples are correctly labeled, the interim decision tree model is updated to make the default labeling more accurate (\textbf{Model Training}).

\subsection{Interactive Machine Learning for Chart Image Reverse Engineering}
\label{sec:mi3}

\begin{figure}[htb]
    \centering
    \includegraphics[width=\linewidth]{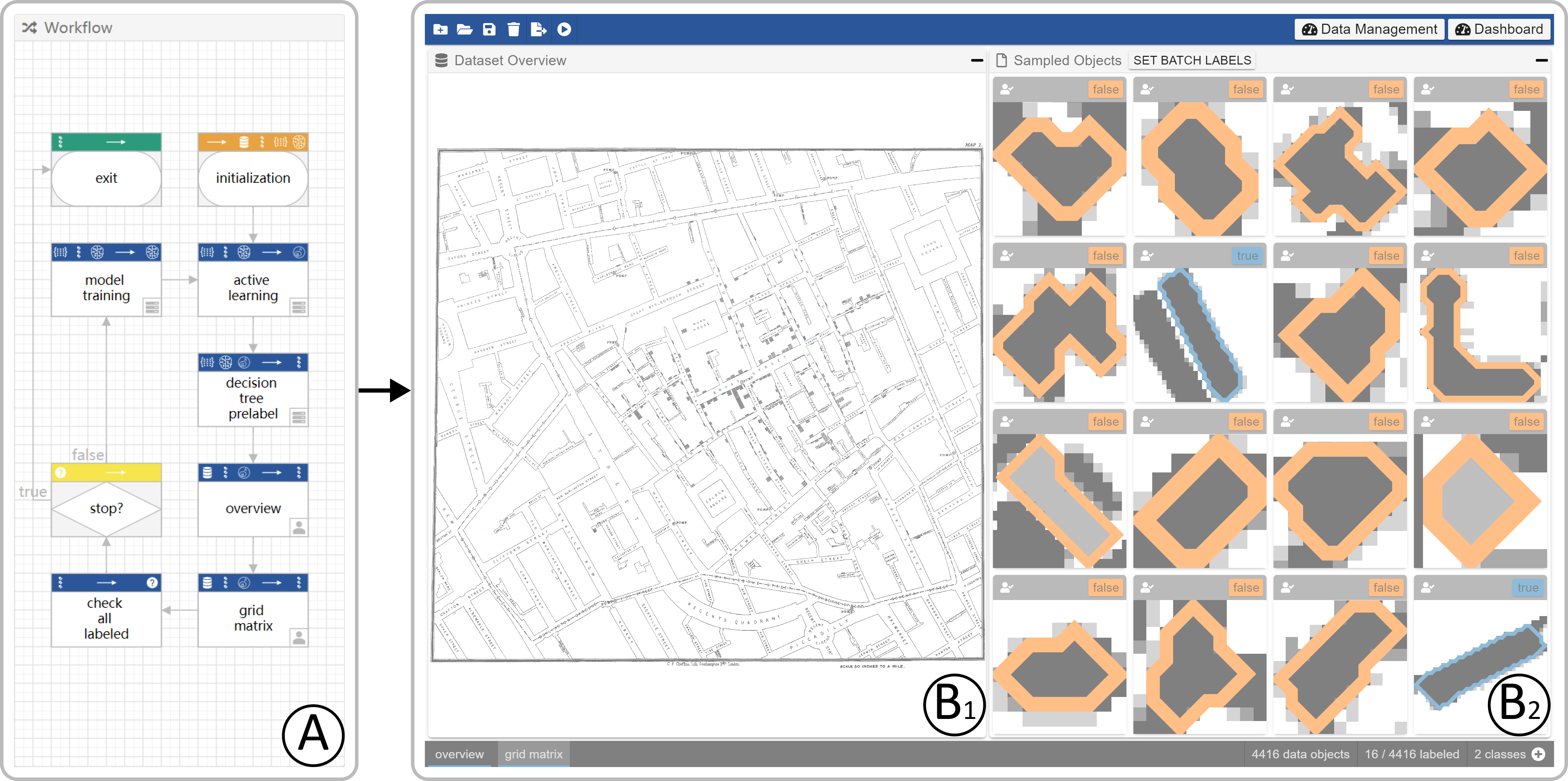}
    \caption{
        Building an interactive machine learning system for visualization reconstruction:
        (A) User-configured workflow for the system.
        The interface of the generated system for visualization reconstruction:
        ($\mathbf{B_1}$) An overview image that allows the annotator to conduct \textbf{interactive labeling} and overview the visualization reconstruction result;
        ($\mathbf{B_2}$) A labeling panel showing the geometric objects for the annotator to conduct \textbf{interactive labeling}.
    }
    \Description[
        This figure shows an interactive machine learning workflow and the created visualization reconstruction tool.
        Subfigure A shows the workflow with modules named initialization, active learning, decision tree prelabel, overview, grid matrix, check all labeled, stop, model training, and exit.
        Subfigures B1 and B2 show the interface components of the visualization reconstruction tool created with the workflow.
    ]{}
    \label{fig:mi3-block}
\end{figure}

Apart from enhancing conventional labeling processes with algorithm and interface modules, OneLabeler can also support fast prototyping of interactive machine learning systems with customized interactive labeling interfaces.
We take the system for visualization reconstruction proposed by Zhang et al.~\cite{Zhang2021MI3} as an example.
Their system aims to extract the dataset generating visualizations, and involves reverse engineering chart images by machine-aided classification of geometric objects.
We demonstrate the reproduction of the system by OneLabeler.

Following the workflow figure in their paper, we construct a data labeling workflow to reproduce the system.
In this application, the concerned data objects are geometric objects (e.g., polygons), and the task is to assign boolean labels (true detection, false detection) to the objects.
The geometric object is not a common data type, and OneLabeler has not built-in support for it.
Thus, the developer needs to implement a customized data-type plugin that specifies the import handler and display handler for geometric object instances.
In the import handler, the image processing algorithm code by Zhang et al.~\cite{Zhang2021MI3} is evoked for extracting geometric objects from images.
In the display handler, we write a simple script for displaying polygons.

Active learning selects informative data objects according to label entropy-based criteria.
The label entropy is computed from label distribution estimated by a label propagation model (\textbf{Data Object Selection}).
A decision tree is used to assign default labels (\textbf{Default Labeling}).
The user can label geometric objects in the grid matrix panel (Fig.~\ref{fig:mi3-block}$\mathrm{B_2}$) or in the overview image (Fig.~\ref{fig:mi3-block}$\mathrm{B_1}$).
The overview image with extracted polygons embedded is provided to OneLabeler as a plugin for interactive labeling (\textbf{Interactive Labeling}).
In the model update phase, the label propagation model used for entropy computation and the decision tree model used for default labeling are both updated (\textbf{Model Training}).

Fig.~\ref{fig:mi3-block} shows the visually programmed workflow for this system and the interface of the resulting system reproduced with similar features as in the literature.
The reproduced system mimics not only the interface design of the original system (an overview plus an annotation panel) but also major algorithm modules for machine-aided labeling (e.g., active learning, algorithmic default labeling).
OneLabeler has the potential to accelerate the research on interactive machine learning by allowing researchers to build research prototypes more efficiently.
The customization required writing in total around 200 lines of TypeScript code for plugins (excluding the image processing code provided in the literature~\cite{Zhang2021MI3}).
Specifically, around 50 lines of code are needed to implement the data type plugin to support polygon as a customized data type.
Around 150 lines of code are needed to implement the overview image that embeds the extracted polygons.
In contrast, the front-end of the original system~\cite{Zhang2021MI3} contains around ten thousand lines of code.

\section{User Study}
\label{sec:user-study}

We have conducted an in-lab user study to evaluate the usability of OneLabeler for prospective developers of data labeling tools.

\subsection{Experimental Design}
\textbf{Participants}:
We recruited eight participants with programming experience between 9 to 12 years and different levels of experience in data labeling.
Five of them (P1, P2, P3, P6, P8) had developed data labeling tools.
One (P4) had an ongoing research project that required developing labeling tools.
The other two (P5, P7) had participated in labeling tool design and conducted labeling tasks as annotators.
The programming expertise of the participants varies.
All of them could write back-end or algorithm code using Python or C$\sharp$.
Three (P1, P2, P8) also had experience in front-end development using JavaScript.

\noindent\textbf{Procedure}:
The study lasted around two hours for each participant.
In each session, we started from asking about the participant's experience on programming and data labeling ($\sim$ 15 minutes);
next, we gave training on the OneLabeler system ($\sim$ 30 minutes);
then we asked the participant to build labeling tools using OneLabeler ($\sim$ 50 minutes);
and finally, we collected feedback on OneLabeler usage experience ($\sim$ 15 minutes).

In training, we showed a documentation website on OneLabeler's usage.
The documentation introduces basic concepts in OneLabeler and its visual programming functionalities, including workflow editing interactions, static checking, and the labeling tool preview.

We designed the following four tasks for each participant to accomplish with OneLabeler:
\begin{itemize}[leftmargin=*]
    \item Task 1: Build an image segmentation tool similar to that shown in Fig.~\ref{fig:gallery}C without using any built-in template.
    \item Task 2: Build an image classification tool similar to Fig.~\ref{fig:gallery}B based on the workflow built for Task 1.
    \item Task 3: Build a machine-aided image classification tool similar to Fig.~\ref{fig:image-classification}A based on the workflow built for Task 2.
    \item Task 4: Either reproduce the participant's own labeling tool or build a webpage classification tool similar to that shown in Fig.~\ref{fig:webpage-classification}.
\end{itemize}

\begin{table*}[!ht]
    \caption{Summary of task completion status and time (in minutes).}
    \label{table:user-study-results}
    \centering
    \scriptsize
    \begin{tabular}{l|ll|ll|ll|ll}
    \toprule
        \textbf{Participant} & \textbf{Task 1} & \textbf{Time} & \textbf{Task 2} & \textbf{Time} & \textbf{Task 3} & \textbf{Time} & \textbf{Task 4} & \textbf{Time} \\
    \midrule
        P1 & complete & 9 & complete & 5 & complete with hint (state initialization) & 13 & complete & 12 \\ 
        P2 & complete & 13 & complete & 4 & complete & 17 & complete & 11 \\ 
        P3 & complete with hint (parameter) & 15 & complete with hint (parameter) & 4 & complete & 12 & partially complete & 14 \\ 
        P4 & complete & 29 & complete & 5 & complete with hint (linting message) & 25 & complete & 4 \\ 
        P5 & complete & 16 & complete & 3 & complete & 14 & partially complete & 12  \\ 
        P6 & complete with hint (parameter) & 15 & complete & 9 & complete with hint (state initialization) & 25 & partially complete & 22 \\ 
        P7 & complete & 9 & complete & 5 & complete & 18 & complete & 4 \\ 
        P8 & complete & 8 & complete & 8 & complete & 13 & complete & 11 \\ 
    \bottomrule
    \end{tabular}
\end{table*}

We designed the first three tasks to evaluate the overall OneLabeler usability.
Specifically, Task 1 evaluates how a user builds new workflows from scratch, which reflects the understanding of basic concepts and usage of OneLabeler.
Task 2 evaluates how a user adapts an existing workflow to a new usage scenario.
Task 3 evaluates how a user builds complex bespoke workflows.
Moreover, we designed Task 4 to further understand if OneLabeler can be used to build a custom labeling tool for real-world usage.
It is an open-ended task.
A participant could either reproduce a labeling tool the participant had built before or extend OneLabeler with custom modules to fulfill the desired labeling scenario.
For each task, we provided the participants a textual specification of the desired functionality of the labeling tool (see Appendix~\ref{sec:supp-study}).

\subsection{Results}

\subsubsection{Task Completion}

Table~\ref{table:user-study-results} summarizes task completion status and time for the four tasks.
For the first three tasks (T1, T2, T3), all the participants were able to complete them.
Out of the 24 trials in total (8 participants $\times$ 3 tasks), the participants could finish the tasks correctly in 18 trials without hints from the experimenter.
Four participants (P1, P3, P4, P6) needed hints in six trials to make their results exactly the same as the specification given in the instructions.
For Task 4, three of the participants (P4, P6, P7) chose to reproduce their own labeling tools: both P4 and P7 developed a text labeling tool similar to Fig.~\ref{fig:gallery}D, and P6 built a tool for pairwise comparison of images.
The other five participants selected the alternative to build a predefined webpage classification tool.
Five out of the eight participants (P1, P2, P4, P7, P8) completed Task 4 independently.
The other three participants (P3, P5, P6) completed Task 4 after receiving help from the experimenter.
The experimenter helped P6 with the coding part as P6's tool required creating a new data type for image pairs, while P6 was unfamiliar with web development skills needed for the customization.
P3 and P5 also had no web development experience.
As a result, they could not complete the coding part when building the webpage classification tool.
After the experimenter helped them develop the required customized module, they constructed the workflow to achieve the desired labeling tool.

\subsubsection{Usability}

As new users, all the participants could use OneLabeler to accomplish the assigned tasks to build labeling tools within a short time.
This indicates that OneLabeler is easy to learn and easy to use overall.
For most participants, building a new workflow from scratch (Task 1) took less than 16 minutes, adapting an existing workflow to a new usage scenario without adding or removing a node (Task 2) took less than 10 minutes, and adding machine assistance to an existing workflow (Task 3) took around 15 minutes.
For Task 4, which requires coding, participants with web development experience (P1, P2, P8) accomplished the task independently and efficiently within 15 minutes.
In the interview, all the participants commented that OneLabeler was flexible and comprehensive.

During the tasks, all the participants had utilized the static checking function to resolve occurred issues, indicating the usefulness of static checking.
During the tasks, participants looked up the documentation website several times.
For example, P1, P2, and P6 looked up the definition of the conceptual modules, while P5 and P7 looked up the detailed illustration for the error code ``no-uninitialized-inputs'' (the Input Initialized rule in Section~\ref{sec:rules}).
It indicates that the documentation website provides helpful information and that OneLabeler requires some learning.

\subsubsection{Potential Improvements}

Through the user study, we have identified several opportunities to improve OneLabeler's usability.
The completion time of the tasks was affected by misoperations.
The fastest participant (P8) spent only 8 minutes on Task 1, while P4 spent 29 minutes on the same task.
The reason for P4's lengthened completion time in Task 1 was that P4 clicked the browser's Refresh button by mistake, which caused the previous progress to be lost, and thus P4 had to rebuild the workflow.
Similarly, P6 refreshed the webpage and spent 25 minutes on Task 3.
This indicates a potential usability improvement by supporting automatic saving of user progress.
Additionally, P1 suggested that OneLabeler should provide a function of refining the workflow layout with one click to save developer's efforts in manually arranging nodes.

In the first three tasks, some participants needed hints from the experimenter to make their workflows consistent with the specification.
Three of the hints (in Task 1 for P3, Task 1 for P6, Task 2 for P3) were given due to wrong parameter choices in their workflows.
For example, the initial workflows created by both P3 and P6 in Task 1 used the default value 48 for the number of data objects sampled by random sampling instead of setting it to 1 as required in the textual specification of the task.
These mistakes were caused by carelessness as they forgot to change the configuration from the default value.
In real-world usages, these mistakes caused by carelessness may be less common, as the specifications for real-world usages are driven by the developer's own needs, and such mistakes would be easy for the developer to spot.
Nevertheless, it may be beneficial to reduce such mistakes via improving the documentation to stress the necessity to try out the labeling tool preview and carefully check if it functions as expected.

The other three hints (in Task 3 for P1, P4, P6) were given as the participants did not recognize that the ``model'' state needs to be initialized before use.
In this case, OneLabeler's static checking provided multiple recommended fixes, as ``model'' can be initialized either by declaring it as an output of the initialization node (described in Section~\ref{sec:workflow-editing}) or by adding a module that outputs model.
The participants either took a wrong path to fix it by adding unnecessary modules, deviating from the specification (P4), or recognized that they needed to initialize the state, but did not realize that the initialization could be done in the initialization node and got stuck (P1 and P6).
This suggests further improvements to provide more understandable error messages to help developers locate and fix errors more effectively.

\subsubsection{Future Usage Scenarios}

According to the participants, existing data labeling tools usually are not suitable for their own tasks, mainly due to the following three reasons:
\begin{enumerate}[leftmargin=*]
    \item \textbf{Unconventional labeling tasks}:
        For example, P1 needed to annotate visual objects with their attributes in visualizations stored as vector and bitmap images.
        There exist no mature labeling tools for this labeling task.
    \item \textbf{Requirement of algorithm support}:
        To save annotation costs, a labeling tool may integrate algorithm support.
        The suitable algorithm largely depends on the usage scenario.
        For example, P3 developed an auto-tracking function to infer bounding boxes of data objects given the bounding box annotations of previous frames.
        While P3 found an existing labeling tool that provides basic auto-tracking support, P3 could not adapt it to meet the requirement, because it was closed source.
    \item \textbf{More control over software changes}:
        The data labeling requirements can change over time.
        For example, P2 said \textit{``Even if there exists a piece of software that temporarily meets my requirements, I would be diffident when using it. I may need to abandon it at some point when the requirements can no longer be accommodated by the existing software.''}
\end{enumerate}

All the participants agreed that OneLabeler could fill these gaps with its build-in features and customization support that enable efficient building and low-level control on the implementation.

\section{Discussion}
\label{sec:discussion}

OneLabeler is an attempt to improve the efficiency of building data labeling tools.
The design of OneLabeler follows the rationale that accelerating the building requires promoting reuse of software modules, which in turn requires identifying commonalities in existing data labeling tools.
Following this rationale, OneLabeler builds on a system architecture based on common data labeling modules in the literature.

\textbf{On the expressiveness of conceptual modules}:
Although there is no guarantee that the primitive modules (i.e., the API definitions) in OneLabeler are capable of representing all possible variations in data labeling tools, there is evidence that they are expressive enough to cover a wide range of interesting variations.
Firstly, the coding process that leads to the modules provides evidence that the modules are expressive enough, as they conceptually can represent at least the labeling-related modules in the 33 related papers.
Moreover, the case study provides additional evidence that OneLabeler can generate diverse labeling tools in action.

\textbf{Build in more module implementations}:
A clear direction to enhance OneLabeler is to extend it with more built-in modules.
We aim to provide more built-in options for the lower frequency modules currently underexplored (e.g., quality assurance, stoppage analysis, and label ideation).
Especially, while quality assurance is not frequently mentioned in the papers we collected, we expect it is important in real-world applications.
Additionally, while the conceptual modules of OneLabeler are sufficient to produce diverse data labeling tools, they may not capture all common techniques related to data labeling emerging in all the related fields.
Meanwhile, the coding process presented in Section~\ref{sec:framework} may be reapplied to a larger corpus of literature to improve the coverage of conceptual modules.
Through this iterative process of extending the expressive power, we move gradually towards the ultimate goal of making OneLabeler \textit{feature complete}, with various built-in implementations (algorithms or interfaces) available for multifarious real-world usages.

\textbf{On orthogonality of conceptual modules}:
While OneLabeler builds on primitive modules, the conceptual modules in the proposed framework are not strictly orthogonal.
For example, both the quality assurance module and the interactive labeling module output labels.
OneLabeler does not and should not forbid developers to use an implementation of quality assurance for the interactive labeling purpose.
We believe that enforcing strict orthogonality in the API definitions is unnecessary.
For example, if the interactive labeling module and the quality assurance module were merged into a single module for strict orthogonality, the discoverability of interactive labeling implementations and quality assurance implementations would both decrease.
A developer who wants to configure the module into an implementation of interactive labeling would have to pick the implementation out of all the interactive labeling implementations and quality assurance implementations, which is cumbersome.

\textbf{On discoverability of modules}:
As the number of built-in module implementations increases, the options provided by OneLabeler may become too many for novice users to discover.
We address this potential issue by allowing new implementations plugged into OneLabeler and built-in modules deleted from OneLabeler, as each module implementation in OneLabeler's source code is an individual script file.
In this way, we can provide multiple versions of customized distributions of OneLabeler, each with a different set of built-in modules suited for different developers.
In addition, OneLabeler can provide advanced search and recommendation functionalities in the module configuration panel to alleviate this issue in the future.

\textbf{Extension to crowd labeling scenarios}:
In this work, we have not explicitly considered the usage scenario of crowd labeling workflows.
Our framework may accommodate some aspects of crowd labeling workflows, e.g., the task scheme design in crowdsourcing~\cite{Li2016Crowdsourced} is related to the task design in interactive labeling and postprocessing of crowdsourced labels is related to quality assurance.
However, others are not captured by the conceptual framework.
To support crowdsourcing, OneLabeler needs to build in additional interface and algorithm modules, such as for administrator's tasks, including monitoring label progress, assessing annotators' reliability, and data management.
OneLabeler's states also need to be extended to accommodate new modules' inputs and outputs.
To integrate with existing infrastructures, we also need to investigate how to distribute a labeling tool created with OneLabeler on crowdsourcing platforms.

\textbf{Exploring other usages of OneLabeler}:
While OneLabeler is designed for building labeling tools, as demonstrated in Section~\ref{sec:mi3}, we envision that OneLabeler may be extended to support interactive machine learning applications other than data labeling.
For example, OneLabeler may be extended to interactive model training if additional interface modules are added to OneLabeler to allow end users to edit the model.
Another example is to apply OneLabeler to build customized information retrieval systems (e.g., photo management systems) by adding more modules for data object selection with information retrieval techniques.
Additionally, OneLabeler is designed for building standalone labeling tools, while it may also be beneficial to explore building UI widgets for data labeling that can embed on other systems.

\textbf{On Customization}:
Customization support is essential to ensure the flexibility of OneLabeler, as revealed by Task 4 in the user study.
OneLabeler allows a developer to add customized implementations, as long as the inputs and outputs of the implementations are within OneLabeler supported states.
Meanwhile, the customizability of OneLabeler may be extended in future work to support customization on workflow execution and UI appearance.
Fine control on parallelizing execution of multiple modules, such as waiting for all or racing, cannot be achieved at the moment.
Besides, each interface module currently appears as a window in OneLabeler's interface, and UI widgets outside a window are not customizable.
As customization requires textual programming, OneLabeler provides CLI (Section~\ref{sec:customization}) to assist customization development, where a developer can start coding based on template code rather than from scratch.
To further reduce textual programming efforts required for customization, it is beneficial to build a marketplace for OneLabeler modules, where developers can share their customized modules for others to reuse.
It may also be beneficial to further decompose common modules into submodules.
Through subdivision, the flexibility for reuse increases, as a developer may customize a submodule instead of an entire module.

\section{Conclusion}
\label{sec:conclusion}

In this paper, we have proposed a conceptual framework for data labeling and the OneLabeler system based on the conceptual framework to support easy building of labeling tools for different usage scenarios.
To build the framework, we have identified common states and modules by coding the literature and summarized constraints in composing the modules to build labeling tools.
Each modular process can be instantiated as a human, machine, or mixed computation procedure.
OneLabeler provides a visual programming interface that uses modules in the conceptual framework as building blocks.
It builds in various implementations for reuse so that developers can create labeling tools with no code or less code.
It provides static checking and preview functionalities to assist development and debugging.
OneLabeler supports customization, allowing developers to extend its capability further.
We have demonstrated OneLabeler's expressiveness through a case study of building ten labeling tools.
We have further conducted a user study to evaluate its usability and collect feedback from potential users.
The user-study results suggest that OneLabeler is easy to learn and enables potential users to build labeling tools efficiently.

\begin{acks}
  This work was conducted when Y. Zhang was an intern at Microsoft Research Asia.
  Y. Wang is the corresponding author.
  S. Chen was partially supported by Shanghai Municipal Science and Technology (Major Project 2018SHZDZX01, Sailing Program No.21YF1402900 and No. 21ZR1403300).
\end{acks}

\balance

\bibliographystyle{style/ACM-Reference-Format}
\bibliography{doc/assets/bibs/main}


\begin{thebibliography}{88}


\ifx \showCODEN    \undefined \def \showCODEN     #1{\unskip}     \fi
\ifx \showDOI      \undefined \def \showDOI       #1{#1}\fi
\ifx \showISBNx    \undefined \def \showISBNx     #1{\unskip}     \fi
\ifx \showISBNxiii \undefined \def \showISBNxiii  #1{\unskip}     \fi
\ifx \showISSN     \undefined \def \showISSN      #1{\unskip}     \fi
\ifx \showLCCN     \undefined \def \showLCCN      #1{\unskip}     \fi
\ifx \shownote     \undefined \def \shownote      #1{#1}          \fi
\ifx \showarticletitle \undefined \def \showarticletitle #1{#1}   \fi
\ifx \showURL      \undefined \def \showURL       {\relax}        \fi
\providecommand\bibfield[2]{#2}
\providecommand\bibinfo[2]{#2}
\providecommand\natexlab[1]{#1}
\providecommand\showeprint[2][]{arXiv:#2}

\bibitem[\protect\citeauthoryear{Amazon}{Amazon}{2005}]%
        {Amazon2005Amazon}
\bibfield{author}{\bibinfo{person}{Amazon}.} \bibinfo{year}{2005}\natexlab{}.
\newblock \bibinfo{title}{Amazon Mechanical Turk}.
\newblock
\newblock
\urldef\tempurl%
\url{https://www.mturk.com/}
\showURL{%
\tempurl}


\bibitem[\protect\citeauthoryear{Andriluka, Uijlings, and Ferrari}{Andriluka
  et~al\mbox{.}}{2018}]%
        {Andriluka2018Fluid}
\bibfield{author}{\bibinfo{person}{Mykhaylo Andriluka}, \bibinfo{person}{Jasper
  R.~R. Uijlings}, {and} \bibinfo{person}{Vittorio Ferrari}.}
  \bibinfo{year}{2018}\natexlab{}.
\newblock \showarticletitle{Fluid Annotation: A Human-Machine Collaboration
  Interface for Full Image Annotation}. In
  \bibinfo{booktitle}{\emph{Proceedings of the ACM International Conference on
  Multimedia}}. \bibinfo{publisher}{ACM}, \bibinfo{pages}{1957--1966}.
\newblock
\showISBNx{9781450356657}


\bibitem[\protect\citeauthoryear{Ayewah, Pugh, Hovemeyer, Morgenthaler, and
  Penix}{Ayewah et~al\mbox{.}}{2008}]%
        {Ayewah2008Using}
\bibfield{author}{\bibinfo{person}{Nathaniel Ayewah}, \bibinfo{person}{William
  Pugh}, \bibinfo{person}{David Hovemeyer}, \bibinfo{person}{J.~David
  Morgenthaler}, {and} \bibinfo{person}{John Penix}.}
  \bibinfo{year}{2008}\natexlab{}.
\newblock \showarticletitle{Using Static Analysis to Find Bugs}.
\newblock \bibinfo{journal}{\emph{IEEE Software}} \bibinfo{volume}{25},
  \bibinfo{number}{5} (\bibinfo{date}{sep} \bibinfo{year}{2008}),
  \bibinfo{pages}{22--29}.
\newblock


\bibitem[\protect\citeauthoryear{B{\"{a}}uerle, Neumann, and
  Ropinski}{B{\"{a}}uerle et~al\mbox{.}}{2020}]%
        {Baeuerle2020Classifier}
\bibfield{author}{\bibinfo{person}{A. B{\"{a}}uerle}, \bibinfo{person}{H.
  Neumann}, {and} \bibinfo{person}{T. Ropinski}.}
  \bibinfo{year}{2020}\natexlab{}.
\newblock \showarticletitle{Classifier-Guided Visual Correction of Noisy Labels
  for Image Classification Tasks}.
\newblock \bibinfo{journal}{\emph{Computer Graphics Forum}}
  \bibinfo{volume}{39}, \bibinfo{number}{3} (\bibinfo{date}{jun}
  \bibinfo{year}{2020}), \bibinfo{pages}{195--205}.
\newblock


\bibitem[\protect\citeauthoryear{Bernard, Hutter, Zeppelzauer, Fellner, and
  Sedlmair}{Bernard et~al\mbox{.}}{2018a}]%
        {Bernard2018Comparing}
\bibfield{author}{\bibinfo{person}{J{\"{u}}rgen Bernard},
  \bibinfo{person}{Marco Hutter}, \bibinfo{person}{Matthias Zeppelzauer},
  \bibinfo{person}{Dieter Fellner}, {and} \bibinfo{person}{Michael Sedlmair}.}
  \bibinfo{year}{2018}\natexlab{a}.
\newblock \showarticletitle{Comparing Visual-Interactive Labeling with Active
  Learning: An Experimental Study}.
\newblock \bibinfo{journal}{\emph{IEEE Transactions on Visualization and
  Computer Graphics}} \bibinfo{volume}{24}, \bibinfo{number}{1}
  (\bibinfo{date}{Jan.} \bibinfo{year}{2018}), \bibinfo{pages}{298--308}.
\newblock
\showISSN{1077-2626}


\bibitem[\protect\citeauthoryear{Bernard, Zeppelzauer, Sedlmair, and
  Aigner}{Bernard et~al\mbox{.}}{2018b}]%
        {Bernard2018VIAL}
\bibfield{author}{\bibinfo{person}{J{\"{u}}rgen Bernard},
  \bibinfo{person}{Matthias Zeppelzauer}, \bibinfo{person}{Michael Sedlmair},
  {and} \bibinfo{person}{Wolfgang Aigner}.} \bibinfo{year}{2018}\natexlab{b}.
\newblock \showarticletitle{{VIAL}: a unified process for visual interactive
  labeling}.
\newblock \bibinfo{journal}{\emph{The Visual Computer}} \bibinfo{volume}{34},
  \bibinfo{number}{9} (\bibinfo{year}{2018}), \bibinfo{pages}{1189--1207}.
\newblock
\showISSN{1432-2315}


\bibitem[\protect\citeauthoryear{Bloodgood and Vijay-Shanker}{Bloodgood and
  Vijay-Shanker}{2009}]%
        {Bloodgood2009Method}
\bibfield{author}{\bibinfo{person}{Michael Bloodgood} {and} \bibinfo{person}{K.
  Vijay-Shanker}.} \bibinfo{year}{2009}\natexlab{}.
\newblock \showarticletitle{A Method for Stopping Active Learning Based on
  Stabilizing Predictions and the Need for User-Adjustable Stopping}. In
  \bibinfo{booktitle}{\emph{Proceedings of the Conference on Computational
  Natural Language Learning}}. \bibinfo{publisher}{Association for
  Computational Linguistics}, \bibinfo{pages}{39--47}.
\newblock
\showISBNx{9781932432299}


\bibitem[\protect\citeauthoryear{Bostock, Ogievetsky, and Heer}{Bostock
  et~al\mbox{.}}{2011}]%
        {Bostock2011D3}
\bibfield{author}{\bibinfo{person}{Michael Bostock}, \bibinfo{person}{Vadim
  Ogievetsky}, {and} \bibinfo{person}{Jeffrey Heer}.}
  \bibinfo{year}{2011}\natexlab{}.
\newblock \showarticletitle{D3: Data-Driven Documents}.
\newblock \bibinfo{journal}{\emph{IEEE Transactions on Visualization and
  Computer Graphics}} \bibinfo{volume}{17}, \bibinfo{number}{12}
  (\bibinfo{date}{Nov} \bibinfo{year}{2011}), \bibinfo{pages}{2301--2309}.
\newblock
\showISSN{1077-2626}


\bibitem[\protect\citeauthoryear{Boykov, Veksler, and Zabih}{Boykov
  et~al\mbox{.}}{2001}]%
        {Boykov2001Fast}
\bibfield{author}{\bibinfo{person}{Yuri Boykov}, \bibinfo{person}{Olga
  Veksler}, {and} \bibinfo{person}{Ramin Zabih}.}
  \bibinfo{year}{2001}\natexlab{}.
\newblock \showarticletitle{Fast Approximate Energy Minimization via Graph
  Cuts}.
\newblock \bibinfo{journal}{\emph{IEEE Transactions on Pattern Analysis and
  Machine Intelligence}} \bibinfo{volume}{23}, \bibinfo{number}{11}
  (\bibinfo{year}{2001}), \bibinfo{pages}{1222--1239}.
\newblock
\showISSN{0162-8828}


\bibitem[\protect\citeauthoryear{Branson, Wah, Schroff, Babenko, Welinder,
  Perona, and Belongie}{Branson et~al\mbox{.}}{2010}]%
        {Branson2010Visual}
\bibfield{author}{\bibinfo{person}{Steve Branson}, \bibinfo{person}{Catherine
  Wah}, \bibinfo{person}{Florian Schroff}, \bibinfo{person}{Boris Babenko},
  \bibinfo{person}{Peter Welinder}, \bibinfo{person}{Pietro Perona}, {and}
  \bibinfo{person}{Serge Belongie}.} \bibinfo{year}{2010}\natexlab{}.
\newblock \showarticletitle{Visual Recognition with Humans in the Loop}.
\newblock In \bibinfo{booktitle}{\emph{Proceedings of the European Conference
  on Computer Vision}}. \bibinfo{publisher}{Springer Berlin Heidelberg},
  \bibinfo{pages}{438--451}.
\newblock


\bibitem[\protect\citeauthoryear{Brinker}{Brinker}{2003}]%
        {Brinker2003Incorporating}
\bibfield{author}{\bibinfo{person}{Klaus Brinker}.}
  \bibinfo{year}{2003}\natexlab{}.
\newblock \showarticletitle{Incorporating Diversity in Active Learning with
  Support Vector Machines}. In \bibinfo{booktitle}{\emph{Proceedings of the
  International Conference on Machine Learning}}. \bibinfo{publisher}{AAAI
  Press}, \bibinfo{pages}{59--66}.
\newblock
\showISBNx{1577351894}


\bibitem[\protect\citeauthoryear{Brodley and Friedl}{Brodley and
  Friedl}{1999}]%
        {Brodley1999Identifying}
\bibfield{author}{\bibinfo{person}{Carla~E. Brodley} {and}
  \bibinfo{person}{Mark~A. Friedl}.} \bibinfo{year}{1999}\natexlab{}.
\newblock \showarticletitle{Identifying Mislabeled Training Data}.
\newblock \bibinfo{journal}{\emph{Journal of Artificial Intelligence Research}}
  \bibinfo{volume}{11}, \bibinfo{number}{1} (\bibinfo{date}{July}
  \bibinfo{year}{1999}), \bibinfo{pages}{131--167}.
\newblock
\showISSN{1076-9757}


\bibitem[\protect\citeauthoryear{Bryan, Mysore, and Wang}{Bryan
  et~al\mbox{.}}{2014}]%
        {Bryan2014ISSE}
\bibfield{author}{\bibinfo{person}{Nicholas~J. Bryan},
  \bibinfo{person}{Gautham~J. Mysore}, {and} \bibinfo{person}{Ge Wang}.}
  \bibinfo{year}{2014}\natexlab{}.
\newblock \showarticletitle{{ISSE}: An Interactive Source Separation Editor}.
  In \bibinfo{booktitle}{\emph{Proceedings of the SIGCHI Conference on Human
  Factors in Computing Systems}}. \bibinfo{publisher}{ACM},
  \bibinfo{pages}{257--266}.
\newblock
\showISBNx{9781450324731}


\bibitem[\protect\citeauthoryear{Cakmak and Thomaz}{Cakmak and Thomaz}{2011}]%
        {Cakmak2011Mixed}
\bibfield{author}{\bibinfo{person}{Maya Cakmak} {and}
  \bibinfo{person}{Andrea~L. Thomaz}.} \bibinfo{year}{2011}\natexlab{}.
\newblock \showarticletitle{Mixed-Initiative Active Learning}. In
  \bibinfo{booktitle}{\emph{Proceedings of ICML Workshop on Combining Learning
  Strategies to Reduce Label Cost}}. \bibinfo{numpages}{5}~pages.
\newblock


\bibitem[\protect\citeauthoryear{Chai, Li, Fan, and Luo}{Chai
  et~al\mbox{.}}{2020}]%
        {Chai2020Crowdsourcing}
\bibfield{author}{\bibinfo{person}{Chengliang Chai}, \bibinfo{person}{Guoliang
  Li}, \bibinfo{person}{Ju Fan}, {and} \bibinfo{person}{Yuyu Luo}.}
  \bibinfo{year}{2020}\natexlab{}.
\newblock \showarticletitle{Crowdsourcing-based Data Extraction from
  Visualization Charts}. In \bibinfo{booktitle}{\emph{Proceedings of the IEEE
  International Conference on Data Engineering}}. \bibinfo{publisher}{IEEE},
  \bibinfo{pages}{1814--1817}.
\newblock
\showISBNx{978-1-7281-2904-4}
\showISSN{1063-6382}


\bibitem[\protect\citeauthoryear{Chang, Amershi, and Kamar}{Chang
  et~al\mbox{.}}{2017}]%
        {Chang2017Revolt}
\bibfield{author}{\bibinfo{person}{Joseph~Chee Chang}, \bibinfo{person}{Saleema
  Amershi}, {and} \bibinfo{person}{Ece Kamar}.}
  \bibinfo{year}{2017}\natexlab{}.
\newblock \showarticletitle{Revolt: Collaborative Crowdsourcing for Labeling
  Machine Learning Datasets}. In \bibinfo{booktitle}{\emph{Proceedings of the
  SIGCHI Conference on Human Factors in Computing Systems}}.
  \bibinfo{publisher}{ACM}, \bibinfo{pages}{2334--2346}.
\newblock
\showISBNx{9781450346559}


\bibitem[\protect\citeauthoryear{Cheng and Bernstein}{Cheng and
  Bernstein}{2015}]%
        {Cheng2015Flock}
\bibfield{author}{\bibinfo{person}{Justin Cheng} {and}
  \bibinfo{person}{Michael~S. Bernstein}.} \bibinfo{year}{2015}\natexlab{}.
\newblock \showarticletitle{Flock: Hybrid Crowd-Machine Learning Classifiers}.
  In \bibinfo{booktitle}{\emph{Proceedings of the ACM Conference on Computer
  Supported Cooperative Work \& Social Computing}}. \bibinfo{publisher}{ACM},
  \bibinfo{pages}{600--611}.
\newblock
\showISBNx{9781450329224}


\bibitem[\protect\citeauthoryear{Choi, Park, Yang, Kim, Choo, and Hong}{Choi
  et~al\mbox{.}}{2019}]%
        {Choi2019AILA}
\bibfield{author}{\bibinfo{person}{Minsuk Choi}, \bibinfo{person}{Cheonbok
  Park}, \bibinfo{person}{Soyoung Yang}, \bibinfo{person}{Yonggyu Kim},
  \bibinfo{person}{Jaegul Choo}, {and} \bibinfo{person}{Sungsoo~Ray Hong}.}
  \bibinfo{year}{2019}\natexlab{}.
\newblock \showarticletitle{{AILA}: Attentive Interactive Labeling Assistant
  for Document Classification through Attention-Based Deep Neural Networks}. In
  \bibinfo{booktitle}{\emph{Proceedings of the SIGCHI Conference on Human
  Factors in Computing Systems}}. \bibinfo{publisher}{ACM},
  \bibinfo{pages}{1--12}.
\newblock
\showISBNx{9781450359702}


\bibitem[\protect\citeauthoryear{Cui, Wen, Xiao, Tian, and Tang}{Cui
  et~al\mbox{.}}{2007}]%
        {Cui2007EasyAlbum}
\bibfield{author}{\bibinfo{person}{Jingyu Cui}, \bibinfo{person}{Fang Wen},
  \bibinfo{person}{Rong Xiao}, \bibinfo{person}{Yuandong Tian}, {and}
  \bibinfo{person}{Xiaoou Tang}.} \bibinfo{year}{2007}\natexlab{}.
\newblock \showarticletitle{EasyAlbum: An Interactive Photo Annotation System
  Based on Face Clustering and Re-ranking}. In
  \bibinfo{booktitle}{\emph{Proceedings of the SIGCHI Conference on Human
  Factors in Computing Systems}}. \bibinfo{publisher}{ACM},
  \bibinfo{pages}{367--376}.
\newblock
\showISBNx{9781595935939}


\bibitem[\protect\citeauthoryear{Cui, Zhang, Wang, Huang, Chen, Fang, Zhang,
  Lou, and Zhang}{Cui et~al\mbox{.}}{2020}]%
        {Cui2020Text}
\bibfield{author}{\bibinfo{person}{Weiwei Cui}, \bibinfo{person}{Xiaoyu Zhang},
  \bibinfo{person}{Yun Wang}, \bibinfo{person}{He Huang}, \bibinfo{person}{Bei
  Chen}, \bibinfo{person}{Lei Fang}, \bibinfo{person}{Haidong Zhang},
  \bibinfo{person}{Jian-Guan Lou}, {and} \bibinfo{person}{Dongmei Zhang}.}
  \bibinfo{year}{2020}\natexlab{}.
\newblock \showarticletitle{Text-to-Viz: Automatic Generation of Infographics
  from Proportion-Related Natural Language Statements}.
\newblock \bibinfo{journal}{\emph{IEEE Transactions on Visualization and
  Computer Graphics}} \bibinfo{volume}{26}, \bibinfo{number}{1}
  (\bibinfo{date}{jan} \bibinfo{year}{2020}), \bibinfo{pages}{906--916}.
\newblock


\bibitem[\protect\citeauthoryear{de~Rooij, van Wijk, and Worring}{de~Rooij
  et~al\mbox{.}}{2010}]%
        {Rooij2010MediaTable}
\bibfield{author}{\bibinfo{person}{Ork de Rooij}, \bibinfo{person}{Jarke~J. van
  Wijk}, {and} \bibinfo{person}{Marcel Worring}.}
  \bibinfo{year}{2010}\natexlab{}.
\newblock \showarticletitle{{MediaTable}: Interactive Categorization of
  Multimedia Collections}.
\newblock \bibinfo{journal}{\emph{IEEE Computer Graphics and Applications}}
  \bibinfo{volume}{30}, \bibinfo{number}{5} (\bibinfo{date}{sep}
  \bibinfo{year}{2010}), \bibinfo{pages}{42--51}.
\newblock


\bibitem[\protect\citeauthoryear{Deng, Russakovsky, Krause, Bernstein, Berg,
  and Fei-Fei}{Deng et~al\mbox{.}}{2014}]%
        {Deng2014Scalable}
\bibfield{author}{\bibinfo{person}{Jia Deng}, \bibinfo{person}{Olga
  Russakovsky}, \bibinfo{person}{Jonathan Krause}, \bibinfo{person}{Michael
  Bernstein}, \bibinfo{person}{Alex Berg}, {and} \bibinfo{person}{Li Fei-Fei}.}
  \bibinfo{year}{2014}\natexlab{}.
\newblock \showarticletitle{Scalable Multi-Label Annotation}. In
  \bibinfo{booktitle}{\emph{Proceedings of the SIGCHI Conference on Human
  Factors in Computing Systems}}. \bibinfo{publisher}{ACM},
  \bibinfo{pages}{3099--3102}.
\newblock
\showISBNx{9781450324731}


\bibitem[\protect\citeauthoryear{Dua and Graff}{Dua and Graff}{2017}]%
        {Dua2017UCI}
\bibfield{author}{\bibinfo{person}{Dheeru Dua} {and} \bibinfo{person}{Casey
  Graff}.} \bibinfo{year}{2017}\natexlab{}.
\newblock \bibinfo{title}{{UCI} Machine Learning Repository}.
\newblock
\newblock
\urldef\tempurl%
\url{http://archive.ics.uci.edu/ml}
\showURL{%
\tempurl}


\bibitem[\protect\citeauthoryear{Dutta and Zisserman}{Dutta and
  Zisserman}{2019}]%
        {Dutta2019VIA}
\bibfield{author}{\bibinfo{person}{Abhishek Dutta} {and}
  \bibinfo{person}{Andrew Zisserman}.} \bibinfo{year}{2019}\natexlab{}.
\newblock \showarticletitle{The VIA Annotation Software for Images, Audio and
  Video}. In \bibinfo{booktitle}{\emph{Proceedings of the ACM International
  Conference on Multimedia}}. \bibinfo{publisher}{ACM},
  \bibinfo{pages}{2276--2279}.
\newblock
\showISBNx{9781450368896}


\bibitem[\protect\citeauthoryear{Evensen, Ge, and Demiralp}{Evensen
  et~al\mbox{.}}{2020}]%
        {Evensen2020Ruler}
\bibfield{author}{\bibinfo{person}{Sara Evensen}, \bibinfo{person}{Chang Ge},
  {and} \bibinfo{person}{Cagatay Demiralp}.} \bibinfo{year}{2020}\natexlab{}.
\newblock \showarticletitle{Ruler: Data Programming by Demonstration for
  Document Labeling}. In \bibinfo{booktitle}{\emph{Findings of the Association
  for Computational Linguistics: {EMNLP} 2020}}.
  \bibinfo{publisher}{Association for Computational Linguistics},
  \bibinfo{pages}{1996--2005}.
\newblock


\bibitem[\protect\citeauthoryear{Explosion}{Explosion}{2017}]%
        {Explosion2017Prodigy}
\bibfield{author}{\bibinfo{person}{Explosion}.}
  \bibinfo{year}{2017}\natexlab{}.
\newblock \bibinfo{title}{Prodigy}.
\newblock
\newblock
\urldef\tempurl%
\url{https://prodi.gy/}
\showURL{%
\tempurl}


\bibitem[\protect\citeauthoryear{Fails and Olsen}{Fails and Olsen}{2003}]%
        {Fails2003Interactive}
\bibfield{author}{\bibinfo{person}{Jerry~Alan Fails} {and}
  \bibinfo{person}{Dan~R. Olsen}.} \bibinfo{year}{2003}\natexlab{}.
\newblock \showarticletitle{Interactive Machine Learning}. In
  \bibinfo{booktitle}{\emph{Proceedings of the International Conference on
  Intelligent User Interfaces}}. \bibinfo{publisher}{ACM},
  \bibinfo{pages}{39--45}.
\newblock
\showISBNx{1581135866}


\bibitem[\protect\citeauthoryear{Felix, Dasgupta, and Bertini}{Felix
  et~al\mbox{.}}{2018}]%
        {Felix2018Exploratory}
\bibfield{author}{\bibinfo{person}{Cristian Felix}, \bibinfo{person}{Aritra
  Dasgupta}, {and} \bibinfo{person}{Enrico Bertini}.}
  \bibinfo{year}{2018}\natexlab{}.
\newblock \showarticletitle{The Exploratory Labeling Assistant:
  Mixed-Initiative Label Curation with Large Document Collections}. In
  \bibinfo{booktitle}{\emph{Proceedings of the ACM Symposium on User Interface
  Software and Technology}}. \bibinfo{publisher}{ACM},
  \bibinfo{pages}{153--164}.
\newblock
\showISBNx{9781450359481}


\bibitem[\protect\citeauthoryear{Garcia~Caballero, Westenberg, Gebre, and van
  Wijk}{Garcia~Caballero et~al\mbox{.}}{2019}]%
        {GarciaCaballero2019V}
\bibfield{author}{\bibinfo{person}{Humberto~S. Garcia~Caballero},
  \bibinfo{person}{Michel~A. Westenberg}, \bibinfo{person}{Binyam Gebre}, {and}
  \bibinfo{person}{Jarke~J. van Wijk}.} \bibinfo{year}{2019}\natexlab{}.
\newblock \showarticletitle{V-Awake: A Visual Analytics Approach for Correcting
  Sleep Predictions from Deep Learning Models}.
\newblock \bibinfo{journal}{\emph{Computer Graphics Forum}}
  \bibinfo{volume}{38}, \bibinfo{number}{3} (\bibinfo{year}{2019}),
  \bibinfo{pages}{1--12}.
\newblock


\bibitem[\protect\citeauthoryear{Halter, Ballester-Ripoll, Flueckiger, and
  Pajarola}{Halter et~al\mbox{.}}{2019}]%
        {Halter2019VIAN}
\bibfield{author}{\bibinfo{person}{Gaudenz Halter}, \bibinfo{person}{Rafael
  Ballester-Ripoll}, \bibinfo{person}{Barbara Flueckiger}, {and}
  \bibinfo{person}{Renato Pajarola}.} \bibinfo{year}{2019}\natexlab{}.
\newblock \showarticletitle{{VIAN}: A Visual Annotation Tool for Film
  Analysis}.
\newblock \bibinfo{journal}{\emph{Computer Graphics Forum}}
  \bibinfo{volume}{38}, \bibinfo{number}{3} (\bibinfo{year}{2019}),
  \bibinfo{pages}{119--129}.
\newblock


\bibitem[\protect\citeauthoryear{H{\"{o}}ferlin, Netzel, H{\"{o}}ferlin,
  Weiskopf, and Heidemann}{H{\"{o}}ferlin et~al\mbox{.}}{2012}]%
        {Hoeferlin2012Inter}
\bibfield{author}{\bibinfo{person}{Benjamin H{\"{o}}ferlin},
  \bibinfo{person}{Rudolf Netzel}, \bibinfo{person}{Markus H{\"{o}}ferlin},
  \bibinfo{person}{Daniel Weiskopf}, {and} \bibinfo{person}{Gunther
  Heidemann}.} \bibinfo{year}{2012}\natexlab{}.
\newblock \showarticletitle{Inter-Active Learning of Ad-Hoc Classifiers for
  Video Visual Analytics}. In \bibinfo{booktitle}{\emph{Proceedings of the IEEE
  Conference on Visual Analytics Science and Technology}}.
  \bibinfo{publisher}{IEEE}, \bibinfo{pages}{23--32}.
\newblock
\showISBNx{9781467347525}


\bibitem[\protect\citeauthoryear{Hoi and Lyu}{Hoi and Lyu}{2005}]%
        {Hoi2005Semi}
\bibfield{author}{\bibinfo{person}{Steven~C.H. Hoi} {and}
  \bibinfo{person}{Michael~R. Lyu}.} \bibinfo{year}{2005}\natexlab{}.
\newblock \showarticletitle{A Semi-Supervised Active Learning Framework for
  Image Retrieval}. In \bibinfo{booktitle}{\emph{Proceedings of the IEEE
  Conference on Computer Vision and Pattern Recognition}},
  Vol.~\bibinfo{volume}{2}. \bibinfo{publisher}{IEEE},
  \bibinfo{pages}{302--309}.
\newblock
\showISBNx{0-7695-2372-2}
\showISSN{1063-6919}


\bibitem[\protect\citeauthoryear{Hua and Qi}{Hua and Qi}{2008}]%
        {Hua2008Online}
\bibfield{author}{\bibinfo{person}{Xian-Sheng Hua} {and}
  \bibinfo{person}{Guo-Jun Qi}.} \bibinfo{year}{2008}\natexlab{}.
\newblock \showarticletitle{Online Multi-Label Active Annotation: Towards
  Large-Scale Content-Based Video Search}. In
  \bibinfo{booktitle}{\emph{Proceedings of the ACM International Conference on
  Multimedia}}. \bibinfo{publisher}{ACM}, \bibinfo{pages}{141--150}.
\newblock
\showISBNx{9781605583037}


\bibitem[\protect\citeauthoryear{Karp}{Karp}{1960}]%
        {Karp1960Note}
\bibfield{author}{\bibinfo{person}{Richard~M. Karp}.}
  \bibinfo{year}{1960}\natexlab{}.
\newblock \showarticletitle{A Note on the Application of Graph Theory to
  Digital Computer Programming}.
\newblock \bibinfo{journal}{\emph{Information and Control}}
  \bibinfo{volume}{3}, \bibinfo{number}{2} (\bibinfo{date}{jun}
  \bibinfo{year}{1960}), \bibinfo{pages}{179--190}.
\newblock


\bibitem[\protect\citeauthoryear{Kelleher and Pausch}{Kelleher and
  Pausch}{2005}]%
        {Kelleher2005Lowering}
\bibfield{author}{\bibinfo{person}{Caitlin Kelleher} {and}
  \bibinfo{person}{Randy Pausch}.} \bibinfo{year}{2005}\natexlab{}.
\newblock \showarticletitle{Lowering the Barriers to Programming: A Taxonomy of
  Programming Environments and Languages for Novice Programmers}.
\newblock \bibinfo{journal}{\emph{Comput. Surveys}} \bibinfo{volume}{37},
  \bibinfo{number}{2} (\bibinfo{date}{jun} \bibinfo{year}{2005}),
  \bibinfo{pages}{83--137}.
\newblock
\showISSN{0360-0300}


\bibitem[\protect\citeauthoryear{Kucher, Paradis, Sahlgren, and Kerren}{Kucher
  et~al\mbox{.}}{2017}]%
        {Kucher2017Active}
\bibfield{author}{\bibinfo{person}{Kostiantyn Kucher}, \bibinfo{person}{Carita
  Paradis}, \bibinfo{person}{Magnus Sahlgren}, {and} \bibinfo{person}{Andreas
  Kerren}.} \bibinfo{year}{2017}\natexlab{}.
\newblock \showarticletitle{Active Learning and Visual Analytics for Stance
  Classification with ALVA}.
\newblock \bibinfo{journal}{\emph{ACM Transactions on Interactive Intelligent
  Systems}} \bibinfo{volume}{7}, \bibinfo{number}{3}, Article
  \bibinfo{articleno}{14} (\bibinfo{date}{Oct.} \bibinfo{year}{2017}),
  \bibinfo{numpages}{31}~pages.
\newblock
\showISSN{2160-6455}


\bibitem[\protect\citeauthoryear{Kulesza, Amershi, Caruana, Fisher, and
  Charles}{Kulesza et~al\mbox{.}}{2014}]%
        {Kulesza2014Structured}
\bibfield{author}{\bibinfo{person}{Todd Kulesza}, \bibinfo{person}{Saleema
  Amershi}, \bibinfo{person}{Rich Caruana}, \bibinfo{person}{Danyel Fisher},
  {and} \bibinfo{person}{Denis Charles}.} \bibinfo{year}{2014}\natexlab{}.
\newblock \showarticletitle{Structured Labeling for Facilitating Concept
  Evolution in Machine Learning}. In \bibinfo{booktitle}{\emph{Proceedings of
  the SIGCHI Conference on Human Factors in Computing Systems}}.
  \bibinfo{publisher}{ACM}, \bibinfo{pages}{3075--3084}.
\newblock
\showISBNx{9781450324731}


\bibitem[\protect\citeauthoryear{{Labelbox, Inc.}}{{Labelbox, Inc.}}{2018}]%
        {L2018Labelbox}
\bibfield{author}{\bibinfo{person}{{Labelbox, Inc.}}}
  \bibinfo{year}{2018}\natexlab{}.
\newblock \bibinfo{title}{Labelbox}.
\newblock
\newblock
\urldef\tempurl%
\url{https://labelbox.com/}
\showURL{%
\tempurl}


\bibitem[\protect\citeauthoryear{Ledo, Houben, Vermeulen, Marquardt, Oehlberg,
  and Greenberg}{Ledo et~al\mbox{.}}{2018}]%
        {Ledo2018Evaluation}
\bibfield{author}{\bibinfo{person}{David Ledo}, \bibinfo{person}{Steven
  Houben}, \bibinfo{person}{Jo Vermeulen}, \bibinfo{person}{Nicolai Marquardt},
  \bibinfo{person}{Lora Oehlberg}, {and} \bibinfo{person}{Saul Greenberg}.}
  \bibinfo{year}{2018}\natexlab{}.
\newblock \showarticletitle{Evaluation Strategies for HCI Toolkit Research}. In
  \bibinfo{booktitle}{\emph{Proceedings of the SIGCHI Conference on Human
  Factors in Computing Systems}}. \bibinfo{publisher}{ACM},
  \bibinfo{pages}{1--17}.
\newblock
\showISBNx{9781450356206}


\bibitem[\protect\citeauthoryear{Lekschas, Peterson, Haehn, Ma, Gehlenborg, and
  Pfister}{Lekschas et~al\mbox{.}}{2020}]%
        {Lekschas2020Peax}
\bibfield{author}{\bibinfo{person}{Fritz Lekschas}, \bibinfo{person}{Brant
  Peterson}, \bibinfo{person}{Daniel Haehn}, \bibinfo{person}{Eric Ma},
  \bibinfo{person}{Nils Gehlenborg}, {and} \bibinfo{person}{Hanspeter
  Pfister}.} \bibinfo{year}{2020}\natexlab{}.
\newblock \showarticletitle{Peax: Interactive Visual Pattern Search in
  Sequential Data Using Unsupervised Deep Representation Learning}.
\newblock \bibinfo{journal}{\emph{Computer Graphics Forum}}
  \bibinfo{volume}{39}, \bibinfo{number}{3} (\bibinfo{date}{jun}
  \bibinfo{year}{2020}), \bibinfo{pages}{167--179}.
\newblock


\bibitem[\protect\citeauthoryear{Lewis and Catlett}{Lewis and Catlett}{1994}]%
        {Lewis1994Heterogeneous}
\bibfield{author}{\bibinfo{person}{David~D. Lewis} {and} \bibinfo{person}{Jason
  Catlett}.} \bibinfo{year}{1994}\natexlab{}.
\newblock \showarticletitle{Heterogeneous Uncertainty Sampling for Supervised
  Learning}. In \bibinfo{booktitle}{\emph{Proceedings of the International
  Conference on Machine Learning}}. \bibinfo{publisher}{Morgan Kaufmann
  Publishers Inc.}, \bibinfo{pages}{148--156}.
\newblock
\showISBNx{978-1-55860-335-6}


\bibitem[\protect\citeauthoryear{Li, Wang, Zheng, and Franklin}{Li
  et~al\mbox{.}}{2016}]%
        {Li2016Crowdsourced}
\bibfield{author}{\bibinfo{person}{Guoliang Li}, \bibinfo{person}{Jiannan
  Wang}, \bibinfo{person}{Yudian Zheng}, {and} \bibinfo{person}{Michael~J.
  Franklin}.} \bibinfo{year}{2016}\natexlab{}.
\newblock \showarticletitle{Crowdsourced Data Management: A Survey}.
\newblock \bibinfo{journal}{\emph{IEEE Transactions on Knowledge and Data
  Engineering}} \bibinfo{volume}{28}, \bibinfo{number}{9} (\bibinfo{date}{sep}
  \bibinfo{year}{2016}), \bibinfo{pages}{2296--2319}.
\newblock


\bibitem[\protect\citeauthoryear{Liao, Chen, Song, and Ming}{Liao
  et~al\mbox{.}}{2016}]%
        {Liao2016Visualization}
\bibfield{author}{\bibinfo{person}{Hongsen Liao}, \bibinfo{person}{Li Chen},
  \bibinfo{person}{Yibo Song}, {and} \bibinfo{person}{Hao Ming}.}
  \bibinfo{year}{2016}\natexlab{}.
\newblock \showarticletitle{Visualization-Based Active Learning for Video
  Annotation}.
\newblock \bibinfo{journal}{\emph{IEEE Transactions on Multimedia}}
  \bibinfo{volume}{18}, \bibinfo{number}{11} (\bibinfo{date}{Nov.}
  \bibinfo{year}{2016}), \bibinfo{pages}{2196--2205}.
\newblock
\showISSN{1520-9210}


\bibitem[\protect\citeauthoryear{Lin}{Lin}{2015}]%
        {Lin2015LabelImg}
\bibfield{author}{\bibinfo{person}{Tzuta Lin}.}
  \bibinfo{year}{2015}\natexlab{}.
\newblock \bibinfo{title}{LabelImg}.
\newblock
\newblock
\urldef\tempurl%
\url{https://github.com/tzutalin/labelImg}
\showURL{%
\tempurl}


\bibitem[\protect\citeauthoryear{Liu, Wang, Hua, and Zhang}{Liu
  et~al\mbox{.}}{2009}]%
        {Liu2009Smart}
\bibfield{author}{\bibinfo{person}{Dong Liu}, \bibinfo{person}{Meng Wang},
  \bibinfo{person}{Xian-Sheng Hua}, {and} \bibinfo{person}{Hong-Jiang Zhang}.}
  \bibinfo{year}{2009}\natexlab{}.
\newblock \showarticletitle{Smart Batch Tagging of Photo Albums}. In
  \bibinfo{booktitle}{\emph{Proceedings of the ACM International Conference on
  Multimedia}}. \bibinfo{publisher}{ACM}, \bibinfo{pages}{809--812}.
\newblock
\showISBNx{9781605586083}


\bibitem[\protect\citeauthoryear{Liu, Chen, Lu, Ouyang, and Wang}{Liu
  et~al\mbox{.}}{2019}]%
        {Liu2019Interactive}
\bibfield{author}{\bibinfo{person}{Shixia Liu}, \bibinfo{person}{Changjian
  Chen}, \bibinfo{person}{Yafeng Lu}, \bibinfo{person}{Fangxin Ouyang}, {and}
  \bibinfo{person}{Bin Wang}.} \bibinfo{year}{2019}\natexlab{}.
\newblock \showarticletitle{An Interactive Method to Improve Crowdsourced
  Annotations}.
\newblock \bibinfo{journal}{\emph{IEEE Transactions on Visualization and
  Computer Graphics}} \bibinfo{volume}{25}, \bibinfo{number}{1}
  (\bibinfo{date}{Jan.} \bibinfo{year}{2019}), \bibinfo{pages}{235--245}.
\newblock
\showISSN{2160-9306}


\bibitem[\protect\citeauthoryear{Maas, Daly, Pham, Huang, Ng, and Potts}{Maas
  et~al\mbox{.}}{2011}]%
        {Maas2011Learning}
\bibfield{author}{\bibinfo{person}{Andrew~L. Maas}, \bibinfo{person}{Raymond~E.
  Daly}, \bibinfo{person}{Peter~T. Pham}, \bibinfo{person}{Dan Huang},
  \bibinfo{person}{Andrew~Y. Ng}, {and} \bibinfo{person}{Christopher Potts}.}
  \bibinfo{year}{2011}\natexlab{}.
\newblock \showarticletitle{Learning Word Vectors for Sentiment Analysis}. In
  \bibinfo{booktitle}{\emph{Proceedings of the Annual Meeting of the
  Association for Computational Linguistics: Human Language Technologies}}.
  \bibinfo{publisher}{Association for Computational Linguistics},
  \bibinfo{pages}{142--150}.
\newblock
\showISBNx{9781932432879}


\bibitem[\protect\citeauthoryear{M{\'{e}}ndez, Nacenta, and
  Vandenheste}{M{\'{e}}ndez et~al\mbox{.}}{2016}]%
        {Mendez2016iVoLVER}
\bibfield{author}{\bibinfo{person}{Gonzalo~Gabriel M{\'{e}}ndez},
  \bibinfo{person}{Miguel~A. Nacenta}, {and} \bibinfo{person}{Sebastien
  Vandenheste}.} \bibinfo{year}{2016}\natexlab{}.
\newblock \showarticletitle{iVoLVER: Interactive Visual Language for
  Visualization Extraction and Reconstruction}. In
  \bibinfo{booktitle}{\emph{Proceedings of the SIGCHI Conference on Human
  Factors in Computing Systems}}. \bibinfo{publisher}{ACM},
  \bibinfo{pages}{4073--4085}.
\newblock
\showISBNx{9781450333627}


\bibitem[\protect\citeauthoryear{Microsoft}{Microsoft}{2017}]%
        {Microsoft2017VoTT}
\bibfield{author}{\bibinfo{person}{Microsoft}.}
  \bibinfo{year}{2017}\natexlab{}.
\newblock \bibinfo{title}{VoTT: Visual Object Tagging Tool}.
\newblock
\newblock
\urldef\tempurl%
\url{https://github.com/microsoft/VoTT}
\showURL{%
\tempurl}


\bibitem[\protect\citeauthoryear{Nakayama, Kubo, Kamura, Taniguchi, and
  Liang}{Nakayama et~al\mbox{.}}{2018}]%
        {Nakayama2018doccano}
\bibfield{author}{\bibinfo{person}{Hiroki Nakayama}, \bibinfo{person}{Takahiro
  Kubo}, \bibinfo{person}{Junya Kamura}, \bibinfo{person}{Yasufumi Taniguchi},
  {and} \bibinfo{person}{Xu Liang}.} \bibinfo{year}{2018}\natexlab{}.
\newblock \bibinfo{title}{{doccano}: Text Annotation Tool for Human}.
\newblock
\newblock
\urldef\tempurl%
\url{https://github.com/doccano/doccano}
\showURL{%
\tempurl}


\bibitem[\protect\citeauthoryear{Northcutt, Jiang, and Chuang}{Northcutt
  et~al\mbox{.}}{2021}]%
        {Northcutt2021Confident}
\bibfield{author}{\bibinfo{person}{Curtis Northcutt}, \bibinfo{person}{Lu
  Jiang}, {and} \bibinfo{person}{Isaac Chuang}.}
  \bibinfo{year}{2021}\natexlab{}.
\newblock \showarticletitle{Confident Learning: Estimating Uncertainty in
  Dataset Labels}.
\newblock \bibinfo{journal}{\emph{Journal of Artificial Intelligence Research}}
   \bibinfo{volume}{70} (\bibinfo{date}{may} \bibinfo{year}{2021}),
  \bibinfo{pages}{1373--1411}.
\newblock
\showISSN{1076-9757}


\bibitem[\protect\citeauthoryear{Oelen, Stocker, and Auer}{Oelen
  et~al\mbox{.}}{2021}]%
        {Oelen2021Crowdsourcing}
\bibfield{author}{\bibinfo{person}{Allard Oelen}, \bibinfo{person}{Markus
  Stocker}, {and} \bibinfo{person}{S\"{o}ren Auer}.}
  \bibinfo{year}{2021}\natexlab{}.
\newblock \showarticletitle{Crowdsourcing Scholarly Discourse Annotations}. In
  \bibinfo{booktitle}{\emph{Proceedings of the International Conference on
  Intelligent User Interfaces}}. \bibinfo{publisher}{ACM},
  \bibinfo{pages}{464--474}.
\newblock
\showISBNx{9781450380171}


\bibitem[\protect\citeauthoryear{Paiva, Schwartz, Pedrini, and Minghim}{Paiva
  et~al\mbox{.}}{2015}]%
        {Paiva2015Approach}
\bibfield{author}{\bibinfo{person}{Jose Gustavo~S. Paiva},
  \bibinfo{person}{William~Robson Schwartz}, \bibinfo{person}{Helio Pedrini},
  {and} \bibinfo{person}{Rosane Minghim}.} \bibinfo{year}{2015}\natexlab{}.
\newblock \showarticletitle{An Approach to Supporting Incremental Visual Data
  Classification}.
\newblock \bibinfo{journal}{\emph{IEEE Transactions on Visualization and
  Computer Graphics}} \bibinfo{volume}{21}, \bibinfo{number}{1}
  (\bibinfo{date}{Jan.} \bibinfo{year}{2015}), \bibinfo{pages}{4--17}.
\newblock
\showISSN{1077-2626}


\bibitem[\protect\citeauthoryear{Panayotov, Chen, Povey, and
  Khudanpur}{Panayotov et~al\mbox{.}}{2015}]%
        {Panayotov2015Librispeech}
\bibfield{author}{\bibinfo{person}{Vassil Panayotov}, \bibinfo{person}{Guoguo
  Chen}, \bibinfo{person}{Daniel Povey}, {and} \bibinfo{person}{Sanjeev
  Khudanpur}.} \bibinfo{year}{2015}\natexlab{}.
\newblock \showarticletitle{Librispeech: An {ASR} corpus based on public domain
  audio books}. In \bibinfo{booktitle}{\emph{Proceedings of the IEEE
  International Conference on Acoustics, Speech and Signal Processing}}.
  \bibinfo{publisher}{IEEE}, \bibinfo{pages}{5206--5210}.
\newblock
\showISSN{1520-6149}


\bibitem[\protect\citeauthoryear{Ratner, Bach, Ehrenberg, Fries, Wu, and
  R\'{e}}{Ratner et~al\mbox{.}}{2017}]%
        {Ratner2017Snorkel}
\bibfield{author}{\bibinfo{person}{Alexander Ratner},
  \bibinfo{person}{Stephen~H. Bach}, \bibinfo{person}{Henry Ehrenberg},
  \bibinfo{person}{Jason Fries}, \bibinfo{person}{Sen Wu}, {and}
  \bibinfo{person}{Christopher R\'{e}}.} \bibinfo{year}{2017}\natexlab{}.
\newblock \showarticletitle{Snorkel: Rapid Training Data Creation with Weak
  Supervision}.
\newblock \bibinfo{journal}{\emph{Proceedings of the VLDB Endowment}}
  \bibinfo{volume}{11}, \bibinfo{number}{3} (\bibinfo{date}{Nov.}
  \bibinfo{year}{2017}), \bibinfo{pages}{269--282}.
\newblock
\showISSN{2150-8097}


\bibitem[\protect\citeauthoryear{Ren, Hollerer, and Yuan}{Ren
  et~al\mbox{.}}{2014}]%
        {Ren2014iVisDesigner}
\bibfield{author}{\bibinfo{person}{Donghao Ren}, \bibinfo{person}{Tobias
  Hollerer}, {and} \bibinfo{person}{Xiaoru Yuan}.}
  \bibinfo{year}{2014}\natexlab{}.
\newblock \showarticletitle{iVisDesigner: Expressive Interactive Design of
  Information Visualizations}.
\newblock \bibinfo{journal}{\emph{IEEE Transactions on Visualization and
  Computer Graphics}} \bibinfo{volume}{20}, \bibinfo{number}{12}
  (\bibinfo{date}{Dec.} \bibinfo{year}{2014}), \bibinfo{pages}{2092--2101}.
\newblock
\showISSN{1077-2626}


\bibitem[\protect\citeauthoryear{Rietz and Maedche}{Rietz and Maedche}{2021}]%
        {Rietz2021Cody}
\bibfield{author}{\bibinfo{person}{Tim Rietz} {and} \bibinfo{person}{Alexander
  Maedche}.} \bibinfo{year}{2021}\natexlab{}.
\newblock \showarticletitle{Cody: An AI-Based System to Semi-Automate Coding
  for Qualitative Research}. In \bibinfo{booktitle}{\emph{Proceedings of the
  SIGCHI Conference on Human Factors in Computing Systems}}.
  \bibinfo{publisher}{ACM}, Article \bibinfo{articleno}{394},
  \bibinfo{numpages}{14}~pages.
\newblock
\showISBNx{9781450380966}


\bibitem[\protect\citeauthoryear{Russakovsky, Li, and Fei-Fei}{Russakovsky
  et~al\mbox{.}}{2015}]%
        {Russakovsky2015Best}
\bibfield{author}{\bibinfo{person}{Olga Russakovsky}, \bibinfo{person}{Li-Jia
  Li}, {and} \bibinfo{person}{Li Fei-Fei}.} \bibinfo{year}{2015}\natexlab{}.
\newblock \showarticletitle{Best of both worlds: Human-machine collaboration
  for object annotation}. In \bibinfo{booktitle}{\emph{Proceedings of IEEE
  Conference on Computer Vision and Pattern Recognition}}.
  \bibinfo{publisher}{IEEE}, \bibinfo{pages}{2121--2131}.
\newblock


\bibitem[\protect\citeauthoryear{Russell, Torralba, Murphy, and
  Freeman}{Russell et~al\mbox{.}}{2008}]%
        {Russell2008LabelMe}
\bibfield{author}{\bibinfo{person}{Bryan~C. Russell}, \bibinfo{person}{Antonio
  Torralba}, \bibinfo{person}{Kevin~P. Murphy}, {and}
  \bibinfo{person}{William~T. Freeman}.} \bibinfo{year}{2008}\natexlab{}.
\newblock \showarticletitle{LabelMe: A Database and Web-Based Tool for Image
  Annotation}.
\newblock \bibinfo{journal}{\emph{International Journal of Computer Vision}}
  \bibinfo{volume}{77}, \bibinfo{number}{1-3} (\bibinfo{year}{2008}),
  \bibinfo{pages}{157--173}.
\newblock


\bibitem[\protect\citeauthoryear{Sch{\"{u}}ldt, Laptev, and
  Caputo}{Sch{\"{u}}ldt et~al\mbox{.}}{2004}]%
        {Schuldt2004Recognizing}
\bibfield{author}{\bibinfo{person}{Christian Sch{\"{u}}ldt},
  \bibinfo{person}{Ivan Laptev}, {and} \bibinfo{person}{Barbara Caputo}.}
  \bibinfo{year}{2004}\natexlab{}.
\newblock \showarticletitle{Recognizing Human Actions: A Local SVM Approach}.
  In \bibinfo{booktitle}{\emph{Proceedings of the International Conference on
  Pattern Recognition}}, Vol.~\bibinfo{volume}{3}. \bibinfo{publisher}{IEEE},
  \bibinfo{pages}{32--36}.
\newblock
\showISBNx{0-7695-2128-2}
\showISSN{1051-4651}


\bibitem[\protect\citeauthoryear{Settles}{Settles}{2009}]%
        {Settles2009Active}
\bibfield{author}{\bibinfo{person}{Burr Settles}.}
  \bibinfo{year}{2009}\natexlab{}.
\newblock \bibinfo{booktitle}{\emph{Active Learning Literature Survey}}.
\newblock \bibinfo{type}{Technical Report}. \bibinfo{institution}{University of
  Wisconsin--Madison}.
\newblock


\bibitem[\protect\citeauthoryear{Settles}{Settles}{2011}]%
        {Settles2011Closing}
\bibfield{author}{\bibinfo{person}{Burr Settles}.}
  \bibinfo{year}{2011}\natexlab{}.
\newblock \showarticletitle{Closing the Loop: Fast, Interactive Semi-supervised
  Annotation with Queries on Features and Instances}. In
  \bibinfo{booktitle}{\emph{Proceedings of the Conference on Empirical Methods
  in Natural Language Processing}}. \bibinfo{publisher}{Association for
  Computational Linguistics}, \bibinfo{pages}{1467--1478}.
\newblock
\showISBNx{978-1-937284-11-4}


\bibitem[\protect\citeauthoryear{Shang, Di, Xiao, Cao, Yang, and Chua}{Shang
  et~al\mbox{.}}{2019}]%
        {Shang2019Annotating}
\bibfield{author}{\bibinfo{person}{Xindi Shang}, \bibinfo{person}{Donglin Di},
  \bibinfo{person}{Junbin Xiao}, \bibinfo{person}{Yu Cao}, \bibinfo{person}{Xun
  Yang}, {and} \bibinfo{person}{Tat-Seng Chua}.}
  \bibinfo{year}{2019}\natexlab{}.
\newblock \showarticletitle{Annotating Objects and Relations in User-Generated
  Videos}. In \bibinfo{booktitle}{\emph{Proceedings of the International
  Conference on Multimedia Retrieval}}. \bibinfo{publisher}{ACM},
  \bibinfo{pages}{279--287}.
\newblock
\showISBNx{9781450367653}


\bibitem[\protect\citeauthoryear{Suh and Bederson}{Suh and Bederson}{2007}]%
        {Suh2007Semi}
\bibfield{author}{\bibinfo{person}{Bongwon Suh} {and}
  \bibinfo{person}{Benjamin~B. Bederson}.} \bibinfo{year}{2007}\natexlab{}.
\newblock \showarticletitle{Semi-Automatic Photo Annotation Strategies Using
  Event Based Clustering and Clothing Based Person Recognition}.
\newblock \bibinfo{journal}{\emph{Interacting with Computers}}
  \bibinfo{volume}{19}, \bibinfo{number}{4} (\bibinfo{date}{July}
  \bibinfo{year}{2007}), \bibinfo{pages}{524--544}.
\newblock
\showISSN{0953-5438}


\bibitem[\protect\citeauthoryear{Tang, Chen, Wang, Yan, Chua, and Jain}{Tang
  et~al\mbox{.}}{2013}]%
        {Tang2013Towards}
\bibfield{author}{\bibinfo{person}{Jinhui Tang}, \bibinfo{person}{Qiang Chen},
  \bibinfo{person}{Meng Wang}, \bibinfo{person}{Shuicheng Yan},
  \bibinfo{person}{Tat-Seng Chua}, {and} \bibinfo{person}{Ramesh Jain}.}
  \bibinfo{year}{2013}\natexlab{}.
\newblock \showarticletitle{Towards Optimizing Human Labeling for Interactive
  Image Tagging}.
\newblock \bibinfo{journal}{\emph{ACM Transactions on Multimedia Computing,
  Communications, and Applications}} \bibinfo{volume}{9}, \bibinfo{number}{4},
  Article \bibinfo{articleno}{29} (\bibinfo{date}{Aug.} \bibinfo{year}{2013}),
  \bibinfo{numpages}{18}~pages.
\newblock
\showISSN{1551-6857}


\bibitem[\protect\citeauthoryear{Tian, Liu, Xiao, Wen, and Tang}{Tian
  et~al\mbox{.}}{2007}]%
        {Tian2007Face}
\bibfield{author}{\bibinfo{person}{Yuandong Tian}, \bibinfo{person}{Wei Liu},
  \bibinfo{person}{Rong Xiao}, \bibinfo{person}{Fang Wen}, {and}
  \bibinfo{person}{Xiaoou Tang}.} \bibinfo{year}{2007}\natexlab{}.
\newblock \showarticletitle{A Face Annotation Framework with Partial Clustering
  and Interactive Labeling}. In \bibinfo{booktitle}{\emph{Proceedings of the
  IEEE Conference on Computer Vision and Pattern Recognition}}.
  \bibinfo{publisher}{IEEE}, \bibinfo{pages}{1--8}.
\newblock
\showISBNx{1-4244-1179-3}
\showISSN{1063-6919}


\bibitem[\protect\citeauthoryear{Tkachenko, Malyuk, Shevchenko, Holmanyuk, and
  Liubimov}{Tkachenko et~al\mbox{.}}{2020}]%
        {Tkachenko2020Label}
\bibfield{author}{\bibinfo{person}{Maxim Tkachenko}, \bibinfo{person}{Mikhail
  Malyuk}, \bibinfo{person}{Nikita Shevchenko}, \bibinfo{person}{Andrey
  Holmanyuk}, {and} \bibinfo{person}{Nikolai Liubimov}.}
  \bibinfo{year}{2020}\natexlab{}.
\newblock \bibinfo{title}{{Label Studio}: Data labeling software}.
\newblock
\newblock
\urldef\tempurl%
\url{https://github.com/heartexlabs/label-studio}
\showURL{%
\tempurl}


\bibitem[\protect\citeauthoryear{von Ahn and Dabbish}{von Ahn and
  Dabbish}{2004}]%
        {Ahn2004Labeling}
\bibfield{author}{\bibinfo{person}{Luis von Ahn} {and} \bibinfo{person}{Laura
  Dabbish}.} \bibinfo{year}{2004}\natexlab{}.
\newblock \showarticletitle{Labeling Images with a Computer Game}. In
  \bibinfo{booktitle}{\emph{Proceedings of the SIGCHI Conference on Human
  Factors in Computing Systems}}. \bibinfo{publisher}{ACM},
  \bibinfo{pages}{319--326}.
\newblock
\showISBNx{1581137028}


\bibitem[\protect\citeauthoryear{von Ahn and Dabbish}{von Ahn and
  Dabbish}{2008}]%
        {Ahn2008Designing}
\bibfield{author}{\bibinfo{person}{Luis von Ahn} {and} \bibinfo{person}{Laura
  Dabbish}.} \bibinfo{year}{2008}\natexlab{}.
\newblock \showarticletitle{Designing Games with a Purpose}.
\newblock \bibinfo{journal}{\emph{Commun. ACM}} \bibinfo{volume}{51},
  \bibinfo{number}{8} (\bibinfo{date}{Aug.} \bibinfo{year}{2008}),
  \bibinfo{pages}{58--67}.
\newblock
\showISSN{0001-0782}


\bibitem[\protect\citeauthoryear{von Ahn, Liu, and Blum}{von Ahn
  et~al\mbox{.}}{2006}]%
        {Ahn2006Peekaboom}
\bibfield{author}{\bibinfo{person}{Luis von Ahn}, \bibinfo{person}{Ruoran Liu},
  {and} \bibinfo{person}{Manuel Blum}.} \bibinfo{year}{2006}\natexlab{}.
\newblock \showarticletitle{Peekaboom: A Game for Locating Objects in Images}.
  In \bibinfo{booktitle}{\emph{Proceedings of the SIGCHI Conference on Human
  Factors in Computing Systems}}. \bibinfo{publisher}{ACM},
  \bibinfo{pages}{55--64}.
\newblock
\showISBNx{1595933727}


\bibitem[\protect\citeauthoryear{von Ahn, Maurer, McMillen, Abraham, and
  Blum}{von Ahn et~al\mbox{.}}{2008}]%
        {Ahn2008reCAPTCHA}
\bibfield{author}{\bibinfo{person}{Luis von Ahn}, \bibinfo{person}{Benjamin
  Maurer}, \bibinfo{person}{Colin McMillen}, \bibinfo{person}{David Abraham},
  {and} \bibinfo{person}{Manuel Blum}.} \bibinfo{year}{2008}\natexlab{}.
\newblock \showarticletitle{{reCAPTCHA}: Human-Based Character Recognition via
  Web Security Measures}.
\newblock \bibinfo{journal}{\emph{Science}} \bibinfo{volume}{321},
  \bibinfo{number}{5895} (\bibinfo{date}{aug} \bibinfo{year}{2008}),
  \bibinfo{pages}{1465--1468}.
\newblock
\showISSN{0036-8075}


\bibitem[\protect\citeauthoryear{Wang, Zuluaga, Li, Pratt, Patel, Aertsen,
  Doel, David, Deprest, Ourselin, and Vercauteren}{Wang et~al\mbox{.}}{2019}]%
        {Wang2019DeepIGeoS}
\bibfield{author}{\bibinfo{person}{Guotai Wang}, \bibinfo{person}{Maria~A.
  Zuluaga}, \bibinfo{person}{Wenqi Li}, \bibinfo{person}{Rosalind Pratt},
  \bibinfo{person}{Premal~A. Patel}, \bibinfo{person}{Michael Aertsen},
  \bibinfo{person}{Tom Doel}, \bibinfo{person}{Anna~L. David},
  \bibinfo{person}{Jan Deprest}, \bibinfo{person}{Sébastien Ourselin}, {and}
  \bibinfo{person}{Tom Vercauteren}.} \bibinfo{year}{2019}\natexlab{}.
\newblock \showarticletitle{{DeepIGeoS}: A Deep Interactive Geodesic Framework
  for Medical Image Segmentation}.
\newblock \bibinfo{journal}{\emph{IEEE Transactions on Pattern Analysis and
  Machine Intelligence}} \bibinfo{volume}{41}, \bibinfo{number}{7}
  (\bibinfo{date}{jul} \bibinfo{year}{2019}), \bibinfo{pages}{1559--1572}.
\newblock


\bibitem[\protect\citeauthoryear{Wang and Hua}{Wang and Hua}{2011}]%
        {Wang2011Active}
\bibfield{author}{\bibinfo{person}{Meng Wang} {and} \bibinfo{person}{Xian-Sheng
  Hua}.} \bibinfo{year}{2011}\natexlab{}.
\newblock \showarticletitle{Active Learning in Multimedia Annotation and
  Retrieval: A Survey}.
\newblock \bibinfo{journal}{\emph{ACM Transactions on Intelligent Systems and
  Technology}} \bibinfo{volume}{2}, \bibinfo{number}{2}, Article
  \bibinfo{articleno}{10} (\bibinfo{date}{Feb.} \bibinfo{year}{2011}),
  \bibinfo{numpages}{21}~pages.
\newblock
\showISSN{2157-6904}


\bibitem[\protect\citeauthoryear{Wu, Song, Khosla, Yu, Zhang, Tang, and
  Xiao}{Wu et~al\mbox{.}}{2015}]%
        {Wu20153D}
\bibfield{author}{\bibinfo{person}{Zhirong Wu}, \bibinfo{person}{Shuran Song},
  \bibinfo{person}{Aditya Khosla}, \bibinfo{person}{Fisher Yu},
  \bibinfo{person}{Linguang Zhang}, \bibinfo{person}{Xiaoou Tang}, {and}
  \bibinfo{person}{Jianxiong Xiao}.} \bibinfo{year}{2015}\natexlab{}.
\newblock \showarticletitle{3D {ShapeNets}: A Deep Representation for
  Volumetric Shapes}. In \bibinfo{booktitle}{\emph{IEEE Conference on Computer
  Vision and Pattern Recognition}}. \bibinfo{publisher}{IEEE},
  \bibinfo{pages}{1912--1920}.
\newblock
\showISBNx{978-1-4673-6963-3}
\showISSN{1063-6919}


\bibitem[\protect\citeauthoryear{Xiang, Ye, Xia, Wu, Chen, and Liu}{Xiang
  et~al\mbox{.}}{2019}]%
        {Xiang2019Interactive}
\bibfield{author}{\bibinfo{person}{Shouxing Xiang}, \bibinfo{person}{Xi Ye},
  \bibinfo{person}{Jiazhi Xia}, \bibinfo{person}{Jing Wu},
  \bibinfo{person}{Yang Chen}, {and} \bibinfo{person}{Shixia Liu}.}
  \bibinfo{year}{2019}\natexlab{}.
\newblock \showarticletitle{Interactive Correction of Mislabeled Training
  Data}. In \bibinfo{booktitle}{\emph{Proceedings of the IEEE Conference on
  Visual Analytics Science and Technology}}. \bibinfo{publisher}{IEEE},
  \bibinfo{pages}{57--68}.
\newblock
\showISBNx{9781728122854}


\bibitem[\protect\citeauthoryear{Xu, Akella, and Zhang}{Xu
  et~al\mbox{.}}{2007}]%
        {Xu2007Incorporating}
\bibfield{author}{\bibinfo{person}{Zuobing Xu}, \bibinfo{person}{Ram Akella},
  {and} \bibinfo{person}{Yi Zhang}.} \bibinfo{year}{2007}\natexlab{}.
\newblock \showarticletitle{Incorporating Diversity and Density in Active
  Learning for Relevance Feedback}. In \bibinfo{booktitle}{\emph{Proceedings of
  the European Conference on IR Research}}.
  \bibinfo{publisher}{Springer-Verlag}, \bibinfo{pages}{246--257}.
\newblock
\showISBNx{9783540714941}


\bibitem[\protect\citeauthoryear{Yang, Zhu, Guo, Yang, Li, and Yu}{Yang
  et~al\mbox{.}}{2008}]%
        {Yang2008Comprehensive}
\bibfield{author}{\bibinfo{person}{Yang Yang}, \bibinfo{person}{Bin~B. Zhu},
  \bibinfo{person}{Rui Guo}, \bibinfo{person}{Linjun Yang},
  \bibinfo{person}{Shipeng Li}, {and} \bibinfo{person}{Nenghai Yu}.}
  \bibinfo{year}{2008}\natexlab{}.
\newblock \showarticletitle{A Comprehensive Human Computation Framework: With
  Application to Image Labeling}. In \bibinfo{booktitle}{\emph{Proceedings of
  the ACM International Conference on Multimedia}}. \bibinfo{publisher}{ACM},
  \bibinfo{pages}{479--488}.
\newblock
\showISBNx{9781605583037}


\bibitem[\protect\citeauthoryear{Ye, Shao, Wang, Ma, Wang, Zheng, and Xue}{Ye
  et~al\mbox{.}}{2016}]%
        {Ye2016Face}
\bibfield{author}{\bibinfo{person}{Hao Ye}, \bibinfo{person}{Weiyuan Shao},
  \bibinfo{person}{Hong Wang}, \bibinfo{person}{Jianqi Ma}, \bibinfo{person}{Li
  Wang}, \bibinfo{person}{Yingbin Zheng}, {and} \bibinfo{person}{Xiangyang
  Xue}.} \bibinfo{year}{2016}\natexlab{}.
\newblock \showarticletitle{Face Recognition via Active Annotation and
  Learning}. In \bibinfo{booktitle}{\emph{Proceedings of the ACM International
  Conference on Multimedia}}. \bibinfo{publisher}{ACM},
  \bibinfo{pages}{1058--1062}.
\newblock
\showISBNx{9781450336031}


\bibitem[\protect\citeauthoryear{Yu and Silva}{Yu and Silva}{2017}]%
        {Yu2017VisFlow}
\bibfield{author}{\bibinfo{person}{Bowen Yu} {and} \bibinfo{person}{Claudio~T.
  Silva}.} \bibinfo{year}{2017}\natexlab{}.
\newblock \showarticletitle{VisFlow - Web-based Visualization Framework for
  Tabular Data with a Subset Flow Model}.
\newblock \bibinfo{journal}{\emph{IEEE Transactions on Visualization and
  Computer Graphics}} \bibinfo{volume}{23}, \bibinfo{number}{1}
  (\bibinfo{date}{Jan.} \bibinfo{year}{2017}), \bibinfo{pages}{251--260}.
\newblock
\showISSN{1077-2626}


\bibitem[\protect\citeauthoryear{Zah{\'{a}}lka and Worring}{Zah{\'{a}}lka and
  Worring}{2014}]%
        {Zahalka2014Towards}
\bibfield{author}{\bibinfo{person}{Jan Zah{\'{a}}lka} {and}
  \bibinfo{person}{Marcel Worring}.} \bibinfo{year}{2014}\natexlab{}.
\newblock \showarticletitle{Towards Interactive, Intelligent, and Integrated
  Multimedia Analytics}. In \bibinfo{booktitle}{\emph{Proceedings of the IEEE
  Conference on Visual Analytics Science and Technology}}.
  \bibinfo{publisher}{IEEE}, \bibinfo{pages}{3--12}.
\newblock


\bibitem[\protect\citeauthoryear{Zah{\'{a}}lka, Worring, and van
  Wijk}{Zah{\'{a}}lka et~al\mbox{.}}{2020}]%
        {Zahalka2020II}
\bibfield{author}{\bibinfo{person}{Jan Zah{\'{a}}lka}, \bibinfo{person}{Marcel
  Worring}, {and} \bibinfo{person}{Jarke~J. van Wijk}.}
  \bibinfo{year}{2020}\natexlab{}.
\newblock \showarticletitle{{II}-20: Intelligent and pragmatic analytic
  categorization of image collections}.
\newblock \bibinfo{journal}{\emph{IEEE Transactions on Visualization and
  Computer Graphics}} \bibinfo{volume}{27}, \bibinfo{number}{2}
  (\bibinfo{year}{2020}), \bibinfo{pages}{422--431}.
\newblock
\showISSN{1077-2626}


\bibitem[\protect\citeauthoryear{Zhang, Fu, and Li}{Zhang
  et~al\mbox{.}}{2018}]%
        {Zhang2018Collaborative}
\bibfield{author}{\bibinfo{person}{Lishi Zhang}, \bibinfo{person}{Chenghan Fu},
  {and} \bibinfo{person}{Jia Li}.} \bibinfo{year}{2018}\natexlab{}.
\newblock \showarticletitle{Collaborative Annotation of Semantic Objects in
  Images with Multi-Granularity Supervisions}. In
  \bibinfo{booktitle}{\emph{Proceedings of the ACM International Conference on
  Multimedia}}. \bibinfo{publisher}{ACM}, \bibinfo{pages}{474--482}.
\newblock
\showISBNx{9781450356657}


\bibitem[\protect\citeauthoryear{Zhang, Feng, Chen, Chen, Zheng, Luo, Huang,
  and Tung}{Zhang et~al\mbox{.}}{2021b}]%
        {Zhang2021ChartNavigator}
\bibfield{author}{\bibinfo{person}{Tianye Zhang}, \bibinfo{person}{Haozhe
  Feng}, \bibinfo{person}{Wei Chen}, \bibinfo{person}{Zexian Chen},
  \bibinfo{person}{Wenting Zheng}, \bibinfo{person}{Xiao-Nan Luo},
  \bibinfo{person}{Wenqi Huang}, {and} \bibinfo{person}{Anthony K.~H. Tung}.}
  \bibinfo{year}{2021}\natexlab{b}.
\newblock \showarticletitle{{ChartNavigator}: An Interactive Pattern
  Identification and Annotation Framework for Charts}.
\newblock \bibinfo{journal}{\emph{IEEE Transactions on Knowledge and Data
  Engineering}} (\bibinfo{year}{2021}), \bibinfo{pages}{1--1}.
\newblock


\bibitem[\protect\citeauthoryear{Zhang, Cai, Wang, and Zhang}{Zhang
  et~al\mbox{.}}{2017}]%
        {Zhang2017Stopping}
\bibfield{author}{\bibinfo{person}{Yexun Zhang}, \bibinfo{person}{Wenbin Cai},
  \bibinfo{person}{Wenquan Wang}, {and} \bibinfo{person}{Ya Zhang}.}
  \bibinfo{year}{2017}\natexlab{}.
\newblock \showarticletitle{Stopping Criterion for Active Learning with Model
  Stability}.
\newblock \bibinfo{journal}{\emph{ACM Transactions on Intelligent Systems and
  Technology}} \bibinfo{volume}{9}, \bibinfo{number}{2}, Article
  \bibinfo{articleno}{19} (\bibinfo{date}{Oct.} \bibinfo{year}{2017}),
  \bibinfo{numpages}{26}~pages.
\newblock
\showISSN{2157-6904}


\bibitem[\protect\citeauthoryear{Zhang, Coecke, and Chen}{Zhang
  et~al\mbox{.}}{2021a}]%
        {Zhang2021MI3}
\bibfield{author}{\bibinfo{person}{Yu Zhang}, \bibinfo{person}{Bob Coecke},
  {and} \bibinfo{person}{Min Chen}.} \bibinfo{year}{2021}\natexlab{a}.
\newblock \showarticletitle{{MI3}: Machine-Initiated Intelligent Interaction
  for Interactive Classification and Data Reconstruction}.
\newblock \bibinfo{journal}{\emph{ACM Transactions on Interactive Intelligent
  Systems}} \bibinfo{volume}{11}, \bibinfo{number}{3--4}, Article
  \bibinfo{articleno}{18} (\bibinfo{date}{Aug.} \bibinfo{year}{2021}),
  \bibinfo{numpages}{34}~pages.
\newblock
\showISSN{2160-6455}


\bibitem[\protect\citeauthoryear{Zhang, Tennekes, de~Jong, Curier, Coecke, and
  Chen}{Zhang et~al\mbox{.}}{2021c}]%
        {Zhang2021Using}
\bibfield{author}{\bibinfo{person}{Yu Zhang}, \bibinfo{person}{Martijn
  Tennekes}, \bibinfo{person}{Tim de Jong}, \bibinfo{person}{Lyana Curier},
  \bibinfo{person}{Bob Coecke}, {and} \bibinfo{person}{Min Chen}.}
  \bibinfo{year}{2021}\natexlab{c}.
\newblock \bibinfo{title}{Using Simulation to Aid the Design and Optimization
  of Intelligent User Interfaces for Quality Assurance Processes in Machine
  Learning}.
\newblock
\newblock
\showeprint[arXiv]{2104.01129}~[cs.HC]


\bibitem[\protect\citeauthoryear{Zhu, Wang, Hovy, and Ma}{Zhu
  et~al\mbox{.}}{2010}]%
        {Zhu2010Confidence}
\bibfield{author}{\bibinfo{person}{Jingbo Zhu}, \bibinfo{person}{Huizhen Wang},
  \bibinfo{person}{Eduard Hovy}, {and} \bibinfo{person}{Matthew Ma}.}
  \bibinfo{year}{2010}\natexlab{}.
\newblock \showarticletitle{Confidence-Based Stopping Criteria for Active
  Learning for Data Annotation}.
\newblock \bibinfo{journal}{\emph{ACM Transactions on Speech and Language
  Processing}} \bibinfo{volume}{6}, \bibinfo{number}{3}, Article
  \bibinfo{articleno}{3} (\bibinfo{date}{April} \bibinfo{year}{2010}),
  \bibinfo{numpages}{24}~pages.
\newblock
\showISSN{1550-4875}


\bibitem[\protect\citeauthoryear{Zhu, Wu, and Chen}{Zhu et~al\mbox{.}}{2003}]%
        {Zhu2003Eliminating}
\bibfield{author}{\bibinfo{person}{Xingquan Zhu}, \bibinfo{person}{Xindong Wu},
  {and} \bibinfo{person}{Qijun Chen}.} \bibinfo{year}{2003}\natexlab{}.
\newblock \showarticletitle{Eliminating Class Noise in Large Datasets}. In
  \bibinfo{booktitle}{\emph{Proceedings of the International Conference on
  Machine Learning}}. \bibinfo{publisher}{AAAI Press},
  \bibinfo{pages}{920--927}.
\newblock
\showISBNx{1577351894}


\end{thebibliography}

\pagebreak
\appendix

\section{Coding Details}
\label{sec:supp-coding}

In the coding process, we first extract phrases from the flowchart figures.
Each phrase forms a preliminary code.
The preliminary codes are first categorized into states and modules.
Then, we categorize preliminary codes into themes.
Finally, we filter out irrelevant themes and merge related themes into the final codes.
In the following, we introduce details of the coding process, including the themes we excluded, the occurrences of final codes in the 36 flowcharts we have coded, and an example of coding a recent labeling tool.

\subsection{Themes}
\label{sec:themes}

\noindent\textbf{State themes}:
The 163 preliminary state codes are grouped into 10 themes.
5 of the themes are directly included in the final codes as described in Section~\ref{sec:framework}, including \textbf{Data Objects}, \textbf{Labels}, \textbf{Samples}, \textbf{Model}, and \textbf{Features}.
The other 5 themes we identified are (marked with frequency):
\begin{itemize}[leftmargin=*]
    \item \textbf{Labeled data}: the dataset together with (partial) labels (22/163).
    \item \textbf{Search related}: the states related to searching (7/163).
    \item \textbf{Knowledge}: the states related to annotators' knowledge (4/163).
    \item \textbf{Data source}: the source where data objects are extracted (2/163).
    \item \textbf{Others}: the other states not fitting previous themes (9/163).
\end{itemize}

The theme ``labeled data'' maps to two final codes ``data objects'' and ``labels'' as ``labeled data'' covers both aspects.
There are 2 cases where we group one preliminary code into two themes, where the preliminary code contains the word ``with'' (e.g., ``selected subsets with default labels'' in Zhang et al.~\cite{Zhang2021MI3}'s Figure 3).

\noindent\textbf{Modules themes}:
The 188 preliminary are grouped into 16 themes.
7 of the themes are directly included in the final codes as described in Section~\ref{sec:framework}, including \textbf{data object selection}, \textbf{model training}, \textbf{feature extraction}, \textbf{default labeling}, \textbf{quality assurance}, \textbf{stoppage analysis}, and \textbf{label ideation}.
The other 9 themes we identified are (marked with frequency):
\begin{itemize}[leftmargin=*]
    \item \textbf{User labeling}: the process where annotators create/edit data labels (33/188).
    \item \textbf{Labeling interface}: the process to present data (possibly with labels) to annotators for creating/editing labels (17/188).
    \item \textbf{Preprocessing}: the application-specific process to precompute auxiliary data structures used in the data labeling tool (11/188).
    \item \textbf{Model understanding}: the process to understand the prediction criteria or quality of the model (5/188).
    \item \textbf{Data exploration}: the process to browse the dataset for the purpose of familiarizing with the dataset (4/188).
    \item \textbf{Postprocessing}: the process to postprocess data labeling results (i.e., labels) for application-specific needs (4/188).
    \item \textbf{Data collection}: the process to collect data or enlarge the dataset (4/188).
    \item \textbf{User modeling}: the process to model user and interpret user intent (2/188).
    \item \textbf{Others}: the other processes not fitting previous themes (13/188).
\end{itemize}

The themes ``user labeling'' and ``labeling interface'' are merged into the final code ``interactive labeling'' as the former theme is related to user interaction in interactive labeling, and the latter is related to the interface supporting interactive labeling.
For module coding, there are 8 edge cases where we group one preliminary code into multiple themes.
It happens typically when the phrase contains the word ``and'', such as ``sampling and annotation'' in Liao et al.~\cite{Liao2016Visualization}'s Figure 1.
For the preliminary codes for both states and modules, we categorize them into ``others'' when the preliminary code is hard to categorize, typically when the code is too verbose (e.g., ``adding to labeled dataset'' in Zhang et al.~\cite{Zhang2021MI3}'s Figure 3) and when the code is too application-specific (e.g., ``superpixels'' in Zhang et al.~\cite{Zhang2018Collaborative}'s Figure 2).

\subsection{State and Module Occurrences}

\begin{table}[!ht]
    \caption{
        Occurrences of states in the literature.
        The final codes are abbreviated by initials (e.g., ``DO'' refers to ``data objects'').
        Slash (/)  is used in final codes when the flowchart contains no states or the flowchart's states are excluded in the coding procedure.
    }
    \label{table:coding-states}
    \centering
    \tiny
    \begin{tabular}{lllccc}
        \toprule
        \textbf{ID} & \textbf{Paper}                                         & \textbf{Venue} & \textbf{Year} & \textbf{Figure Index} & \textbf{Final Codes} \\
        \midrule
        1           & Fails2003Interactive~\cite{Fails2003Interactive}       & IUI            & 2003          & 2                     & /                    \\
        2           & Hoi2005Semi~\cite{Hoi2005Semi}                         & CVPR           & 2005          & 1                     & DO, L, S             \\
        3           & Tian2007Face~\cite{Tian2007Face}                       & CVPR           & 2007          & 1                     & DO, L, S, F          \\
        4           & Cui2007EasyAlbum~\cite{Cui2007EasyAlbum}               & CHI            & 2007          & 13                    & DO, L, F             \\
        5           & Hua2008Online~\cite{Hua2008Online}                     & MM             & 2008          & 2                     & DO                   \\
        6           & Rooij2010MediaTable~\cite{Rooij2010MediaTable}         & CG\&A          & 2010          & 3                     & DO, L, S             \\
        7           & Wang2011Active~\cite{Wang2011Active}                   & TIST           & 2011          & 3                     & DO, L, M             \\
        8           & Wang2011Active~\cite{Wang2011Active}                   & TIST           & 2011          & 5                     & DO, L, M             \\
        9           & Wang2011Active~\cite{Wang2011Active}                   & TIST           & 2011          & 7                     & DO                   \\
        10          & Hoeferlin2012Inter~\cite{Hoeferlin2012Inter}           & VAST           & 2012          & 1(b)                  & DO, L, M             \\
        11          & Tang2013Towards~\cite{Tang2013Towards}                 & TOMM           & 2013          & 2                     & DO, L, S             \\
        12          & Zahalka2014Towards~\cite{Zahalka2014Towards}           & VAST           & 2014          & 9                     & DO, L, M, F          \\
        13          & Bryan2014ISSE~\cite{Bryan2014ISSE}                     & CHI            & 2014          & 5                     & DO                   \\
        14          & Paiva2015Approach~\cite{Paiva2015Approach}             & TVCG           & 2015          & 1                     & DO                   \\
        15          & Russakovsky2015Best~\cite{Russakovsky2015Best}         & CVPR           & 2015          & 2                     & L, S                 \\
        16          & Liao2016Visualization~\cite{Liao2016Visualization}     & TMM            & 2016          & 1                     & DO, L, F             \\
        17          & Ye2016Face~\cite{Ye2016Face}                           & MM             & 2016          & 1                     & DO, L, S             \\
        18          & Kucher2017Active~\cite{Kucher2017Active}               & TIIS           & 2017          & 1                     & DO, L, M             \\
        19          & Ratner2017Snorkel~\cite{Ratner2017Snorkel}             & VLDB           & 2017          & 2                     & DO, L, M             \\
        20          & Bernard2018VIAL~\cite{Bernard2018VIAL}                 & TVC            & 2018          & 1                     & DO, L, S             \\
        21          & Bernard2018VIAL~\cite{Bernard2018VIAL}                 & TVC            & 2018          & 2                     & DO, L, S             \\
        22          & Felix2018Exploratory~\cite{Felix2018Exploratory}       & UIST           & 2018          & 1                     & L                    \\
        23          & Zhang2018Collaborative~\cite{Zhang2018Collaborative}   & MM             & 2018          & 2                     & DO, L                \\
        24          & Shang2019Annotating~\cite{Shang2019Annotating}         & ICMR           & 2019          & 2                     & /                    \\
        25          & Xiang2019Interactive~\cite{Xiang2019Interactive}       & VAST           & 2019          & 2                     & DO, L, S             \\
        26          & Liu2019Interactive~\cite{Liu2019Interactive}           & TVCG           & 2019          & 3                     & DO, L                \\
        27          & Choi2019AILA~\cite{Choi2019AILA}                       & CHI            & 2019          & 6                     & /                    \\
        28          & Wang2019DeepIGeoS~\cite{Wang2019DeepIGeoS}             & TPAMI          & 2019          & 1                     & DO, L                \\
        29          & Halter2019VIAN~\cite{Halter2019VIAN}                   & CGF            & 2019          & 2                     & DO, L, S, F          \\
        30          & Evensen2020Ruler~\cite{Evensen2020Ruler}               & EMNLP          & 2020          & 2                     & DO, S, M             \\
        31          & Baeuerle2020Classifier~\cite{Baeuerle2020Classifier}   & CGF            & 2020          & 1                     & DO, L                \\
        32          & Lekschas2020Peax~\cite{Lekschas2020Peax}               & CGF            & 2020          & 3                     & DO, L                \\
        33          & Oelen2021Crowdsourcing~\cite{Oelen2021Crowdsourcing}   & IUI            & 2021          & 2                     & DO, L, M             \\
        34          & Rietz2021Cody~\cite{Rietz2021Cody}                     & CHI            & 2021          & 4                     & DO, L, M             \\
        35          & Zhang2021ChartNavigator~\cite{Zhang2021ChartNavigator} & TKDE           & 2021          & 1                     & DO, F                \\
        36          & Zhang2021MI3~\cite{Zhang2021MI3}                       & TIIS           & 2021          & 3                     & DO, L, S, M          \\
        \bottomrule
    \end{tabular}
\end{table}

Table~\ref{table:coding-states} summarizes the occurrences of final codes for states in the 36 flowchart figures.
In the table, for each flowchart figure, if the flowchart contains at least one phrase that corresponds to a specific final code, we add the final code to the ``Final Codes'' column.
When a flowchart figure contains no phrases that refer to inputs/outputs, or that the phrases that refer to inputs/outputs are excluded in the processing of generating final codes from the themes, we mark the figure's ``Final Codes'' empty, represented by a slash (/).
Similarly, Table~\ref{table:coding-modules} summarizes the occurrences of final codes for modules in the 36 flowchart figures.

\begin{table}[!ht]
    \caption{
        Occurrences of modules in the literature.
        The final codes are abbreviated by initials (e.g., ``IL'' refers to ``interactive labeling'').
    }
    \label{table:coding-modules}
    \centering
    \tiny
    \begin{tabular}{lllccc}
        \toprule
        \textbf{ID} & \textbf{Paper}                                         & \textbf{Venue} & \textbf{Year} & \textbf{Figure Index} & \textbf{Final Codes}        \\
        \midrule
        1           & Fails2003Interactive~\cite{Fails2003Interactive}       & IUI            & 2003          & 2                     & IL, MT, DL                  \\
        2           & Hoi2005Semi~\cite{Hoi2005Semi}                         & CVPR           & 2005          & 1                     & IL, DOS, MT                 \\
        3           & Tian2007Face~\cite{Tian2007Face}                       & CVPR           & 2007          & 1                     & IL, DOS, FE                 \\
        4           & Cui2007EasyAlbum~\cite{Cui2007EasyAlbum}               & CHI            & 2007          & 13                    & IL, DOS, FE                 \\
        5           & Hua2008Online~\cite{Hua2008Online}                     & MM             & 2008          & 2                     & DOS                         \\
        6           & Rooij2010MediaTable~\cite{Rooij2010MediaTable}         & CG\&A          & 2010          & 3                     & IL, DOS, LI                 \\
        7           & Wang2011Active~\cite{Wang2011Active}                   & TIST           & 2011          & 3                     & IL, DOS, MT                 \\
        8           & Wang2011Active~\cite{Wang2011Active}                   & TIST           & 2011          & 5                     & IL, DOS, MT                 \\
        9           & Wang2011Active~\cite{Wang2011Active}                   & TIST           & 2011          & 7                     & IL, DOS, MT, QA             \\
        10          & Hoeferlin2012Inter~\cite{Hoeferlin2012Inter}           & VAST           & 2012          & 1(b)                  & IL, DOS, MT                 \\
        11          & Tang2013Towards~\cite{Tang2013Towards}                 & TOMM           & 2013          & 2                     & IL, DOS, SA                 \\
        12          & Zahalka2014Towards~\cite{Zahalka2014Towards}           & VAST           & 2014          & 9                     & IL, MT, FE, LI              \\
        13          & Bryan2014ISSE~\cite{Bryan2014ISSE}                     & CHI            & 2014          & 5                     & IL, MT                      \\
        14          & Paiva2015Approach~\cite{Paiva2015Approach}             & TVCG           & 2015          & 1                     & IL, DOS, MT, FE, DL         \\
        15          & Russakovsky2015Best~\cite{Russakovsky2015Best}         & CVPR           & 2015          & 2                     & IL, DOS, DL                 \\
        16          & Liao2016Visualization~\cite{Liao2016Visualization}     & TMM            & 2016          & 1                     & IL, DOS, MT                 \\
        17          & Ye2016Face~\cite{Ye2016Face}                           & MM             & 2016          & 1                     & MT                          \\
        18          & Kucher2017Active~\cite{Kucher2017Active}               & TIIS           & 2017          & 1                     & IL                          \\
        19          & Ratner2017Snorkel~\cite{Ratner2017Snorkel}             & VLDB           & 2017          & 2                     & MT                          \\
        20          & Bernard2018VIAL~\cite{Bernard2018VIAL}                 & TVC            & 2018          & 1                     & IL, DOS, FE, DL, SA         \\
        21          & Bernard2018VIAL~\cite{Bernard2018VIAL}                 & TVC            & 2018          & 2                     & IL, DOS, MT, FE, DL         \\
        22          & Felix2018Exploratory~\cite{Felix2018Exploratory}       & UIST           & 2018          & 1                     & IL, QA, LI                  \\
        23          & Zhang2018Collaborative~\cite{Zhang2018Collaborative}   & MM             & 2018          & 2                     & IL, DL                      \\
        24          & Shang2019Annotating~\cite{Shang2019Annotating}         & ICMR           & 2019          & 2                     & IL                          \\
        25          & Xiang2019Interactive~\cite{Xiang2019Interactive}       & VAST           & 2019          & 2                     & IL, DOS, DL                 \\
        26          & Liu2019Interactive~\cite{Liu2019Interactive}           & TVCG           & 2019          & 3                     & IL, DOS, FE, DL, QA         \\
        27          & Choi2019AILA~\cite{Choi2019AILA}                       & CHI            & 2019          & 6                     & IL, DOS, FE                 \\
        28          & Wang2019DeepIGeoS~\cite{Wang2019DeepIGeoS}             & TPAMI          & 2019          & 1                     & IL, DL, SA                  \\
        29          & Halter2019VIAN~\cite{Halter2019VIAN}                   & CGF            & 2019          & 2                     & IL, FE, DL                  \\
        30          & Evensen2020Ruler~\cite{Evensen2020Ruler}               & EMNLP          & 2020          & 2                     & IL, DOS, MT                 \\
        31          & Baeuerle2020Classifier~\cite{Baeuerle2020Classifier}   & CGF            & 2020          & 1                     & MT, DL, QA                  \\
        32          & Lekschas2020Peax~\cite{Lekschas2020Peax}               & CGF            & 2020          & 3                     & IL, DOS                     \\
        33          & Oelen2021Crowdsourcing~\cite{Oelen2021Crowdsourcing}   & IUI            & 2021          & 2                     & IL, DOS, DL                 \\
        34          & Rietz2021Cody~\cite{Rietz2021Cody}                     & CHI            & 2021          & 4                     & IL, MT, DL                  \\
        35          & Zhang2021ChartNavigator~\cite{Zhang2021ChartNavigator} & TKDE           & 2021          & 1                     & DOS, FE                     \\
        36          & Zhang2021MI3~\cite{Zhang2021MI3}                       & TIIS           & 2021          & 3                     & IL, DOS, MT, FE, DL, QA, SA \\
        \bottomrule
    \end{tabular}
\end{table}

\subsection{A Coding Example}
\label{sec:coding-example}

\begin{table}[!ht]
    \caption{
        Coding the flowchart in AILA~\cite{Choi2019AILA}.
        The final codes are abbreviated by initials.
        A slash (/) in the final code refers to the case that the theme is excluded from the final code.
    }
    \label{table:example-aila}
    \centering
    \tiny
    \begin{tabular}{lllccc}
    \toprule
        \textbf{Preliminary Code} & \textbf{Theme} & \textbf{Final Code} \\
    \midrule
        preprocessing - stemming & feature extraction & FE  \\ 
        preprocessing - bag of words & feature extraction & FE  \\ 
        preprocessing - term-document matrix & feature extraction & FE  \\ 
        preprocessing - word vector & feature extraction & FE  \\ 
        preprocessing - sentence vector & feature extraction & FE \\ 
        document analysis - re-ordering - selecting & data object selection & DOS \\ 
        document analysis - re-ordering - sorting & data object selection & DOS  \\ 
        document classifier - interactive attentive module - attention weight & preprocessing & / \\ 
        document classifier - interactive attentive module - prediction score & preprocessing & / \\ 
        labeling interface - document embedding & data object selection & DOS  \\ 
        labeling interface - document visualization & labeling interface & IL  \\ 
    \bottomrule
    \end{tabular}
\end{table}

AILA~\cite{Choi2019AILA} supports labeling document classification.
Figure 6 in its paper is a flowchart.
The figure contains nested blocks, and we use a hyphen to denote the nesting relation in the preliminary code.
For example, ``preprocessing - stemming'' refers to a block named ``stemming'' placed in a block named ``preprocessing''.
We decompose the nested blocks to the lowest level instead of treating the root block, such as ``preprocessing'', as a single phrase because different nested blocks may correspond to different themes.
The flowchart contains 11 phases, all categorized as preliminary codes for modules because they describe actions.
The 11 phrases are grouped into themes and final codes as shown in Table~\ref{table:example-aila}.
We categorize ``attention weight'' and ``prediction score'' as preprocessing because they prepare the data visualized in the labeling interface.
``Labeling interface - document embedding'' is categorized as ``data object selection'' because it refers to an interactive projection where the annotator can select data objects to label.

\section{Graph-Theoretic Constraints on Data Labeling Workflows}
\label{sec:supp-workflow-constraints}

\subsection{Notations}

\begin{itemize}[leftmargin=*]
    \item $ G = (V, E) $: a directed graph modeling a data labeling tool.
    \item $ V $: a set of nodes, each corresponding to a software module, or initialization, decision, and exit node.
    \item $ E $: a set of edges specifying the execution order.
    \item $ type(v) $: the type of a node $ v $, which takes one of the following values: initialization, process, decision, exit.
    \item $ fun(v) $: the function of a node $ v $.
        When $ type(v) $ takes value in $ \{ $ initialization, decision, exit $ \} $, $ fun(v) = type(v) $.
        When $ type(v) = process $, $ fun(v) $ takes one of the following values: interactiveLabeling, dataObjectSelection, modelTraining, featureExtraction, defaultLabeling, qualityAssurance, stoppageAnalysis, labelIdeation.
    \item $ input(v) $: the set of input states to a node $ v $.
    \item $ output(v) $: the output state of a node $ v $.
    \item $ Exec(G) $: the set of all the directed walks on the graph from the initialization node to the exit node.
\end{itemize}

\subsection{Constraints on the Workflow Graph}

\textbf{Valid Flowchart:}
\begin{itemize}[leftmargin=*]
    \item The graph contains no parallel edges (i.e., $ E $ is not a multi-set).
    \item The graph contains one initialization node (denote as $ v_s $).
        $$ \exists! v (v \in V \wedge type(v) = initialization) $$
    \item The graph contains one exit node (denote as $ v_e $).
        $$ \exists! v (v \in V \wedge type(v) = exit) $$
    \item A process node has outdegree 1.
        $$ \forall v (v \in V \wedge type(v) = process \rightarrow deg^+(v) = 1) $$
    \item A decision node has outdegree 2.
        $$ \forall v (v \in V \wedge type(v) = decision \rightarrow deg^+(v) = 2) $$
    \item An initialization node has indegree 0 and outdegree 1.
        $$ \forall v (v \in V \wedge type(v) = initialization \rightarrow deg^+(v) = 1 \wedge deg^-(v) = 0) $$
    \item An exit node has outdegree 0.
        $$ \forall v (v \in V \wedge type(v) = exit \rightarrow deg^+(v) = 0) $$
    \item All the nodes can be reached from the initialization node.
        The exit node can be reached from all the nodes.
        $$ \forall v (v \in V \rightarrow \exists exec (exec \in Exec(G) \wedge exec = (v_s, ..., v, ..., v_e)))  $$
    \item No self loops.
        $$ \forall v (v \in V \rightarrow (v, v) \notin E) $$
\end{itemize}

\noindent \textbf{Input Initialized:}
$$ \forall exec = (v_s, ..., v_e) \in Exec(G) \forall v_i \in exec \forall input \in input(v_i) $$
$$ \exists j (j < i \wedge output(v_j) = input) $$

\noindent \textbf{No Redundancy:}
\begin{itemize}[leftmargin=*]
    \item After a module is visited, it should not be revisited until at least one of its inputs' value has been changed.
        $$ \forall exec = (v_1, ..., v_k) \in Exec(G) $$
        $$ ((\exists 1 \leq i < j \leq k (v_i = v_j)) \rightarrow $$
        $$ (\exists i < l < j (output(v_l) \in input(v_j)))) $$
    \item After a module is visited, its output(s) should be used by a module.
        $$ \forall exec = (v_1, ..., v_k) \in Exec(G) $$
        $$ \forall 1 \leq i \leq k (output \in output(v_i) \rightarrow $$
        $$ (\exists i < j \leq k (output \in input(v_j) \wedge \nexists i < l < j (output \in output(v_l)) ))) $$
\end{itemize}

\noindent \textbf{Involve Labeling:}
$$ \forall exec = (v_s, ..., v_e) \in Exec(G) \exists i (fun(v_i) = interactiveLabeling) $$

\section{Customized Workflow Template}
\label{sec:customized-workflow-template}

Below is the code snippet to declare a customized workflow template for webpage classification.

\begin{lstlisting}[language=TypeScript]
/** The declaration of a minimal webpage classification workflow. */
export default {
  nodes: [
    {
      label: 'initialization',
      type: 'Initialization',
      value: {
        dataType: 'Webpage',
        labelTasks: ['Classification'],
      },
      layout: { x: 40, y: 40 },
    },
    {
      label: 'random sampling',
      type: 'DataObjectSelection',
      value: ['Random', { params: { nBatch: { value: 2 } } }],
      layout: { x: 160, y: 40 },
    },
    {
      label: 'grid matrix',
      type: 'InteractiveLabeling',
      value: ['GridMatrix', {
        params: {
          nRows: { value: 1 },
          nColumns: { value: 2 },
        }
      }],
      layout: { x: 280, y: 40 },
    },
    {
      label: 'check all labeled',
      type: 'StoppageAnalysis',
      value: 'AllLabeled',
      layout: { x: 400, y: 40 },
    },
    {
      label: 'stop?',
      type: 'Decision',
      layout: { x: 400, y: 130 },
    },
    {
      label: 'exit',
      type: 'Exit',
      layout: { x: 410, y: 220 },
    },
  ],
  edges: [
    { source: 'initialization', target: 'random sampling' },
    { source: 'random sampling', target: 'grid matrix' },
    { source: 'grid matrix', target: 'check all labeled' },
    { source: 'check all labeled', target: 'stop?' },
    { source: 'stop?', target: 'exit', condition: true },
    { source: 'stop?', target: 'random sampling', condition: false },
  ],
}
\end{lstlisting}

\section{Customized Algorithm Module}
\label{sec:customized-algorithm-module}

Below is the code snippet to declare a customized default labeling module for span annotation by detecting nouns.
The module is implemented in the form of an algorithm server.

\begin{lstlisting}[language=Python]
# A customized default labeling module
# for span annotation by detecting nouns.
import json
import nltk
import tornado.httpserver
import tornado.ioloop
import tornado.options
import tornado.web
import uuid
from typing import List

# Core function
def predict(data_object: dict) -> List[dict]:
    """ Create default spans by detecting nouns. """
    sentence = data_object['content']
    pos_tag = nltk.pos_tag(sentence.split())
    start = 0
    end = 0
    spans = []
    for i, (segment, tag) in enumerate(pos_tag):
        start = end + 1 if i != 0 else end
        end = start + len(segment)
        # filter tags that are not nouns
        if tag not in ['NN', 'NNP', 'NNS']:
            continue
        spans.append({
            'text': segment,
            'start': start,
            'end': end,
            'category': 'subject',
            'uuid': str(uuid.uuid4()),
        })
    return spans

class DefaultLabelingHandler(tornado.web.RequestHandler):
    def post(self):
        self.set_header('Access-Control-Allow-Origin', '*')
        data_objects = json.loads(self.request.body)['dataObjects']
        labels = [{ 'spans': predict(d) } for d in data_objects]
        self.write({ 'labels': labels })

def main():
    tornado.options.parse_command_line()
    http_server = tornado.httpserver.HTTPServer(
        tornado.web.Application(
            handlers=[(r'/defaultLabels', DefaultLabelingHandler)],
            debug=True,
        )
    )
    http_server.listen(8007)
    print('Serve at http://localhost:8007/defaultLabels')
    tornado.ioloop.IOLoop.instance().start()

if __name__ == "__main__":
    main()
\end{lstlisting}

\pagebreak

\section{User Study Instructions}
\label{sec:supp-study}

The following are the instructions we gave the participants for the 4 labeling tool development tasks.

\subsection{Task 1 (Image Segmentation)}

Please create a labeling tool for image segmentation according to the following specification of the labeling tool's functionality.
Please do not use OneLabeler's built-in templates for this task.

\noindent Specification:
\begin{enumerate}[leftmargin=*]
    \item In the image segmentation tool, data objects are iteratively selected for annotators to label.
        1 data object is randomly selected each time.
    \item The interface displays the 1 data object in a single object display.
    \item After annotating the 1 data object, the system checks whether all the data objects are labeled.
    \item If all the data objects are labeled, the labeling STOPS.
    \item Otherwise, goto step 1.
\end{enumerate}

\noindent Please tell the experimenter when you have finished.

\subsection{Task 2 (Image Classification)}

Based on the previous image segmentation workflow in Task 1, please modify the workflow to create another labeling tool for image classification according to the following specification of the labeling tool's functionality.

\noindent Specification:
\begin{enumerate}[leftmargin=*]
    \item In the image classification tool, data objects are iteratively selected for annotators to label.
        16 data objects are randomly selected each time.
    \item The interface displays the 16 data objects in a grid matrix with 4 rows and 4 columns.
    \item After annotating the 16 data objects, the system checks whether all the data objects are labeled.
    \item If all the data objects are labeled, the labeling STOPS.
    \item Otherwise, goto step 1.
\end{enumerate}

\noindent Please tell the experimenter when you have finished.

\subsection{Task 3 (Machine-aided Image Classification)}

Based on the previous image classification workflow in Task 2, please modify the workflow to create another labeling tool for image classification according to the following specification of the labeling tool's functionality.

\noindent Specification:
\begin{enumerate}[leftmargin=*]
    \item At the beginning, SVD features are extracted for the data objects.
    \item Then, data objects are iteratively selected for annotators to label.
        16 data objects are randomly selected each time.
    \item The selected data objects are then assigned default labels by a decision tree classifier.
    \item The interface displays the 16 data objects in a grid matrix with 4 rows and 4 columns.
    \item After annotating the 16 data objects, the system checks whether all the data objects are labeled.
    \item If all the data objects are labeled, the labeling STOPS.
    \item Otherwise, the decision tree classifier used for default labeling is updated by retraining.
    \item Then, goto step 2.
\end{enumerate}

\noindent Please tell the experimenter when you have finished.

\subsection{Task 4 - Option 1 (Customize Data Type)}

Based on the given image classification workflow template, please modify the workflow to create another labeling tool for webpage classification according to the following specification of the labeling tool's functionality.

\noindent Specification:
\begin{enumerate}[leftmargin=*]
    \item In the webpage classification tool, data objects are iteratively selected for annotators to label.
        2 data objects are randomly selected each time.
    \item The interface displays the 2 data objects in a grid matrix with 1 row and 2 columns.
    \item After annotating the 2 data objects, the system checks whether all the data objects are labeled.
    \item If all the data objects are labeled, the labeling STOPS.
    \item Otherwise, goto step 1.
\end{enumerate}

\noindent The webpage data type is not currently supported in OneLabeler.
Before you start to build the labeling tool in the visual programming interface, you will need to first create the new webpage data type with some coding.
To create this new data type, please follow the instructions below.

\begin{enumerate}[leftmargin=*]
    \item Enter the ``./client'' folder of OneLabeler's source code.
    \item Execute ``npm run customize'' in the command-line, and choose suitable options in the command-line interface to create template code files for the webpage data type.
    \item Revise template code files to create the webpage data type.
\end{enumerate}

\noindent The webpage data type should satisfy this specification:

When using a grid matrix to display webpage data objects, each grid should display the webpage content.
Each webpage data object should fill the grid that contains it.

\noindent Additional Tips:

\begin{itemize}[leftmargin=*]
    \item You may first read the ``README.md'' file in the created template code files, which gives you an overview of what the template code files are.
    \item When editing the code, you may refer to the HTML iframe tag API (https://developer.mozilla.org/docs/Web/HTML/Element/iframe) if you are unfamiliar with it.
    \item In this task, you will not need to edit any code files other than the files created by the ``npm run customize'' command.
    \item You may refer to the ``Customization'' section of the documentation website.
    \item If you do not have web development experience, you may ask the experimenter to help you with the programming part.
\end{itemize}

\noindent Please tell the experimenter when you have finished.

\subsection{Task 4 - Option 2 (Reproduce Your Tool)}

Please use OneLabeler to reproduce your labeling tool.
The reproduction should have similar major functionalities as your labeling tool but does not need to be the same.
For this task, you may ask the experimenter for help.

\end{document}